%% file: main_bayes.tex
\RequirePackage{etoolbox}
\csdef{input@path}{%
 {sty/}
 {img/}
}%
\csgdef{bibdir}{bib/}

\documentclass[ba,preprint]{imsart}
\pubyear{0000}
\volume{00}
\issue{0}
\doi{0000}
\firstpage{1}
\lastpage{1}

\usepackage{amsthm}
\usepackage{amsfonts}
\usepackage{amsmath}
\usepackage{float}
\usepackage[table]{xcolor}
\newcolumntype{s}{>{\columncolor[HTML]{DDE1E0}} p{1.8cm}}
\usepackage{natbib}
\usepackage[colorlinks,citecolor=blue,urlcolor=blue,filecolor=blue,backref=page]{hyperref}
\usepackage{graphicx}
\usepackage{algorithm2e}

\usepackage{booktabs}       
\usepackage[title]{appendix}
\usepackage{nicefrac}       
\usepackage{microtype}      
\usepackage{wrapfig}
\usepackage{multirow}

\usepackage{enumerate}
\usepackage[font=small,labelfont=bf,tableposition=top]{caption}
\usepackage[font=footnotesize]{subcaption}
\usepackage{algorithmic}
\usepackage{bbm}

\startlocaldefs
\newcommand{\xhdr}[1]{\vspace{0.1mm}\noindent{{\bf #1.}}}
\endlocaldefs

\begin{document}


\begin{frontmatter}
\title{Constrained Bayesian ICA for Brain Connectome Inference}

\runtitle{Constrained Bayesian ICA for Brain Connectome Inference}

\begin{aug}

\author{\fnms{Claire} \snm{Donnat}\thanksref{addr1}\ead[label=e1]{cdonnat@stanford.edu}},
\author{\fnms{Leonardo} \snm{Tozzi}\thanksref{addr2}\ead[label=e2]{ltozzi@stanford.edu}}
\and
\author{\fnms{Susan} \snm{Holmes}\thanksref{addr1}\ead[label=e3]{susan@stat.stanford.edu}}

\runauthor{Donnat et a.}

 \address[addr1]{Department of Statistics, Stanford University,
 390 Jane Stanford Way,
 Stanford CA 94305, USA
cdonnat@stanford.edu; susan@stat.stanford.edu
 }
 \address[addr2]{Department of Psychiatry and Behavioral Sciences, Stanford University,
 401 Quarry Road,
 Stanford CA 94305, USA
ltozzi@stanford.edu. 
 }

\end{aug}

\begin{abstract}
\input{abstract}
\end{abstract}


\end{frontmatter}


%
%
	\section{Introduction}\label{sec:intro}
	\input{introduction}

	\section{A Bayesian ICA model}\label{sec:bayesian_ica}
	\input{model}

	\section{Synthetic Experiments} \label{sec:exp}
	\input{synthetic_exp}

	\section{Results on fMRI study} \label{sec:reallife}
	\input{reallife_experiments.tex}

	\section{Conclusion} \label{sec:conclusion}
	\input{conclusion}
	




%

\bibliographystyle{ba}

	\appendix
	\appendixpage
    \addappheadtotoc
	\section{Numerical approximation}\label{appendix:numerical} 
	\vspace{-0.1cm}
	\input{EM_full}


    \addappheadtotoc
	\section{fMRI-Studies}\label{appendix:scan} 
	\vspace{-0.1cm}
	\input{appendix_scan.tex}

\end{document}

%% file: abstract.tex

Brain connectomics is a developing field in neurosciences which strives to understand cognitive processes and psychiatric diseases through the analysis of interactions between brain regions. 
However, in the high-dimensional, low-sample, and noisy regimes that typically characterize fMRI data, the recovery of such interactions remains an ongoing challenge: how can we discover patterns of co-activity between brain regions that could then be associated to cognitive processes or psychiatric disorders?
In this paper, we investigate a constrained Bayesian ICA approach  which, in comparison to current methods, simultaneously allows (a) the flexible integration of multiple sources of information (fMRI, DWI, anatomical, etc.), (b) an automatic and  parameter-free selection of the appropriate sparsity level and number of connected submodules and (c) the provision of estimates on the uncertainty of the recovered interactions. 
Our experiments, both on synthetic and real-life data, validate the flexibility of our method and highlight the benefits of integrating anatomical information for connectome inference.

%% file: introduction.tex

\xhdr{Motivation} Over the recent years, the study of brain connectomics \citep{bullmore2009complex, fornito2015connectomics,  fornito2016fundamentals} has gained increased interest amidst the neuroscience and cognitive psychology communities. In this framework, the brain is modeled as a graph in which nodes denote voxels or regions of interest (ROIs), while edges represent some notion of functional connectivity, typically inferred from fMRI scans. Examples of such connectivity measures include Pearson correlations, partial correlations or mutual information between the Blood Oxygenation Level Dependent (BOLD) signals of each region of interest. The driving hypothesis behind brain connectomics is that the identification of interactions between modules of nodes is crucial to our understanding of cognitive processes and psychological diseases.  Central to the field is the analysis of resting-state functional MRI (rs-fMRI), believed to capture the brain's default activity \citep{fornito2010can,greicius2003functional,raichle2001default,raichle2007default} and the diverse functional interactions between ROIs. Mathematically speaking, the recovery of these interactions boils down to an estimation of the correlation (or precision) matrix between brain regions.  Yet, in addition to the intensive pre-processing that fMRI data typically requires (denoising, unringing, motion correction, transformation to a standard template space, etc), in most studies, the number of potential edges is significantly greater than the number of time points in the BOLD time series---thus compelling a heavy filtering of the sample correlation matrix to get rid of spurious correlations.
The variability between sessions and individuals further hinders the generalizability of the analysis and complicates the recovery of precise estimates of these interactions. 

Meanwhile, structural connectivity has been a subject of increased interest in the brain connectomics literature over the past few years. Structural connectivity reflects the physical wiring of the brain by inferring white matter tracts from Diffusion Weighted Images (DWI). Since structural connectivity is widely believed to restrict functional connectivity \citep{damoiseaux2009greater,honey2009predicting,ombao2016handbook,sui2012review,uludaug2014general}, the incorporation of DWI data as supplementary information seems like an appealing way of making the estimation of functional connectivity more robust. However, structural data does not come without its fair share of caveats---thus perhaps explaining why more emphasis has been given to structural connectivity to guide (rather than strictly constrain) functional analysis. In particular, several studies have shown that, while structural connectomes typically did not contain many false negative edges \citep{honey2009predicting,rubinov2010complex}---in other words, we can assume that all the long-range anatomical connections are recovered,---they might however contain many phantom tracts.
Moreover, by design of the recovery process, longer-range connection weights are typically inflated compared to their shorter counterparts, thus yielding additional uncertainty in the actual significance of the strength of the recovered structural connections. 
In spite of these drawbacks, the joint use of different imaging modalities could nonetheless allow an increased prediction accuracy in the study of clinical or behavioral outcomes, as well as a deeper understanding of the interplay between neurophysiological signals and cognitive processes. This new multimodal setting calls for the development of new statistical methods \citep{ombao2016handbook}  tailored to the integration of these sources of information. \\

\xhdr{Prior Work: Single-modality methods} There is an extremely rich literature on single modality methods for the estimation of functional connectivity and subsequent network analyses. In this setting, only fMRI BOLD time series are considered, and the inference boils down to a filtering of the sample correlation matrix to detect significant interactions between brain regions and limit the number of false discoveries --- a setting which thus transcends the field of brain connectomics. In this broader setting, in order to ``filter" the correlation matrix, practioners have typically chosen one of three mutually exclusive paths. \\
\textit{Shrinkage-based methods.} When it comes to correlation estimation, the first of these paths consists in using shrinkage methods \citep{chen2011robust,couillet2014large,efron1976multivariate,kourtis2012parameter,ledoit2012nonlinear,schafer2005shrinkage}  and Random Matrix Theory \citep{bai2010spectral,yao2015sample} to correct for the spectrum deviations due to the data's high dimensional regime. However, these approaches are typically completely agnostic to either anatomical constraints or sparsity assumptions --- perhaps explaining why such approaches are seldom used in fMRI studies. \\
\textit{Threshold-based methods.} Another popular branch of analysis uses thresholding as a way to constrain and recover estimates of the inverse covariance matrix \citep{bickel2008covariance,bien2011sparse,cai2011adaptive,friedman2008sparse,scheinberg2010sparse,zhang2018large}. Such estimates are typically more organically aligned with our prior assumptions on structural connectivity, as we only expect a subset of regions to be actively involved in a given cognitive process. Yet, while the regularization and inclusion of a prior on the sparsity in the estimation of the connectome is a useful --- if not necessary--- starting point, all of these methods remain agnostic to the underlying anatomy of the brain. When applied to brain connectomics, such threshold-based approaches exhibit several deficiencies:  
\vspace{-0.2cm}
\begin{enumerate}
  \setlength\itemsep{0.01cm}
	\item They rely on an arbitrary definition of the cutoff threshold. This is usually achieved by controlling for the level of sparsity of the recovered graph, which has itself to be selected---thus introducing additional sources of variability in the analysis. 
	\item They provide no way of accommodating for the uncertainty of the edge weights. 
	\item Mainly, these methods do not allow the inclusion of complementary sources of information, such as DWI data. 
\end{enumerate}
\vspace{-0.2cm}
\textit{Factor Analysis} The last popular pipeline for filtering the sample correlation matrix in spite of the high-dimensionality of the data consists in using a flavor of factor analysis. Broadly speaking, factor analysis has been suggested as a useful way to ``clean up'' correlation matrices in a number of of applications ranging from neuroscience to empirical finance, where it has been shown to accurately and efficiently recover precision matrix \citep{bun2017cleaning}. However, in the case of brain connectomics, we have no a priori model for the factors, and $K$ must thus be inferred  from the data. In neuroscience in particular, Independent Component Analysis (ICA) has been a long-time favorite amongst factor methods, as it allows the discovery of sparse interacting regions of the brain \citep{daubechies2009independent}. In its most widely adopted form, ``Vanilla'' ICA \citep{hyvarinen1999fast,hyvarinen2000independent,mackay1996maximum} is a matrix decomposition technique which utilizes the non-Gaussianity of the distribution of the multivariate time series $Y \in \mathbb{R}^{T \times N}$ to decompose it as a linear combination of non-Gaussian sources:
 \begin{equation}\label{eq:ICA1}
 \setlength\abovedisplayskip{1pt}
 \setlength\belowdisplayskip{0pt}
 	  	Y = SA 
 \end{equation}
where $S \in \mathbb{R}^{T \times K}$ denotes the time series associated to each independent component (assumed to be centered and  non-Gaussian, such that $\mathbb{E}[\frac{S^TS}{T}] =I$) and $A \in \mathbb{R}^{K \times N}$, their corresponding loadings. Each loading $A_{k\cdot}$ can thus be associated to a subnetwork of co-activated nodes. In other words, this assumes that the time series observed at each node $i$ is a linear combination of independent cognitive processes (each associated to a given source). The coefficient $|A_{ki}|$ thus quantify how much node $i$ responds to the activation of source $k$, so that when $|A_{ki}|$ is high, we say that node $i$ is activated by source $k$. The study of the components thus allows to identify regions of the graph that are ``co-activated" by the same source and that thus potentially contribute to the same cognitive process.
Bayesian models for ICA have been proposed in the past, but to the best of our knowledge \citep{chan2003variational,choudrey2003variational,valpola2000fast}, such models have seldom been used in conjunction with structural information to analyze resting-state fMRI data \citep{beckmann2004probabilistic,groves2009bayesian,groves2011linked,kim2008hybrid,li2011large}. \\


\xhdr{Prior work: Multimodal methods} Recently, there has been a surge of ``multi-modal'' methods \citep{ombao2016handbook,xue2015multimodal} for analyzing fMRI data. Most of these methods define probability models that allow the integration of the anatomical information to guide connectome inference. Another set of methods favors joint ICA \citep{franco2008multimodal,ombao2016handbook,sui2011discriminating}, where a traditional ICA model is fit on horizontally concatenated structural and functional data. However, to the best of our knowledge, these methods make at most one assumption for the prior of the data (i.e, selects from either structural or sparsity constraints), and none of these methods enforce the recovery of connected components.\\

 \xhdr{Objectives of the paper} We propose a Bayesian Independent Component Analysis-based model (Eq. \ref{eq:ICA1}) of the covariance matrix that leads to the inference robust functional connectomes. Using DWI structural information to build the prior on the ICA components, we devise a sparse Bayesian hierarchical model that permits the automatic selection of the appropriate number of components and sparsity level through the use of an Automatic Relevance Determination (ARD) prior. In contrast to existing methods, our approach simultaneously leverages  the following natural assumptions on the data:
 \begin{itemize}
  \setlength\abovedisplayskip{1pt}
 \setlength\belowdisplayskip{0pt}
  \setlength\itemsep{1pt}
 	\item \textbf{Hypothesis 1:} The loadings are assumed to be sparse: $||A||_0 \leq s$, where $s$ denotes a sparsity level. From a biological perspective, this is equivalent to assuming that only a few processes over a subset of brain regions are involved in the activity captured by each fMRI scan.
 	\item \textbf{Hypothesis 2:}  Each loading (or component) $A_{k \cdot}$ is assumed to correspond to an anatomically-connected subset of nodes.  
 \end{itemize}
 Under these constraints, our formulation of Bayesian ICA simultaneously achieves (a) the flexible integration of multiple sources of information (fMRI, DWI, anatomical, etc.), (b) an automatic and  parameter-free selection of the appropriate sparsity level and number of connected submodules and (c) the provision of estimates on the uncertainty of the recovered interactions.
 




%% file: model.tex

To fulfill these objectives, we build upon the standard Bayesian ICA model \citep{beckmann2004probabilistic,choudrey2001variational,roberts2004bayesian}. We assume that $Y$ comes from a noisy mixture of independent components, that is:
\begin{equation}\label{eq:ICA_model}
Y =SA +\epsilon \hspace{0.5cm} \text{with } \quad \mathbb{E}[Y]=0,\quad \mathbb{V}\text{ar}[Y]=1 
\end{equation}
Note that we assume here that, following standard ICA procedure, the data $Y$ has been centered and scaled. This is also consistent with the bulk of the connectome literature, which typically focuses on studying correlations --- rather than covariances --- between Regions of Interest (ROIs).

Eq. \ref{eq:ICA1} induces the following correlation structure on the time series:
\vspace{0.2cm}

\begin{equation}\label{eq:ICA_cov}
 \setlength\abovedisplayskip{1pt}
 \setlength\belowdisplayskip{0pt}
\begin{split}
\epsilon &\sim N \big(\textbf{0}, \text{Diag}(\gamma)\big)\\
\frac{1}{T}\mathbb{E}[Y^TY] &=     \frac{1}{T}\mathbb{E}[A^TS^TSA] + \frac{1}{T} \text{Diag}(T \gamma) =     \mathbb{E}\big[\mathbb{E}[A^T\frac{S^TS}{T}A|A]\big] + \text{Diag}(\gamma) \\ 
 &=     \mathbb{E}\big[ A^TA \big] + \text{Diag}(\gamma), \hspace{1cm} \text{assuming } \frac{1}{T}\mathbb{E}[S^TS] = 1 \\ 
\end{split}
\end{equation}
\vspace{0.2cm}

Let us now turn to the connectedness and sparsity assumptions which constitute the bulk of our contribution. 
These additional modelling assumptions are closely aligned with the biology and guide the selection of our priors on $A$:
\begin{enumerate}
    \item \textbf{Sparsity}: each component $A_k$ should be non-zero on a small subset of nodes. Indeed, as explained in the introduction, $A_{k\cdot}$ can be interpreted as a ``cognitive" subnetwork, as the magnitude of the coefficients in $A_{k\cdot}$ indicate which node is activated by source $k$. From a biological perspective, we expect specific cognitive processes to involve only a fraction of the brain. From a mathematical viewpoint, ICA relies on the non-Gaussianity of the timeseries $Y$. Since the number of components $K$  here can be arbitrarily large and since each timepoint $Y_{ti}$ is a mixture of coefficients with finite fourth moment, by virtue of the central limit theorem, our non-Gaussian mixture can only involve a few components. 
    \item \textbf{Connectedness:} each component $A_k$ should span a subset of anatomically connected nodes. The connectedness assumption also stems from biogical considerations: the activation of the different ROIs originates from neurophysiological signals which travel along the white-matter tracts in the brain. As such, it is natural to assume that cognitive processes activate pathways, rather than disconnected sets of nodes.
    \item \textbf{Non-negativity:}  The components $A$ are assumed to be non-negative. From a biological viewpoint, this assumes that we are only considering positive excitation mechanisms (and discarding inhibitory ones). From a modeling perspective, this also alleviates some of identifiability issue of  the traditional ICA components, which are only determined up to a sign flip.
\end{enumerate}
To meet criteria (1-3), we model the components $A_k$ as independent multivariate Gaussians with a carefully selected precision matrix. Let $L = D -W$ be the graph's combinatorial Laplacian (where $D$ is a diagonal matrix in which each diagonal entry $D_{ii}$ is the degree of node $i$, and $W$ is the graph's adjacency matrix). We define the regularized Laplacian $L_{\alpha}$ as: $L_{\alpha} = L + \alpha I$. Here, $\alpha$ is a small regularization coefficient which ensures $L_{\alpha}$ to be positive semi definite. With these notations, we propose the following generative mechanism for the components:

\begin{equation}
    \forall k\leq K, \quad A_{k \cdot} \sim   \Big|  \mathcal{N}\Big(0, \Sigma_{\alpha} \Big)   \Big|
\end{equation}\label{eq:gaussian_comp2}
where $\Sigma_{\alpha}^{-1}=\Omega_{\alpha} =L_{\alpha}$.   Under these assumptions, the negative log-likelihood of our model thus includes a penalty on a term of the form:
\begin{equation}\label{eq:penalty}.
    \begin{split}
   A_{k \cdot} L_{\alpha}A_{k \cdot}^T &=  \frac{1}{2} \sum_{i}\sum_{j \sim i} \omega_{ij}\Big(A_{ki}- A_{kj} \Big)^2 \\
   &+ \alpha \sum_{i}A_{ki}^2.     
    \end{split}
\end{equation}
This effectively pushes the values of the activations of neighboring nodes to be close and thus enforces the connectivity constraint of assumption (2). While the underlying motivation behind this model is clearly explained with the graph combinatorial Laplacian, we also note in passing that, under this model, any version of the Laplacian (normalized, etc.) can be used in lieu of $L$, with similar effects: the only difference lies in the value of the penalty $\omega_{ij}$ in Eq. \ref{eq:penalty}.\\

The multivariate normal model creates components that are smooth and dense over the graph. To enforce localization, we convolve these components with binary mask $D$, such that:
\begin{equation}
A = \tilde{A} \odot D 
\end{equation}\label{eq:gaussian_comp3}
where $\tilde{A} \sim \Big|  \mathcal{N}\Big(0, \Sigma \Big)   \Big|$ are our multivariate Gaussian components and $D$ is roughly $\{0,1\}$-valued and effectively creates a localization effect.

Let us now turn to the value of the coefficients $D$ and the selection of the number of components. The latter can be inferred using automatic relevance determination if we further decomposing the loadings $A$ as:
\begin{equation}
    A = \Lambda \tilde{A} \hspace{0.5cm} \text{ with } \Lambda = \text{Diag}((\lambda_{i})_{i=1}^K)
\end{equation} where the rows of $\tilde{A}$ are the convolved multivariate Gaussians of Eq. \ref{eq:gaussian_comp3} and each  $\lambda_k$ roughly comes from a mixture with point mass at $0$ so as to effectively ``switch on'' or ``off'' the different components. In this model, the component $A_k$ is  activated if   $\lambda_k$ is significantly greater than 0.\\

Before going any further in our selection of a prior for $D$, let us consider the identifiability of our Bayesian model. As underlined in \cite{choudrey2001variational}, the Bayesian ICA model in Eq. \ref{eq:ICA_model} is  not identifiable as such, as it is invariant to row and column permutations, as well as rotations. To get rid of this permutation invariability, we order the $\lambda_k$s by increasing order. The additional sparsity assumption on the components $A$ also alleviates some of the rotational identifiability issues of our model. In our current formulation of the problem, there remains the problem of the coupling of $D$ and $\Lambda$: a scenario in which $\lambda_k$ is almost 0 amounts the same model if we had instead $\lambda_k>\delta$ but $\forall i, D_{ki} \approx 0$. We thus constrain the columns of the mask $D$ to have fixed norm equal to $1$ in order to avoid any such scaling issues. This motivates modelling $D_{ \cdot i}^2$ as a Dirichlet distribution:
$$ \forall i \in [1, N],\quad D_{\cdot i}^2 \sim \text{Dirichlet}[\frac{1}{K}] \in \mathbb{R}^{K}.$$ 
Indeed, from the modelling perspective, $D_{ki}$  can be regarded as an indicator variable which selects which component node  $i$ is responsive to: $D_{ki}  \neq 0$ if node $i$ is in the component, and $0$ everywhere else. By independently switching ``on" and ``off" the coefficients $A_{ki}$ within each component $k$, we allow the component to be localized on the graph.  Since we are considering voxels' scaled time series, by design, we must have:
$$ \frac{1}{T}\text{diag}(Y^TY) = 1$$
This implies that the variance of each node must be such that:
\begin{equation}\label{eq:var}
\begin{split}
\forall i \in [1, N],\quad  1 &= \frac{1}{T}Y_{\cdot i}^T Y_{\cdot i}= \sum_{k=1}^K\lambda_k^2 D_{ki}^2 A_{ki}^2 \frac{(S^TS)_{kk}}{T}  +\sigma^2 \\
& \approx \sum_{k=1}^K \lambda_k^2 D_{ki}^2\Sigma^{(A)}_{ii} +\sigma^2\\
\end{split}
\end{equation} 
 Each entry $
 \lambda_k^2 D_{ k i}^2$ can thus be understood as a measure of the proportion of the variance attributed to component $k$ in node $i$.  

The full model is summarized in the plate model in Figure \ref{fig:model_summary}.
%
%

\begin{figure}\label{fig:model_summary}
\centering
\begin{subfigure}[b]{.4\textwidth}
\begin{minipage}{\textwidth}
        \begin{equation*}
        \begin{split}
            S_t & \sim \text{Laplace}(0, \frac{1}{\sqrt{2}})\\
           \sigma^2 & \sim 1/\Gamma_{1,1} \\
         \forall k, A_k & \sim  \Big| \mathcal{N}(\mathbf{0}, \Sigma) \Big| \\
              \Sigma&: \text{from structural }\\
              &\text{ information.} \\
            \lambda_k & \sim \text{Mixture}_{\pi_{\Lambda}}(\Gamma_{1,1},\Gamma_{10,10}  )\\
            D^2_k & \sim \text{Dirichlet}(\frac{1}{K} )
        \end{split}
    \end{equation*}
\end{minipage}
\caption{Explicit Bayesian model}\label{eq:model}
\end{subfigure}
\begin{subfigure}[b]{.57\textwidth}\label{eq:model}
\begin{minipage}{\textwidth}\label{eq:model}
\centering
\includegraphics[width=7.6cm,height=5cm]{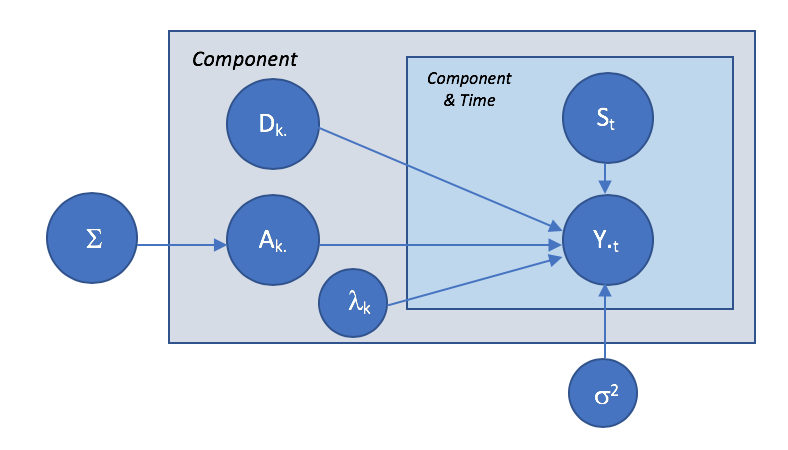}
\end{minipage}
\caption{Full plate model}\label{fig:plate}\label{fig:plate}
\end{subfigure}
\caption{Summary of the Bayesian Model used in this paper.}\label{fig:model_summary}
\end{figure}

\vspace{0.5cm}

\xhdr{Solving the problem} In order to solve model \ref{fig:model_summary}, we propose two different methods:
\begin{itemize}
    \item \textit{Exact inference:} We use the Stan  programming suite to perform MCMC sampling and get both modes and posterior confidence intervals for the distributions \citep{gelman2015stan}\footnote{The code is publicly available at \url{https://github.com/donnate/ConstrainedBayesianICA}}.  Stan uses Hamiltonian Monte Carlo to solve for the different components. This is particularly suited to our problem, which, with a total of roughly $K \times (2\times  N + T +1)$ parameters to be estimated, is high-dimensional and thus difficult to solve using standard Monte Carlo methods. Despite its high-dimensionality, we underline that Bayesian ICA still represents an efficient compression  of the original time series as long as $K<<T$ and $K<<N$. 
   \item \textit{Numerical Approximation:} The MCMC-based procedure is unfortunately difficult to scale up to more than a few hundred nodes. For larger-scale problems, we thus resort to a  numerical approximation to Model \ref{fig:model_summary}. This approximation solves directly for the subnetworks $A$ and is well-suited for fast approximations of numerical point estimates consistent with model \ref{fig:model_summary}. The full derivation of this method is provided in  Appendix \ref{appendix:numerical}.
\end{itemize}

 Finally, further stability and faster convergence of the algorithm are also achieved by warm-starting it in a region of plausible solutions, obtained via numerical optimization.

%% file: synthetic_exp.tex

As a proof of concept, we begin by checking the ability of our method to recover signals over synthetic graphs: the goal of this subsection is to show (a) the ability of our method to recover \textbf{ sparse, localized, connected} components, and (b) to show that the recovered components are indeed more reliable than for any other method.

In order to mimic our underlying assumptions on cognitive brain mechanisms, we generate the data according to the following process:
\begin{enumerate}
    \item \textbf{Step 1: Backbone graph generation:} we begin by creating the structural graph which will dictate the partial correlations between the different parts of the brain.  To do so:
    \begin{enumerate}
        \item  We generate 5 random structural graphs (i.e,  white matter pathways), each representing a separate component (i.,e, a set of co-activated nodes).
        \item We then fill the edges of these subgraphs matrix by uniformly sampling in $[0.3, 0.4]$. This allows the creation of an invertible precision matrix-- that is, the partial correlations -- corresponding to each submodule.
        \item The full graph is created by patching together the adjacency matrices of these subgraphs (some nodes are shared across components, in order to ensure that the final graph is truly fully connected).
    \end{enumerate}
    \item \textbf{Step 2: Creating the loadings:} We sample each component from a folded multivariate normal distribution, where the precision matrix is the partial correlation of each of the submodules from Step 1.
    \item \textbf{Step 3: Sampling the sources:} we then sample $S$ from a Laplace distribution, with scale parameter $\frac{1}{\sqrt{2}}$ (each entry is identically distributed, there is no temporal auto-correlation in the data).
\end{enumerate}

An illustration of one of such synthetic graph and its corresponding five different components is shown in Fig. \ref{fig:little_graphs}. Note that in this setting, the components can overlap and some nodes can be involved in several processes.
\begin{figure}\label{fig:little_graphs}
    \centering
    \includegraphics[width=\textwidth]{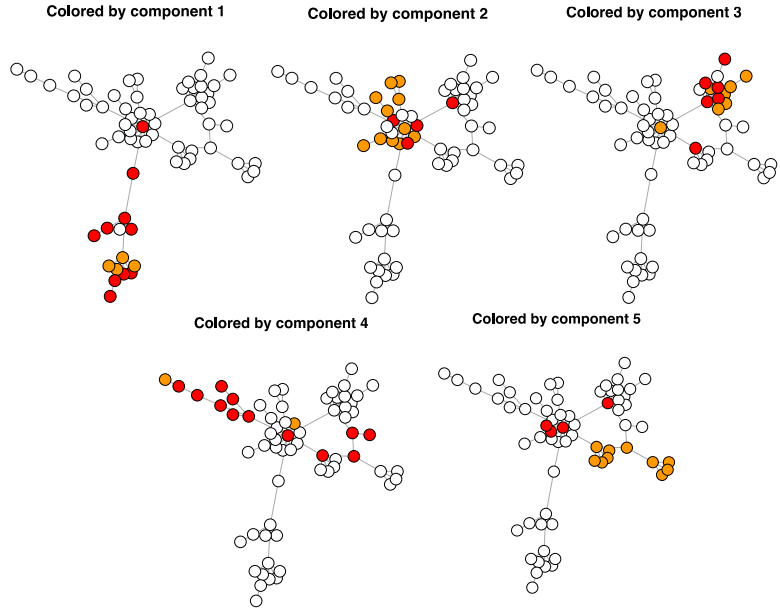}
    \caption{Examples of independent sparse, localized loadings over the graphs.}
    \label{fig:little_graphs}
\end{figure}

\xhdr{Model Analysis} We begin by testing the ability of our model to recover five different components which are consistent with the desired specifications (hypotheses 1-3): sparse, localized and connected. Fig. \ref{fig:single_graph} displays the graphs that are recovered for a low-noise level ($\sigma =0.01$), which we use to start our discussion.

\begin{figure}\label{fig:single_graph}
\begin{subfigure}{0.3\textwidth}
    \centering
    \includegraphics[width=\textwidth, height=4cm]{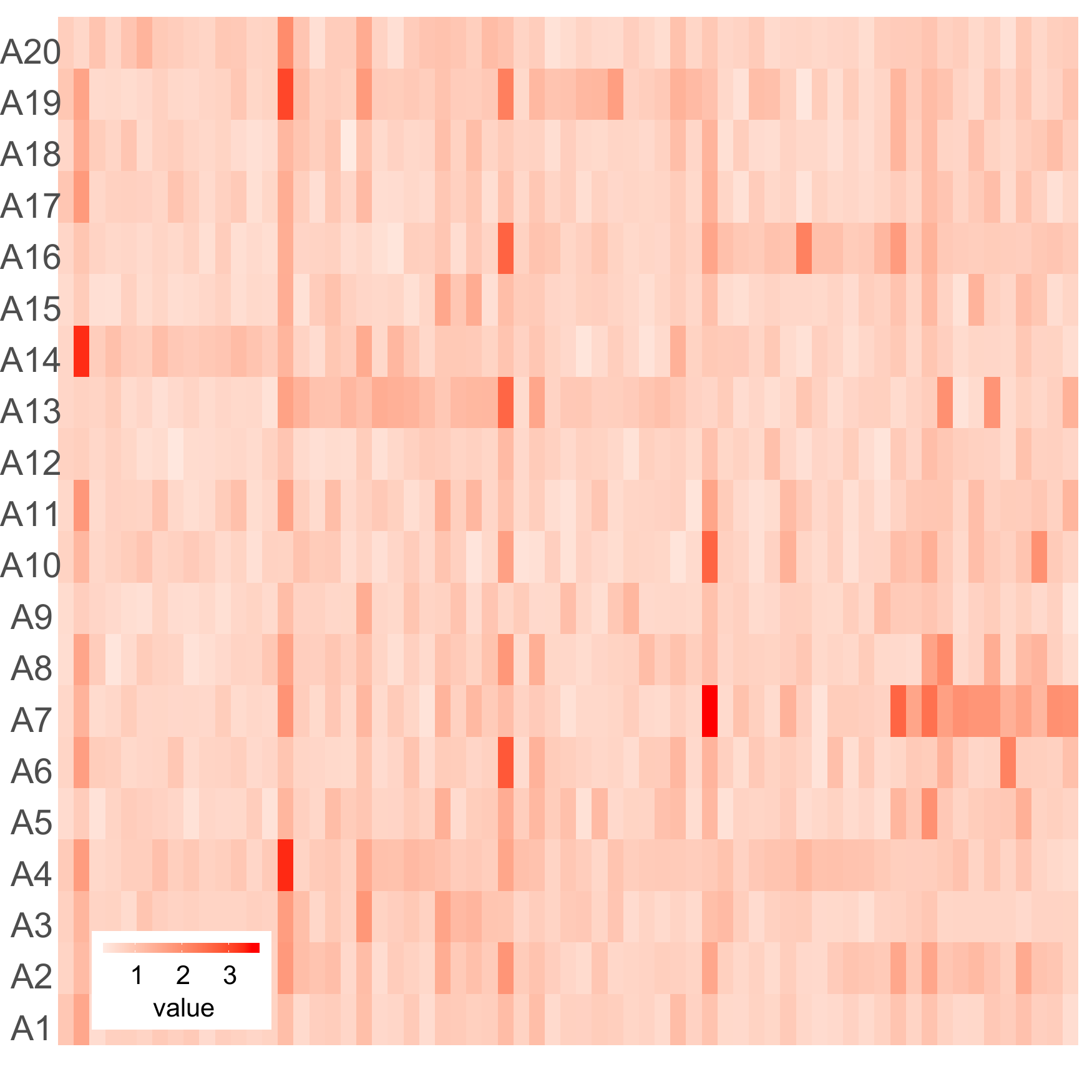}
    \caption{Components A}
    \label{fig:single_graph:A}
\end{subfigure}
\begin{subfigure}{0.3\textwidth}
    \centering
    \includegraphics[width=\textwidth,height=4cm]{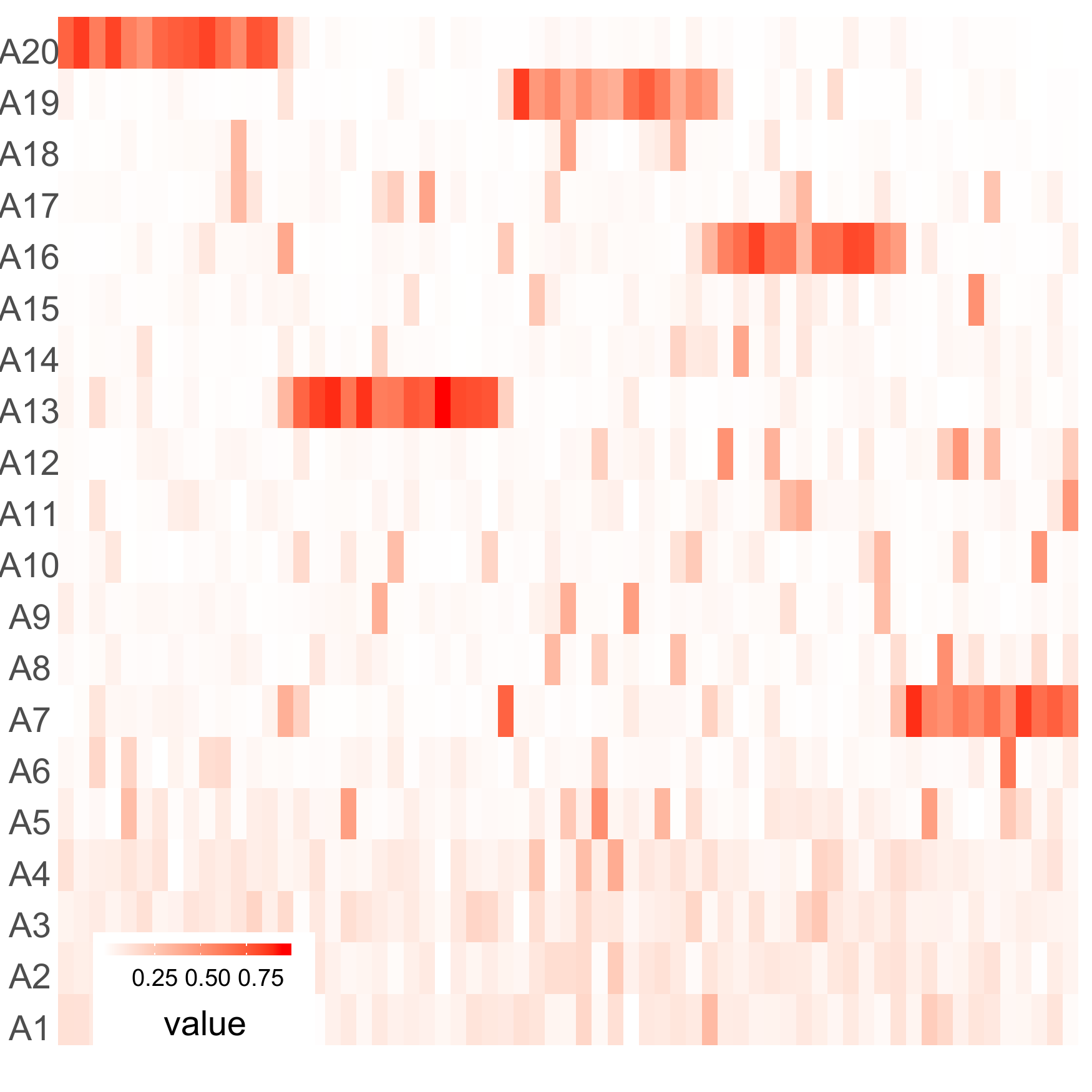}
    \caption{Components D}
    \label{fig:single_graph:D}
\end{subfigure}
\begin{subfigure}{0.3\textwidth}
    \centering
    \includegraphics[width=\textwidth,height=4cm]{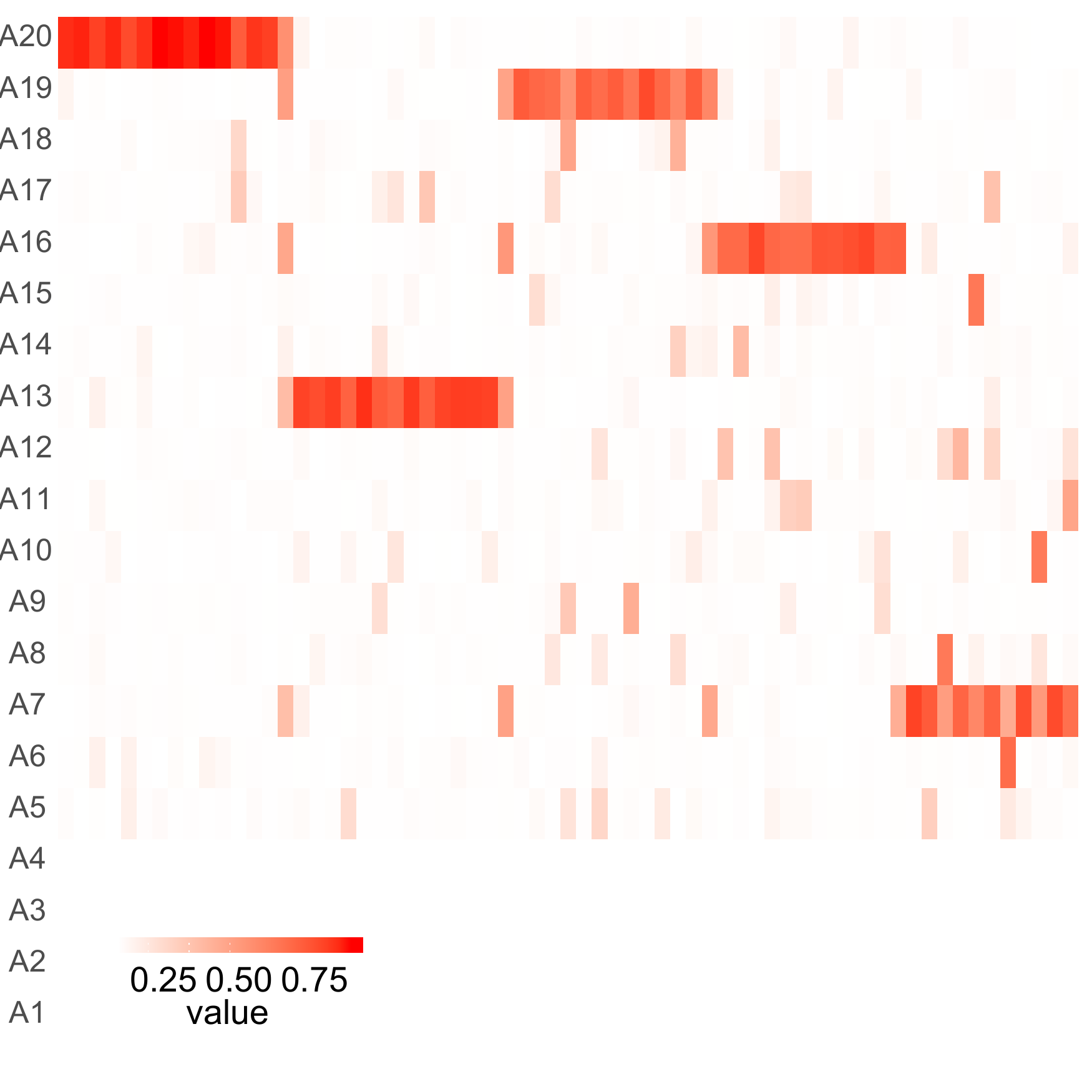}
    \caption{Full Loadings}
    \label{fig:single_graph:L}
\end{subfigure}

\begin{subfigure}[t]{0.49\textwidth}
    \centering
    \includegraphics[width=\textwidth,height=4cm]{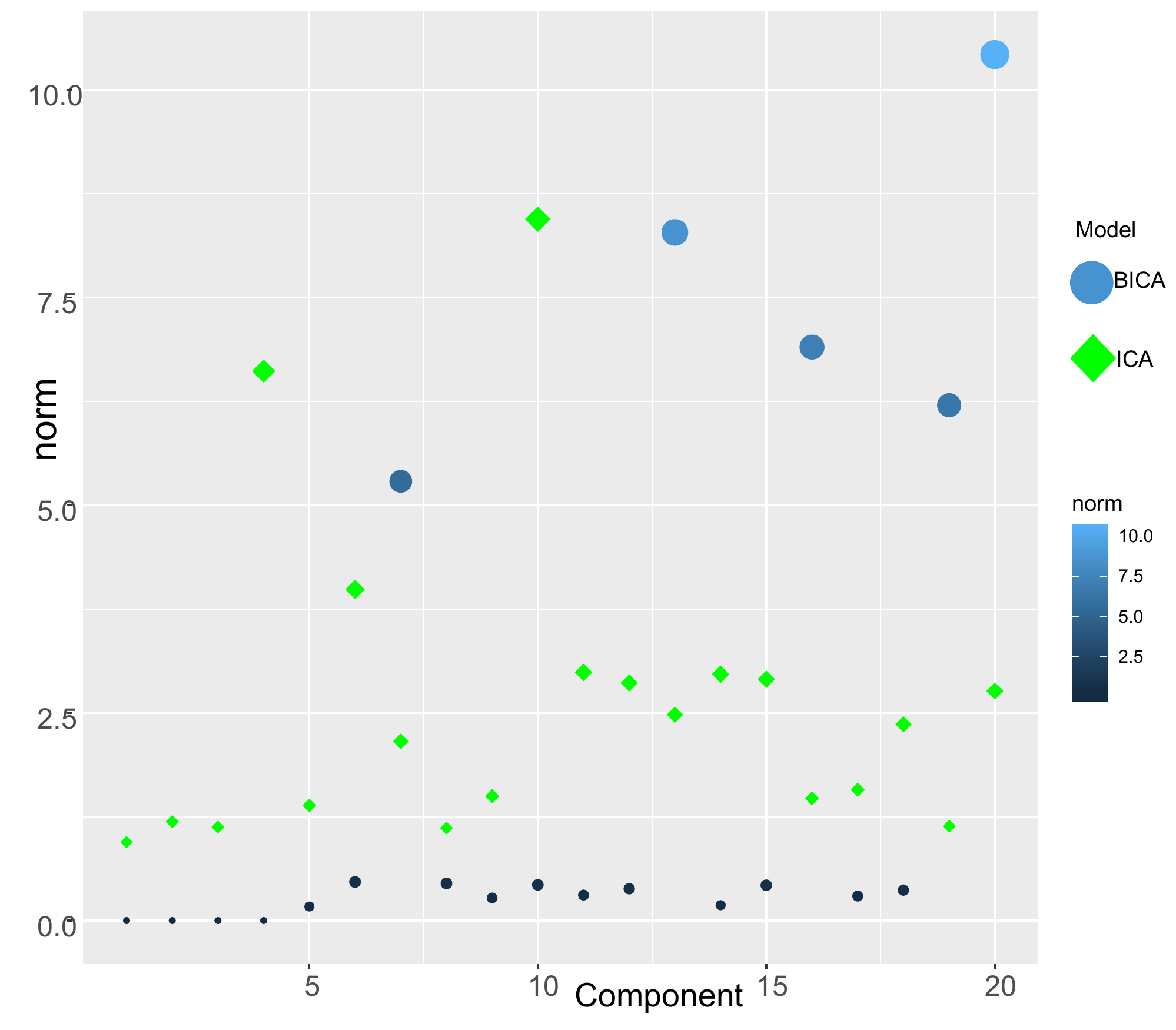}
    \caption{Norm of the recovered coefficients}
    \label{fig:single_graph:histl}
\end{subfigure}
\begin{subfigure}[t]{0.49\textwidth}
    \centering
    \includegraphics[width=\textwidth,height=4cm]{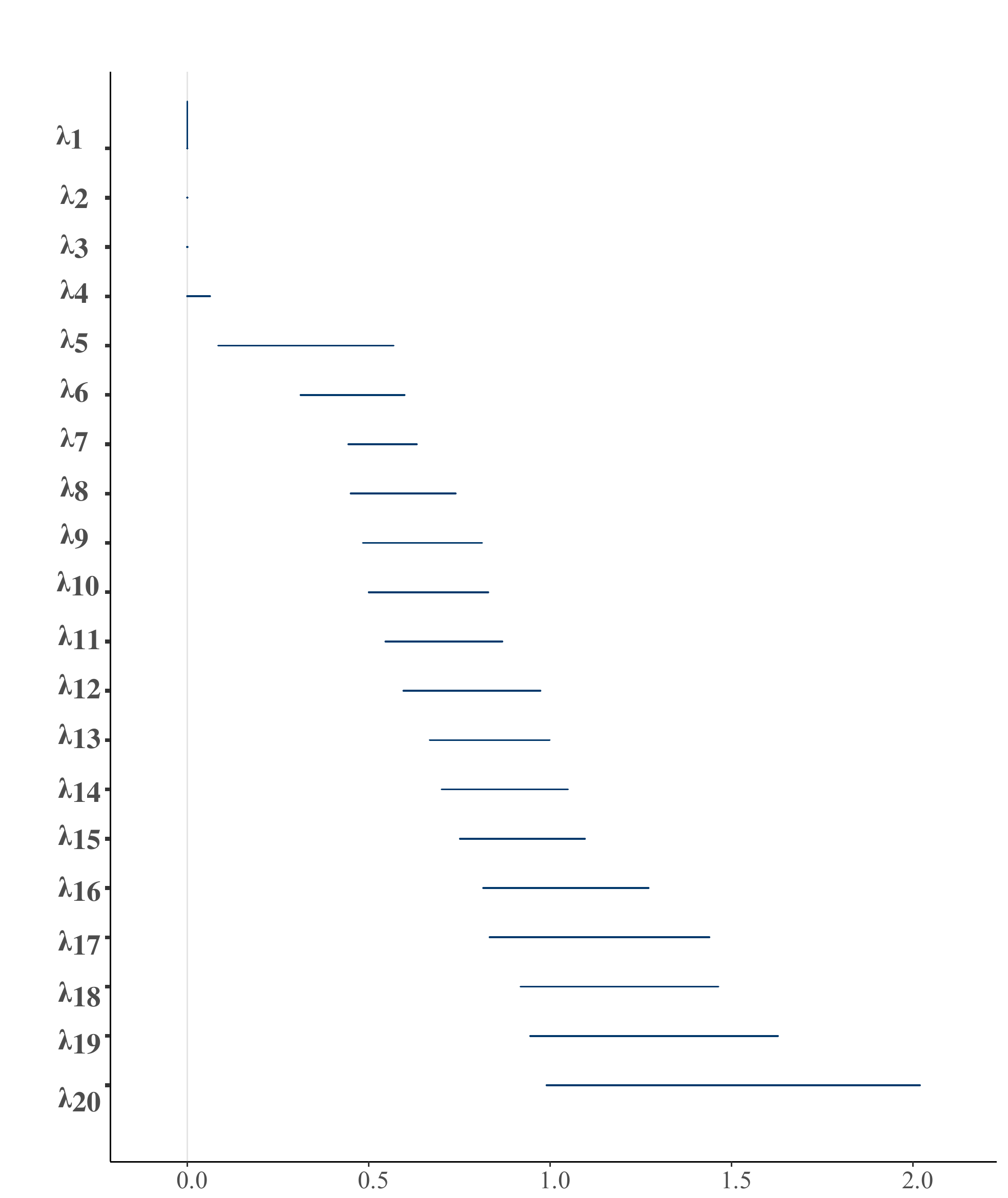}
    \caption{Posterior distribution for the $\lambda_k$ with median and 80\% intervals.}
    \label{fig:single_graph:credl}
\end{subfigure}

\begin{subfigure}[t]{0.49\textwidth}
    \centering
    \includegraphics[width=\textwidth, height=5cm]{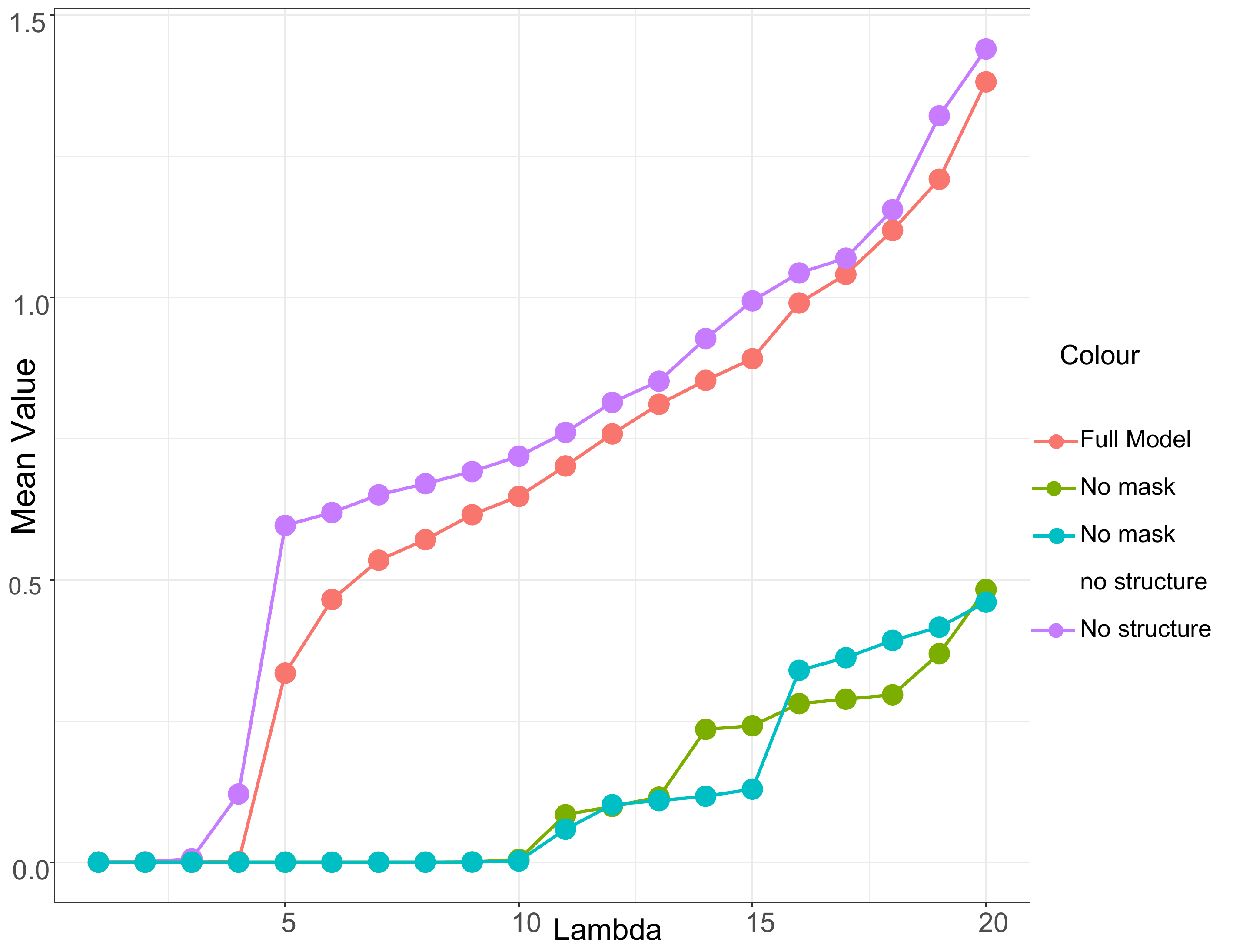}
    \caption{Recovered $\lambda$s}
    \label{fig:single_graph:lambd}
\end{subfigure}
\begin{subfigure}[t]{0.49\textwidth}
    \centering
    \includegraphics[width=\textwidth,height=5.5cm]{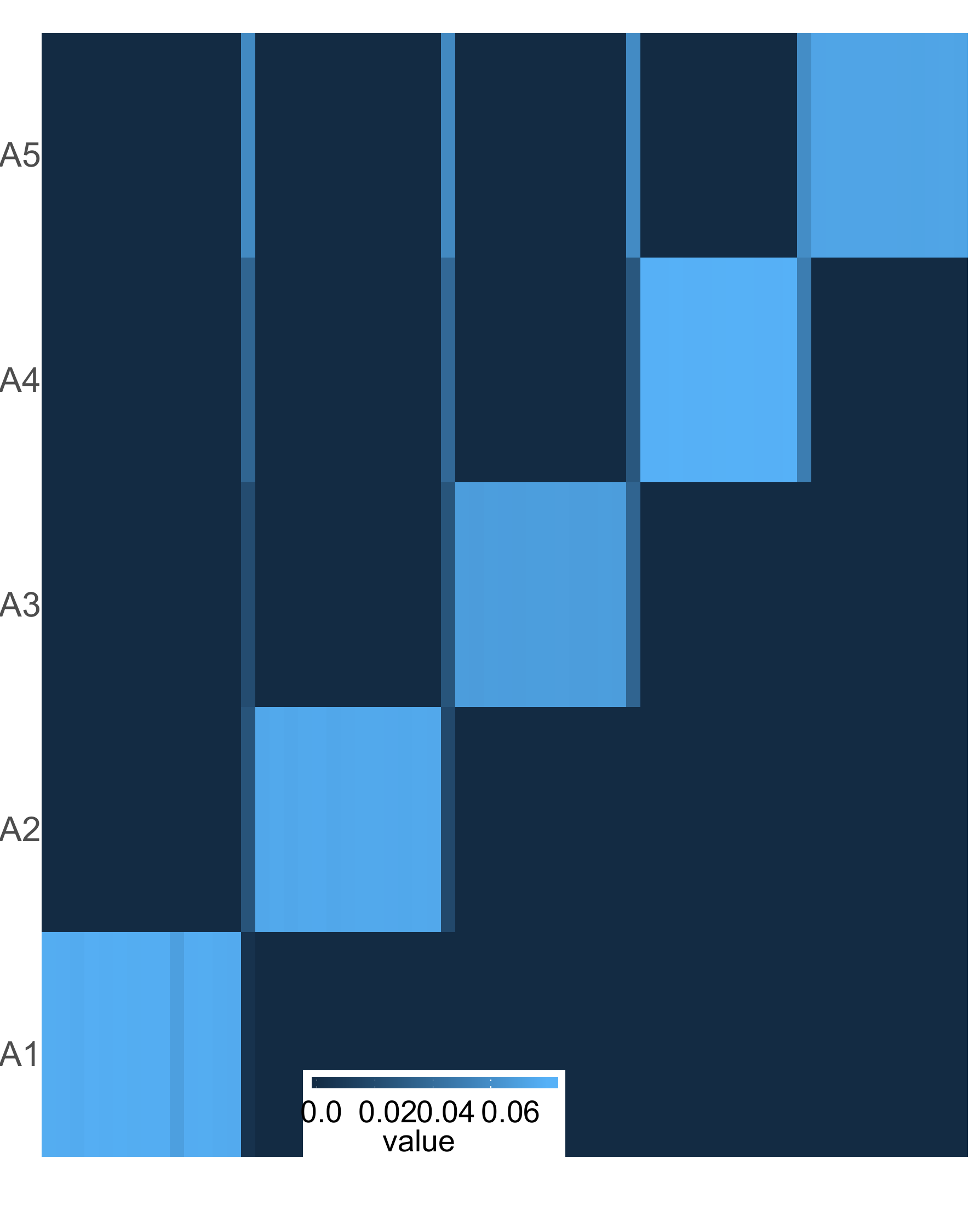}
    \caption{Ground truth components}
    \label{fig:single_graph:gt}
\end{subfigure}

\caption{Results for the Bayesian model on the recovery of components on a single graph} \label{fig:single_graph}
\end{figure}

\begin{figure}\label{fig:single_graph2}
\begin{subfigure}[t]{0.49\textwidth}
    \centering
    \includegraphics[width=\textwidth,height=4.5cm]{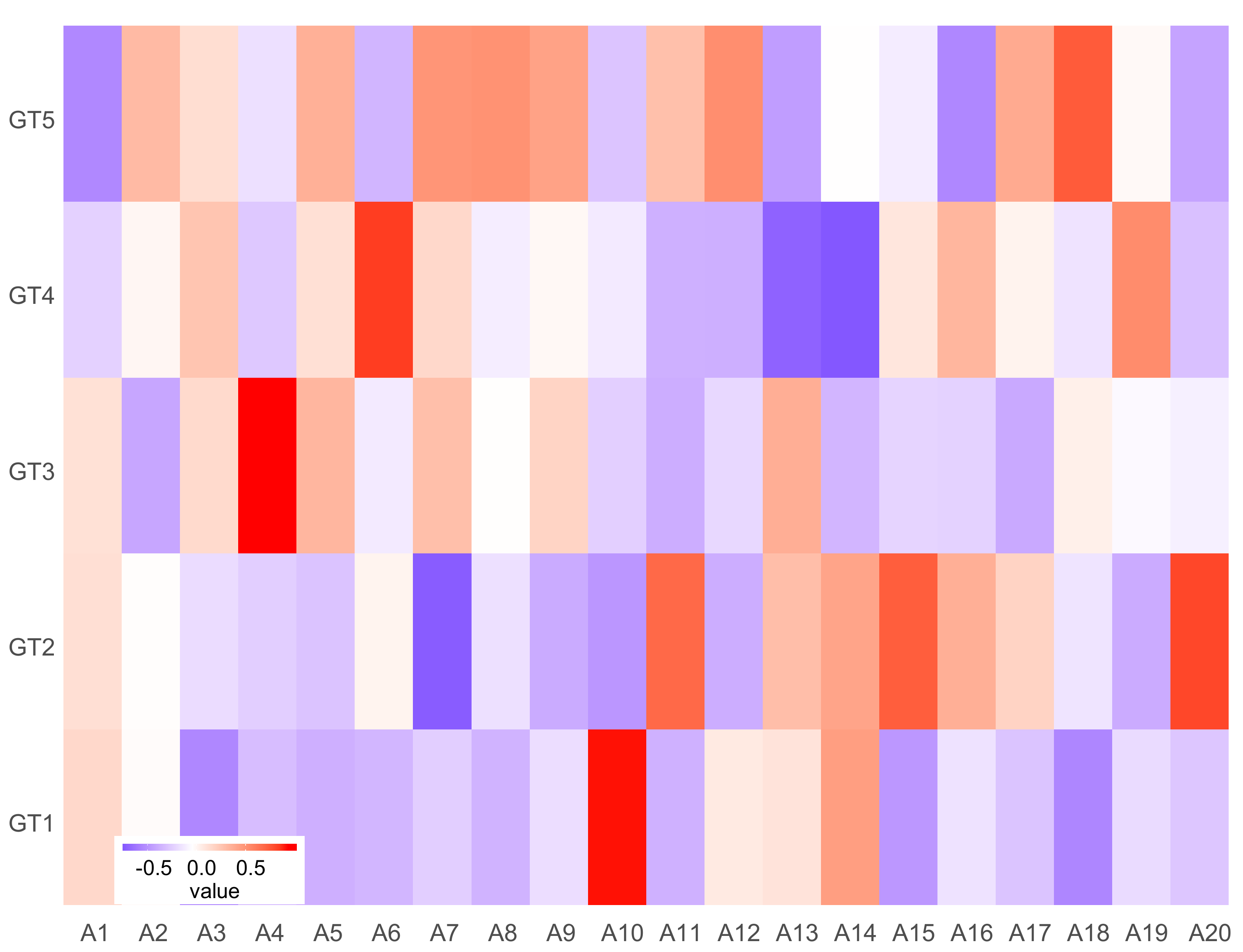}
    \caption{Correlation between ground-truth components and vanilla ICA components.}
    \label{fig:single_graph:gtica}
    \end{subfigure}
\begin{subfigure}[t]{0.49\textwidth}
    \centering
    \includegraphics[width=\textwidth,height=4.5cm]{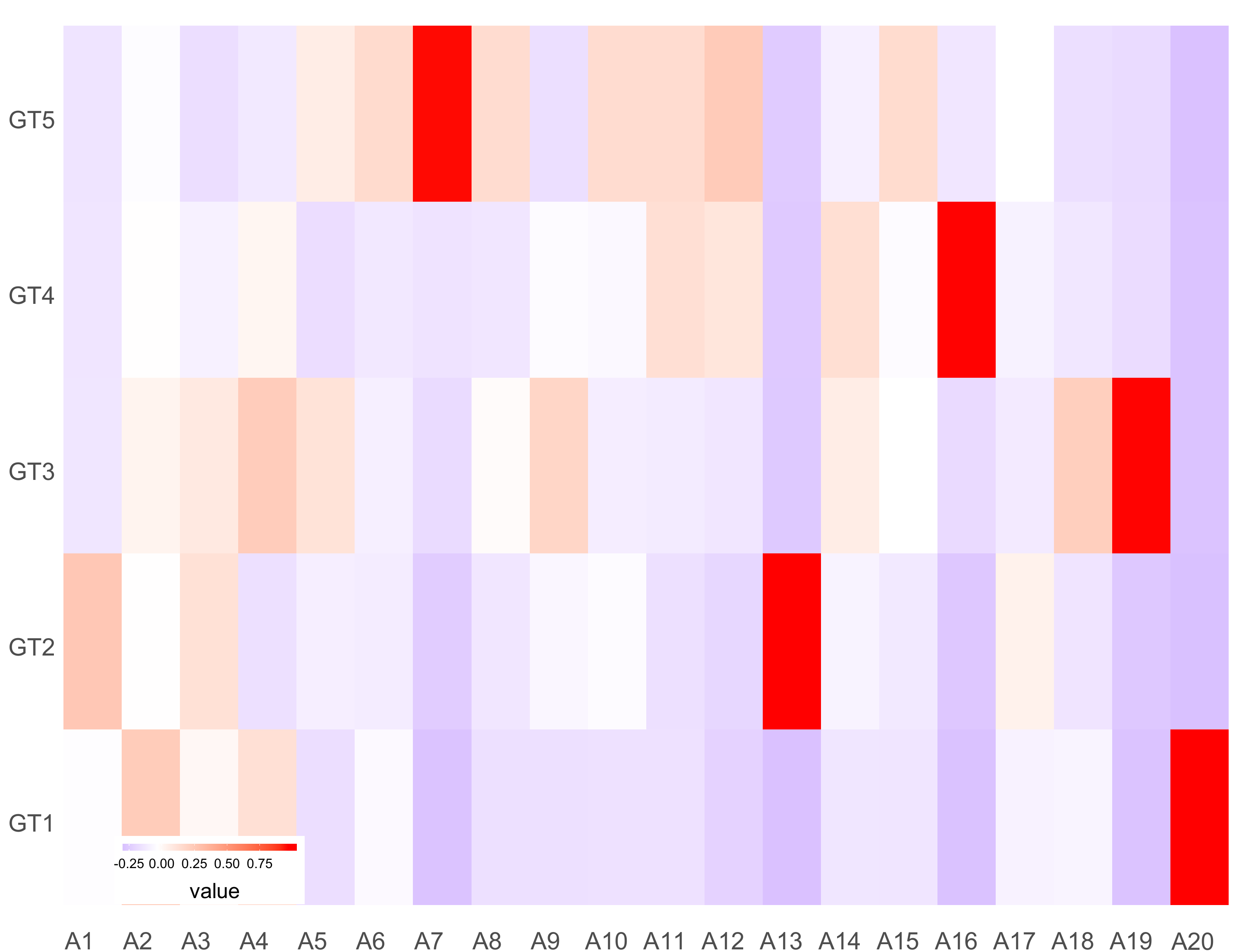}
    \caption{Correlation between ground-truth components and Bayesian recovery.}
    \label{fig:single_graph:gtcorr}
\end{subfigure}
\caption{Comparison accuracy of the recovery} \label{fig:single_graph}
\end{figure}

\begin{figure}  \label{fig:single_res_cred}
    \centering
\begin{subfigure}{0.475\textwidth}
    \centering
    \includegraphics[width=\textwidth,height=5cm]{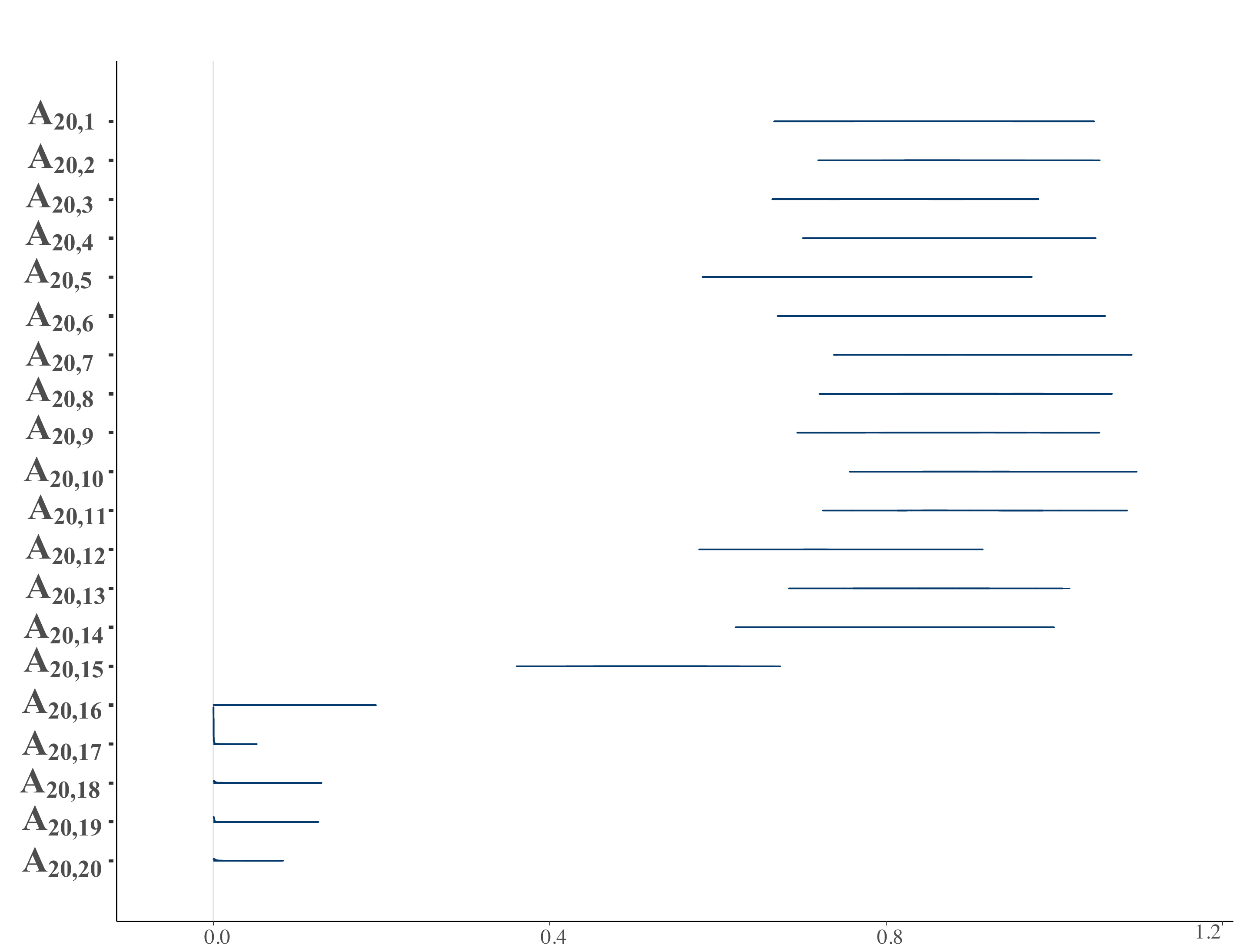}
    \caption{Credible intervals for Node 20.}
    \label{fig:single_res_credL20_1}
\end{subfigure}
\begin{subfigure}{0.475\textwidth}
    \centering
    \includegraphics[width=\textwidth,height=5cm]{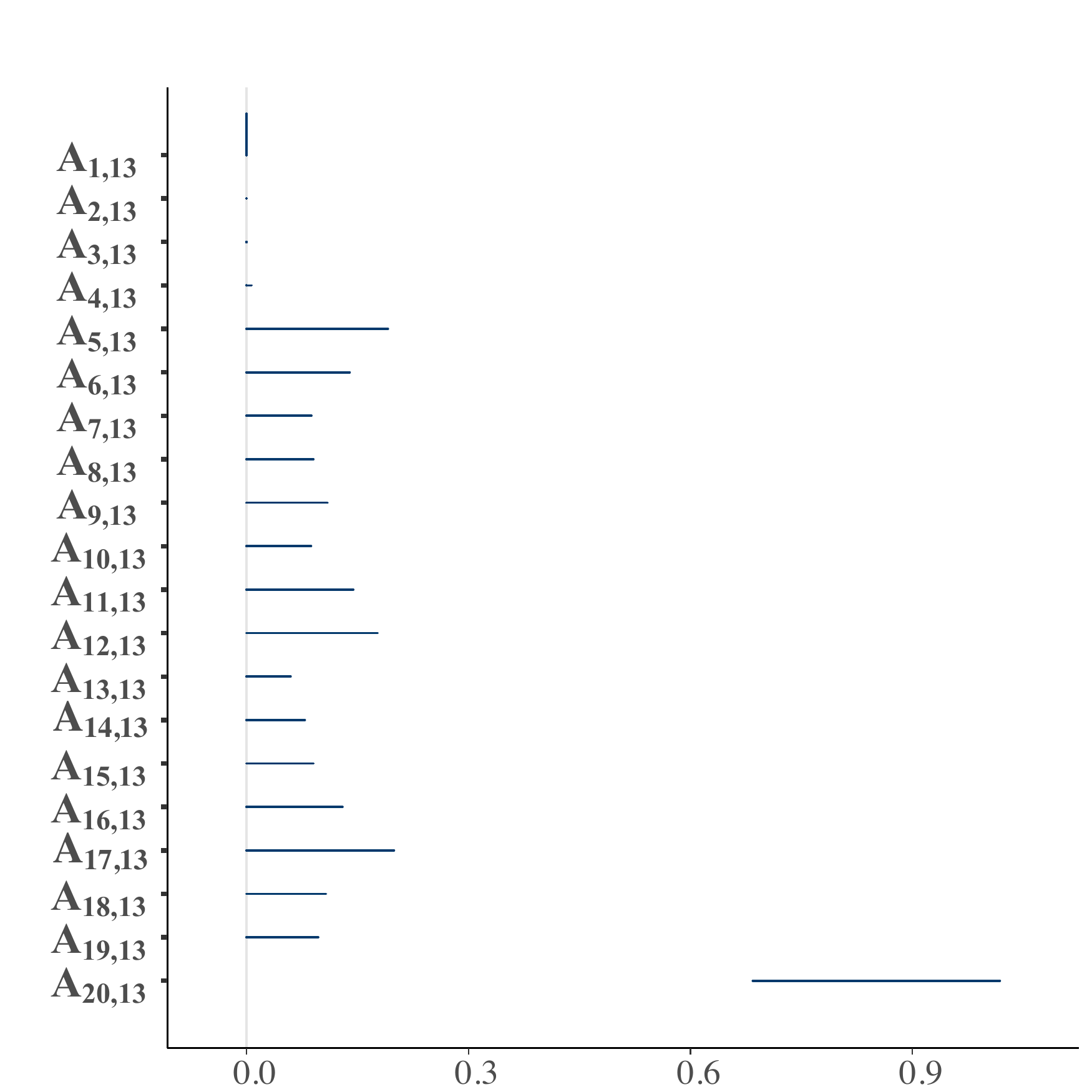}
    \caption{Credible intervals for Component 20.}
 \label{fig:single_res_credL20_2}
\end{subfigure}
    \caption{Credible intervals (row-wise and column-wise)}\label{fig:single_res_cred}

\end{figure}
.

We see that the model correctly identifies 5 distinct components (Fig.{\ref{fig:single_graph:L}). The norm for the different components (blue dots in Fig. \ref{fig:single_graph:histl}) show indeed that only 5 of those are above 0. By contrast, for the regular ICA model (green diamonds in Fig. \ref{fig:single_graph:histl}) show many of these components being distinct from 0, thus highlighting the difficulty of assessing the correct number of components in the traditional ICA setting. 
In particular, the correlation with the Bayesian components with the ground truth components is almost perfect (Fig.\ref{fig:single_graph:gtcorr}), while substantially more diffuse for the traditional ICA model (Fig.\ref{fig:single_graph:gtica}).
We also see that our model's recovered coefficients are both sparse and localized, as most of the coefficients of the loadings matrix $\Lambda A \odot D$ are around 0. (Fig. \ref{fig:single_graph:L},\ref{fig:single_graph:histl}, Fig.\ref{fig:single_res_cred}). Finally, the correlation between the Bayesian ICA components and the ground-truth (Fig \ref{fig:single_graph:gtcorr}) is extremely high, highlighting an accurate recover of the components. By comparison, the standard ICA model (Fig \ref{fig:single_graph:gtica}) fails to capture the appropriate number of components and to accurately recover them.\\


\xhdr{Assessing the impact of the different assumptions} Building upon the previous visual inspection of the results, we quantify the performance of the method by running this experiment multiple times with the same setting. We also quantify how much our recovered coefficients abide with our structural assumptions:
\begin{itemize}
    \item \textit{spatial localization:} to quantify the spatial localization of the different subgraphs, we evaluate two quantities:
    \begin{itemize}
        \item The average spread of the components, defined as $s = \frac{1}{K}\text{Trace}[A_k L A_k^T]$. The smaller the contrast, the smoother the component.
        \item A ``localization'' coefficient, defined a the ratio of the within norms of the outside and inside activations associated to each ground truth component. That is, letting  $\mathcal{I}$ be the set of all nodes  and $\mathcal{I}_k$ be the indices of the nodes in the ground-truth component that is the most correlated with $A_k$,  $\rho = \sum_{k=1}^K \frac{||A_{\mathcal{I} \setminus \mathcal{I}_k}||^2 } {||A_{ \mathcal{I}_k}||^2}$. Thus, the smaller the coefficient, the more localized the component.
    \end{itemize}
    \item \textit{sparsity:} We measure the sparsity of the recovered components by assessing:
        \begin{itemize}
        \item the $\ell_1$-norm of the components, $s = ||A||_1$
        \item the number of recovered components, which we define as the number of components that have correlation above 0.8 with the ground-truth components. For perfect recovery, we expect this number to be 5. Higher number indicate redundancy in the components that are recovered, whereas lower number indicate that too few components were actually recovered.
    \end{itemize}
    \item \textit{accuracy:} finally, we assess the accuracy of the recovered components by comparing the correlation of the ground-truth loadings with the recovered ones. We quantify this accuracy by providing the mean of the top-5 correlations: the higher the mean, the more accurately each component is recovered.
\end{itemize}

We display the results in Tables \ref{tab:synthetic_results} and \ref{tab:synthetic_results2}, by providing the average and standard deviation of these metrics over 55 independent trials. In particular, these tables highlight the performance of the Bayesian algorithm: a higher number of factors are recovered compared to the vanilla counterpart. 

\definecolor{Gray}{gray}{0.85}

\begin{table}[ht]
\centering
\begin{tabular}{|>{\cellcolor{Gray}}c|cccc|}
  \hline
  {\bf{Algorithm}}& \cellcolor{Gray} {\bf  Model} & \cellcolor{Gray} {\bf Sparsity} ($\times 10^{-2}$)   & \cellcolor{Gray}{\bf Spread $s$} & \cellcolor{Gray}{\bf Localization $\rho$} \\ 
  \hline
 \hspace{0.2cm} \textbf{ICA}& {\small Vanilla} & 14.11 $\pm$ 0.66  & 34.46 $\pm$ 1.94& 16.6 $\pm$ 4.45  \\ \hline
&   Full &  $5.04 \pm 0.39$   & 22.68 $\pm$ 2.69  & 2.51 $\pm$ 1.73 \\ 
 \multirow{-1}{*}{\rotatebox[origin=c]{00}{\textbf{BICA}}}&Vanilla & 7.11 $\pm$ 0.71  & 22.36 $\pm$ 2.57 & 3.04 $\pm$ 1.60\\ 
& No localization & 6.97 $\pm$ 0.70&  22.03 $\pm$ 2.66 & 2.75 $\pm$  1.40 \\ 
&No structure & 5.07 $\pm$ 0.36 &  22.83 $\pm  $ 2.54 & 2.45 $\pm$ 1.46\\ 
   \hline
\end{tabular}
    \caption{Comparison of the different algorithms' results on our synthetic problem}\label{tab:synthetic_results}
\end{table}

\begin{table}[ht]
\centering
\begin{tabular}{|>{\cellcolor{Gray}}c|cccc|}
  \hline
  {\bf{Algorithm}}& \cellcolor{Gray} {\bf  Model} & \cellcolor{Gray} {\bf Nb Recovered }  & \cellcolor{Gray} {\bf Mean top-5}  & \cellcolor{Gray} {\bf Accuracy} \\ 
    & \cellcolor{Gray}  & \cellcolor{Gray}{ \bf Factors} & \cellcolor{Gray} {\bf $\lambda$s}  & \cellcolor{Gray}  \\ 
  \hline
 \hspace{0.2cm} \textbf{ICA}& {\small Vanilla} & 2.02 $\pm$ 1.13  & NA   & 0.76 $\pm$ 0.094   \\ \hline
&   Full &  5.00 $\pm$ 0.00 & $1.20 \pm  0.09$   & 1.00 $\pm$ 0.001    \\ 
 \multirow{-1}{*}{\rotatebox[origin=c]{00}{\textbf{BICA}}}&Vanilla & 5.04 $\pm$ 0.19 & 0.37 $\pm$  0.03 & 0.99 $\pm$ 0.002\\ 
& No localization & 5.04 $\pm$ 0.19 & 0.35 $\pm$ 0.04 & 0.99 $\pm$ 0.002 \\ 
&No structure & 5.07 $\pm$ 0.13 & 1.28 $\pm$ 0.09 & 1.00 $\pm $ 0.001\\ 
   \hline
\end{tabular}
    \caption{Comparison of the different algorithms on synthetic problem: Evaluation of the different components}\label{tab:synthetic_results2}
\end{table}

We observe that the accuracy of the recovery is perfect for the Bayesian ICA model (both in the number of recovered factors and in their accuracy). On the other hand, the recovery of these factors is much noisier with traditional ICA, with an average best top-5 correlation hovering around 0.76. Similarly, the BICA components are 2.5 times more sparse and 1.56 more spread than ICA, thus varying more smoothly over the graph.



\xhdr{Robustness to Noise} We also evaluate the performance of the different models under varying levels of noise (the $\epsilon$ in Eq. \ref{eq:ICA_model}).
The results are displayed in Figure \ref{fig:multiplegraphs}. In particular, this setting, we see that, up to reasonable amounts of noise, the Bayesian ICA model is far more accurate that the Vanilla ICA: the top-5 Recovered/GT pairs have a correlation that is up to 20\% bigger than the Vanilla ICA model, indicating an extremely accurate detection of the underlying factors. We also see that the inclusion of structural priors (red and purple curves) yield components that are overall more accurate under varying levels of noise. The inclusion of both localization and structural priors (red curve) yields  increased mean and max accuracies.

\begin{figure}
    \centering
\begin{subfigure}{0.49\textwidth}
    \centering
    \includegraphics[width=\textwidth]{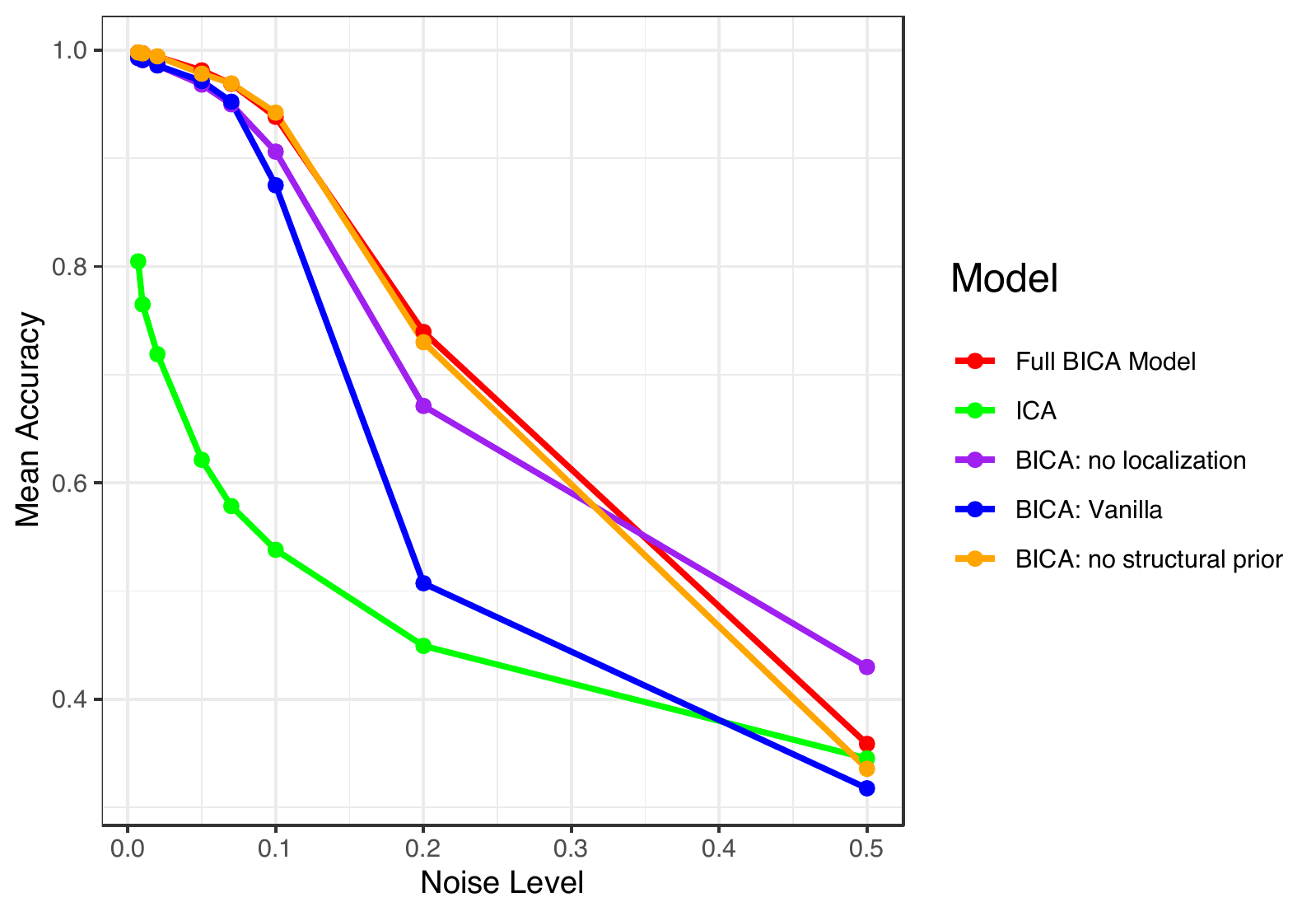}
    \caption{Mean of the top-5 pairwise correlations between recovered components and ground-truth factors. }
    \label{fig:my_label}
\end{subfigure}
\begin{subfigure}{0.49\textwidth}
    \centering
    \includegraphics[width=\textwidth]{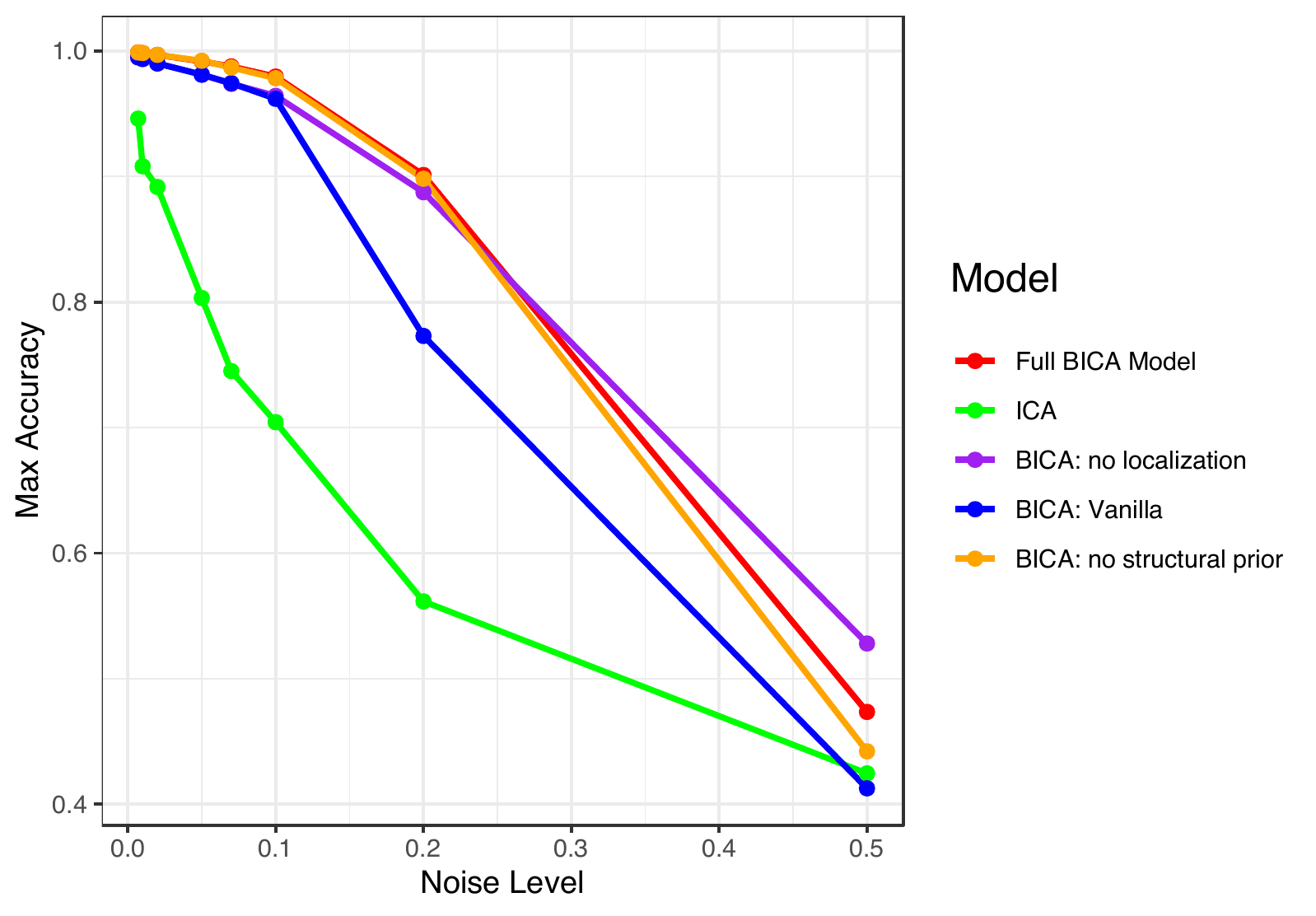}
    \caption{Max correlation between recovered component/ground-truth factors pairs. }
    \label{fig:my_label}
\end{subfigure}

\begin{subfigure}{0.49\textwidth}
    \centering
        \includegraphics[width=\textwidth]{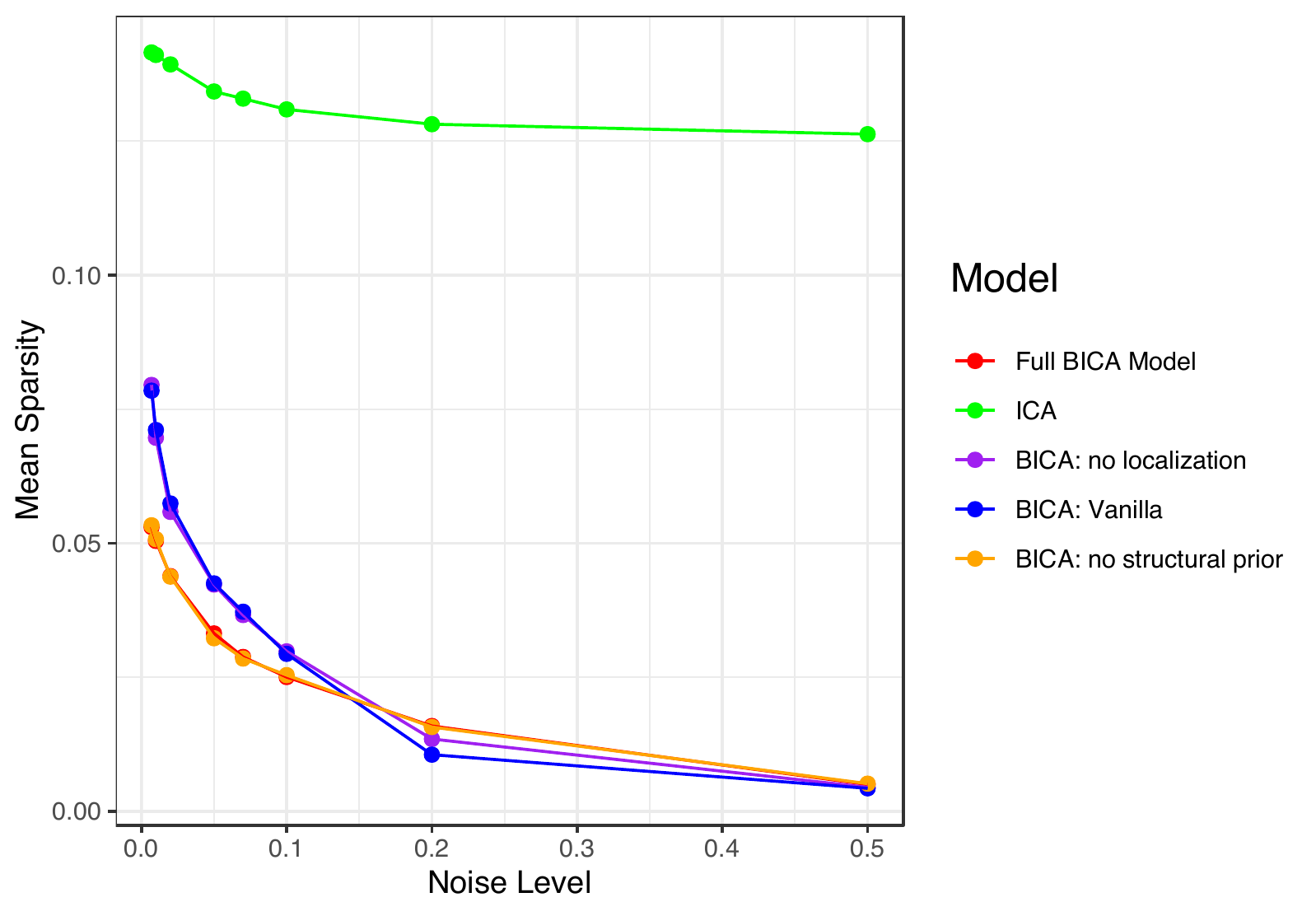}
    \caption{Sparsity ($\ell_1$ norm).}
    \label{fig:my_label}
\end{subfigure}
\begin{subfigure}{0.49\textwidth}
    \centering
    \includegraphics[width=\textwidth]{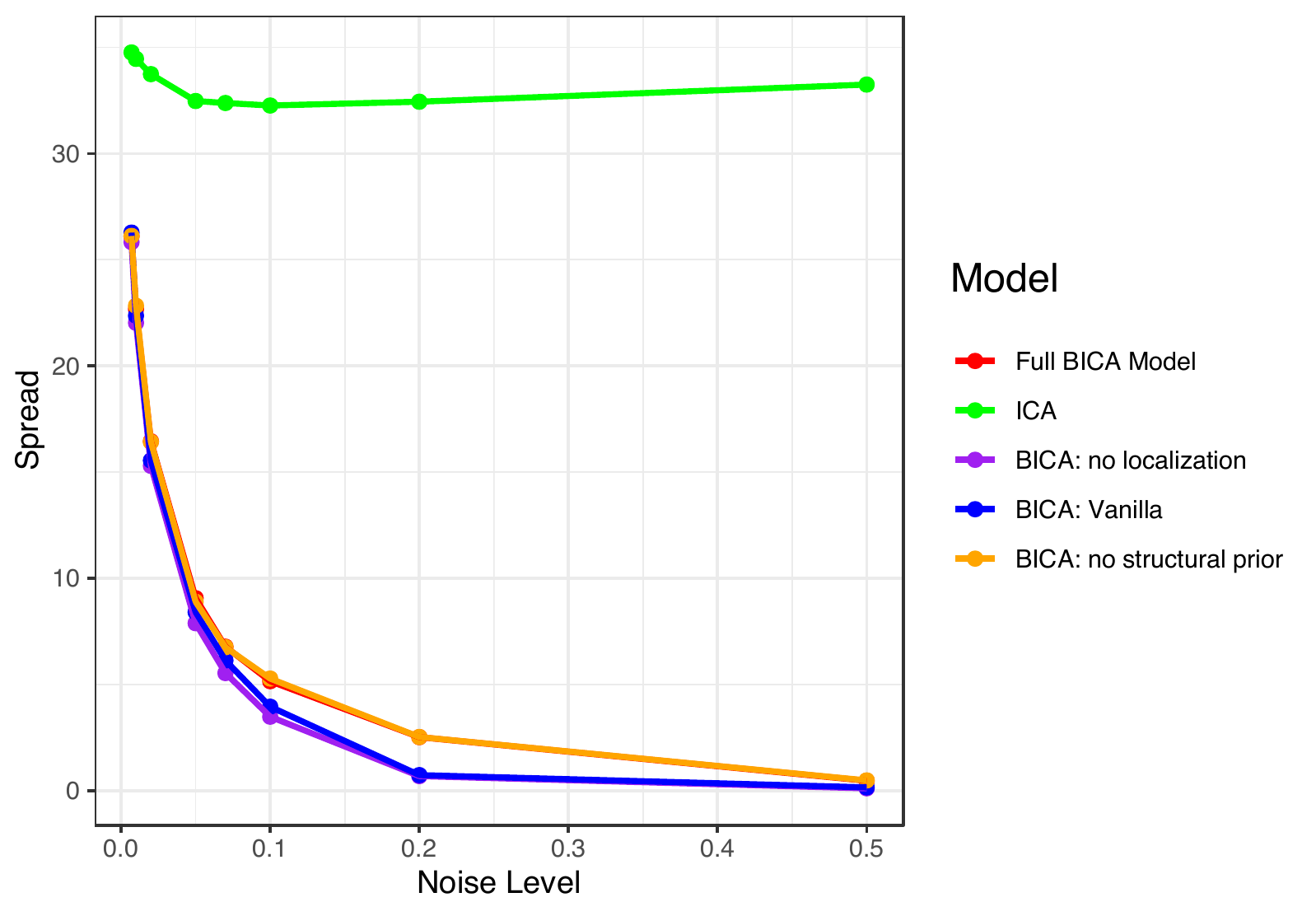}
    \caption{Spread }
    \label{fig:my_label}
\end{subfigure}
    \caption{Analysis of the impact of the noise on the recovery of the components}
    \label{fig:multiplegraphs}
\end{figure}

%% file: reallife_experiments.tex

Our method having shown good promise on controlled synthetic examples, the goal of this section is to assess its behavior on real data, and if it can be used to process and produce interesting results for time series on graphs. We focus here on one particular test/retest fMRI study: the The Hangzhou Normal University (HNU1) dataset \citep{zuo2014open}.

\xhdr{The Hangzhou Normal University dataset}\citep{zuo2014open} This dataset was gathered as part of a one-month Test-Retest Reliability and Dynamical Resting-State study. It consists of the fMRI of 30 healthy patients taken every 3 days across one month (each patient thus has roughly a total of ten scans). Five modalities (EPI/ASL/T1/DTI/T2) of brain images were acquired for all subjects. As per the dataset's website\footnote{\url{http://fcon_1000.projects.nitrc.org/indi/CoRR/html/hnu_1.html}}, "during functional scanning, subjects were presented with a fixation cross and were instructed to keep their eyes open, relax and move as little as possible while observing the fixation cross. Subjects were also instructed not to engage in breath counting or meditation.". For the purpose of this study, we chose to use the Craddock 200-node atlas, which provides a manageable, yet relatively fine-scale representation of the brain. More details on the properties of this dataset, the study, the parameters of the scan  as well as the fMRI pre-processing can be found in Appendix \ref{appendix:functional}. The structural matrices were obtained using pre-processed connectomes using NeuroData's MR graphs package, \textbf{NDMG} \citep{10.1093/gigascience/gix013,kiar2017comprehensive}\footnote{The data and more information are available on NeuroData's website: \url{https://neurodata.io/mri/}}. More details can also be found in Appendix \ref{appendix:structural}. This dataset provides an extremely convenient framework to test the validity of our model: not only do we have functional and structural information for each scan, we can test for subject effects and robustness across subjetcs and sessions.

\begin{figure}
    \begin{subfigure}{0.49\textwidth}
        \includegraphics[width=\textwidth,height=6cm]{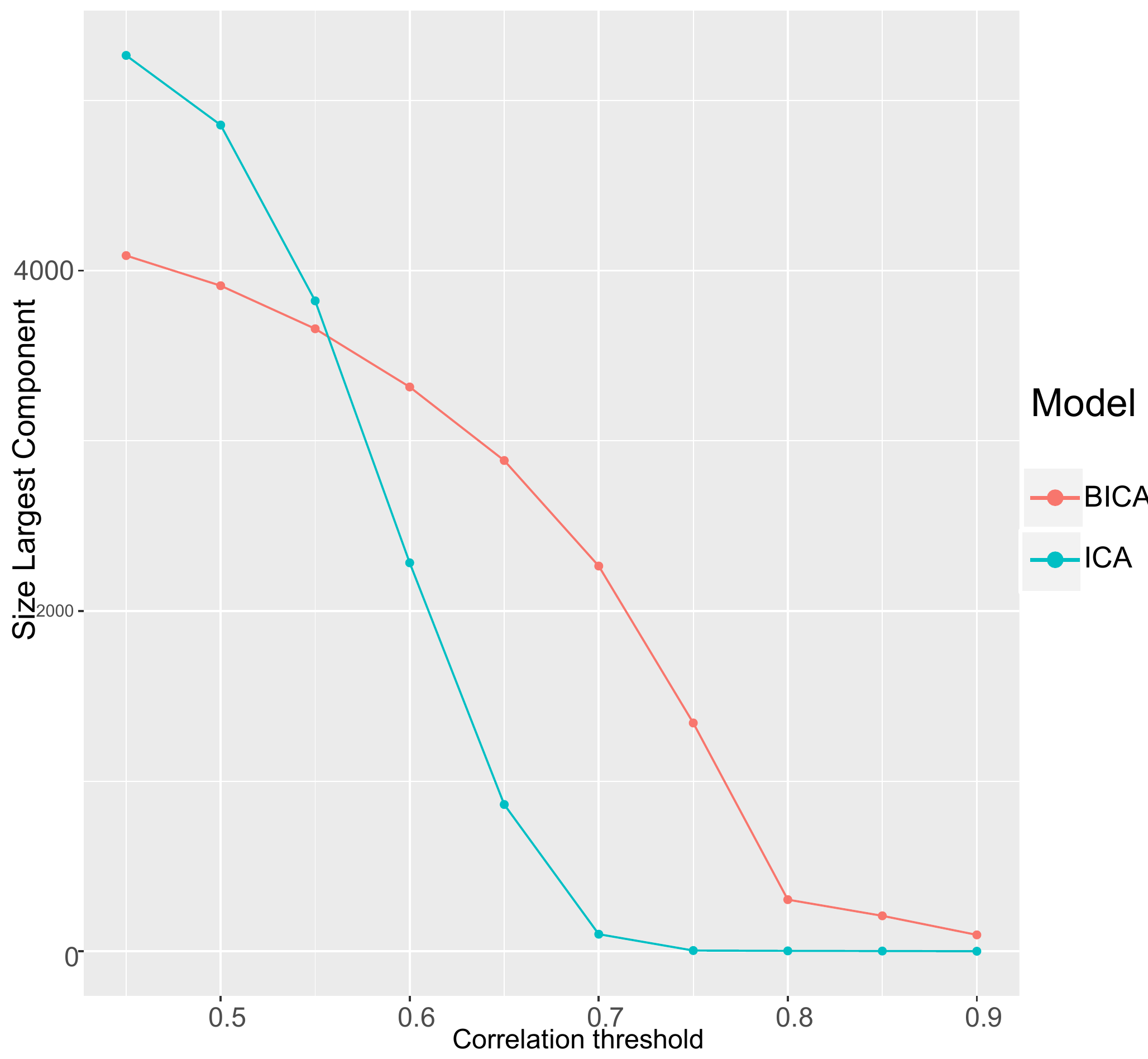}
        \caption{Sizes of the connected components }\label{fig:HNU1:full_model}
    \end{subfigure}
    \begin{subfigure}{0.5\textwidth}
        \includegraphics[width=\textwidth, height=6cm]{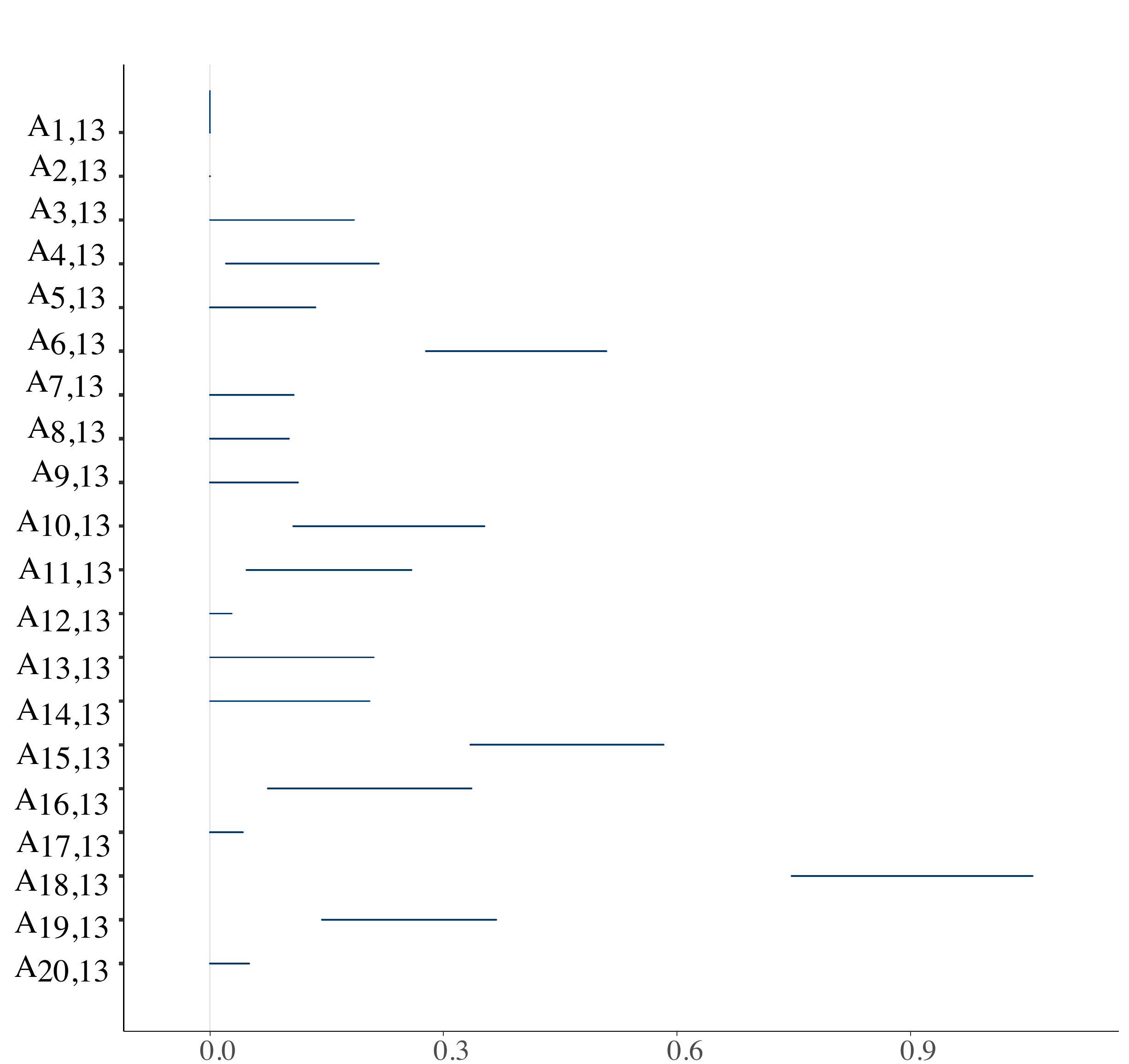}
        \caption{Posterior Distribution of the activations of Node 13 per component
(Right Superior Frontal Gyrus), with medians and 80\% intervals.}\label{fig:HNU1:credintL}
    \end{subfigure}

    \begin{subfigure}{0.49\textwidth}
        \includegraphics[width=\textwidth, height=6cm]{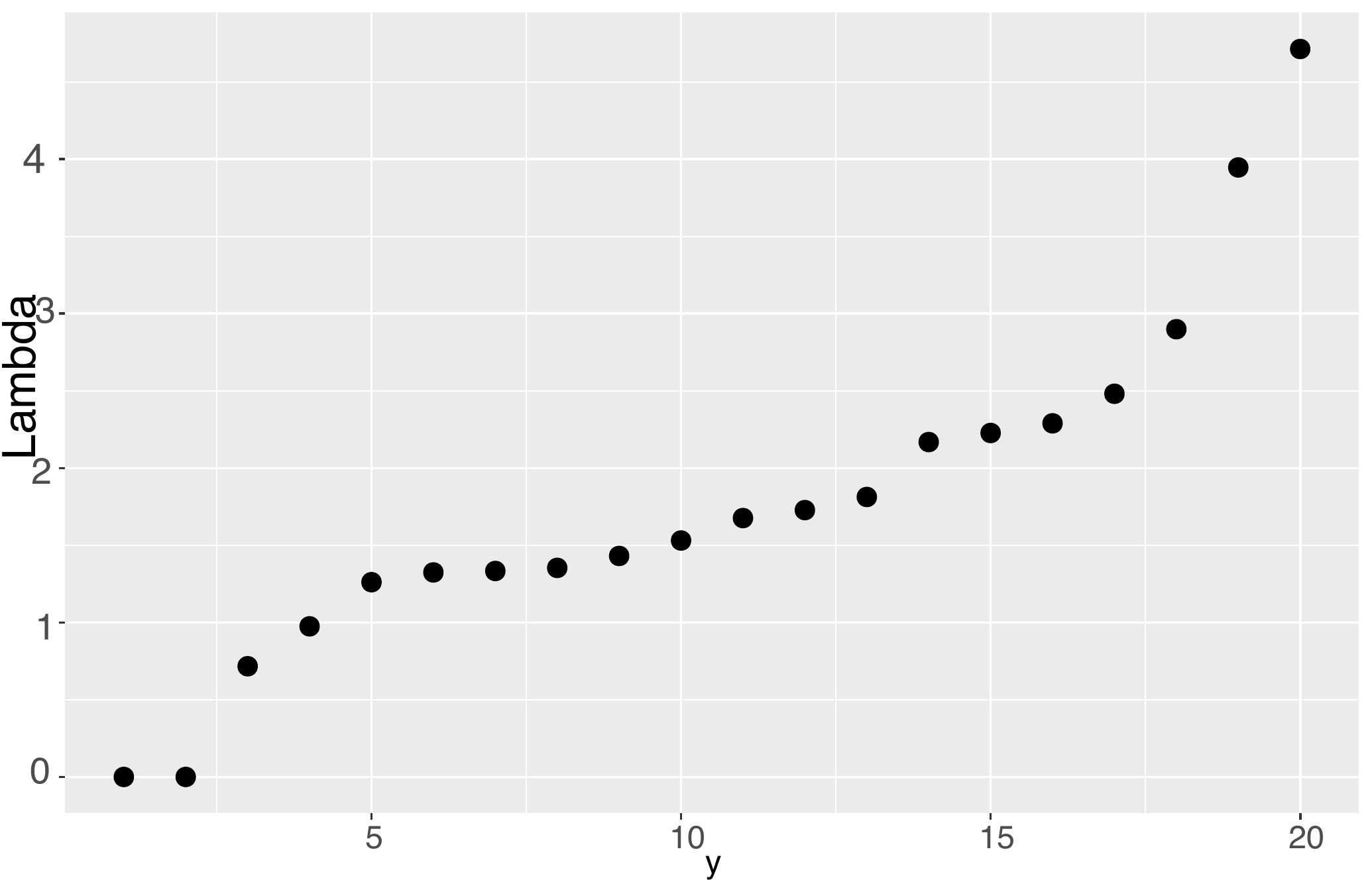}
        \caption{Recovered Lambdas}
    \end{subfigure}
    \begin{subfigure}{0.5\textwidth}
        \includegraphics[width=\textwidth, height=6cm]{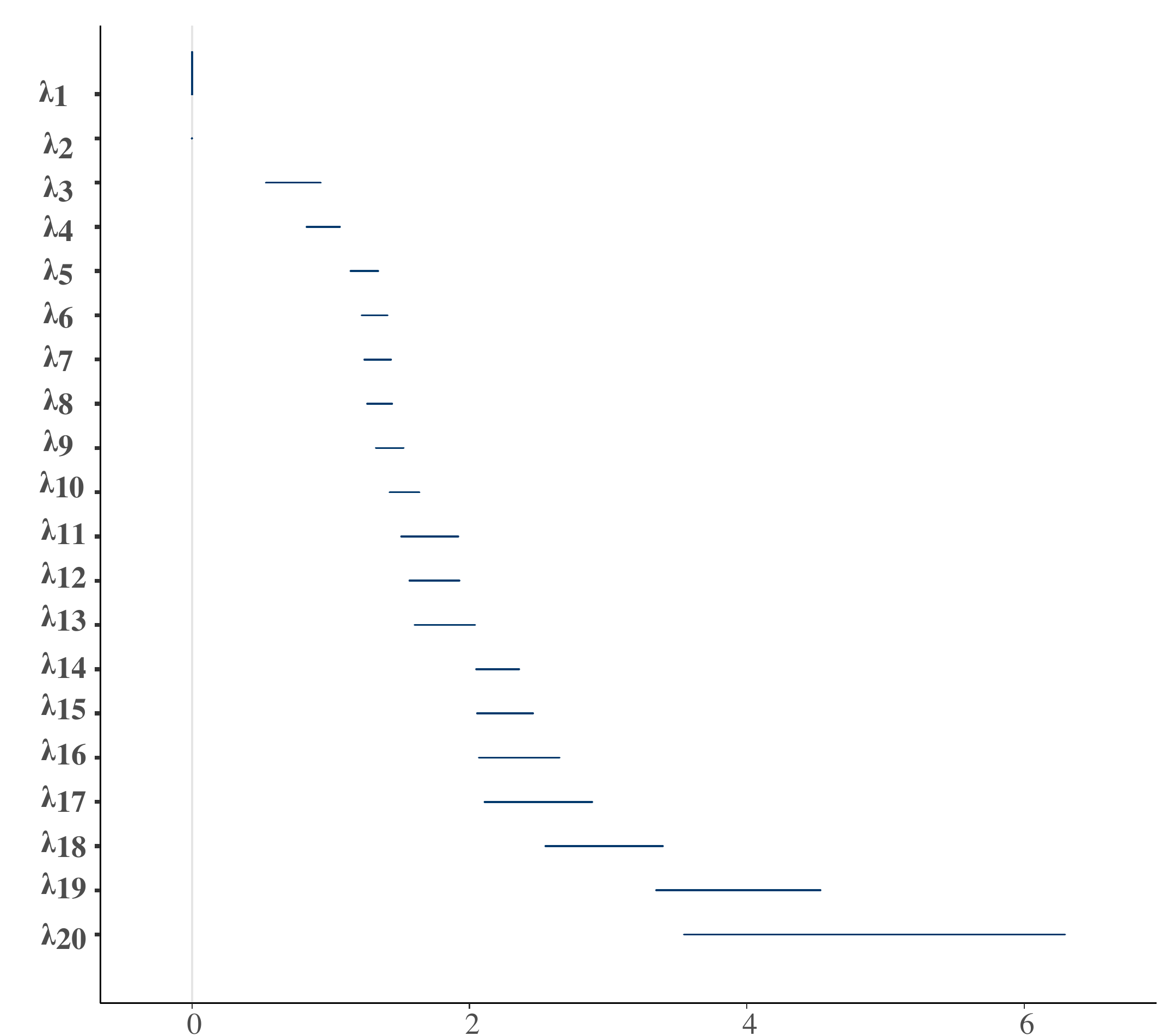}
         \caption{Credible intervals for the $\Lambda$s, with medians and 80\% intervals.}\label{fig:HNU1:credintlambda}
    \end{subfigure}

        \begin{subfigure}{0.3\textwidth}
        \includegraphics[width=\textwidth, height=4cm]{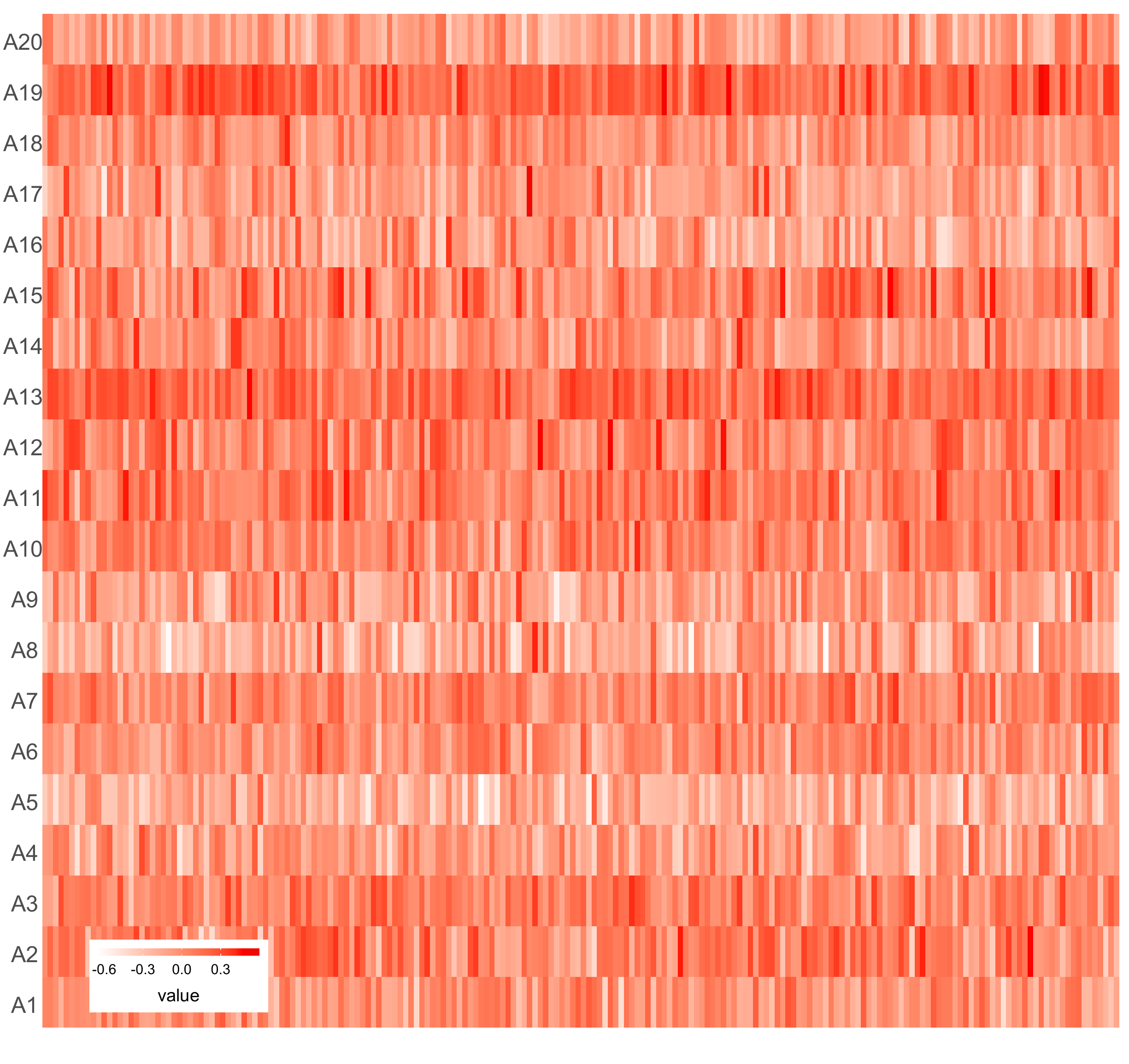}
        \caption{"Vanilla ICA" components}\label{fig:HNU1:ica}
    \end{subfigure}
        \begin{subfigure}{0.3\textwidth}
        \includegraphics[width=\textwidth, height=4cm]{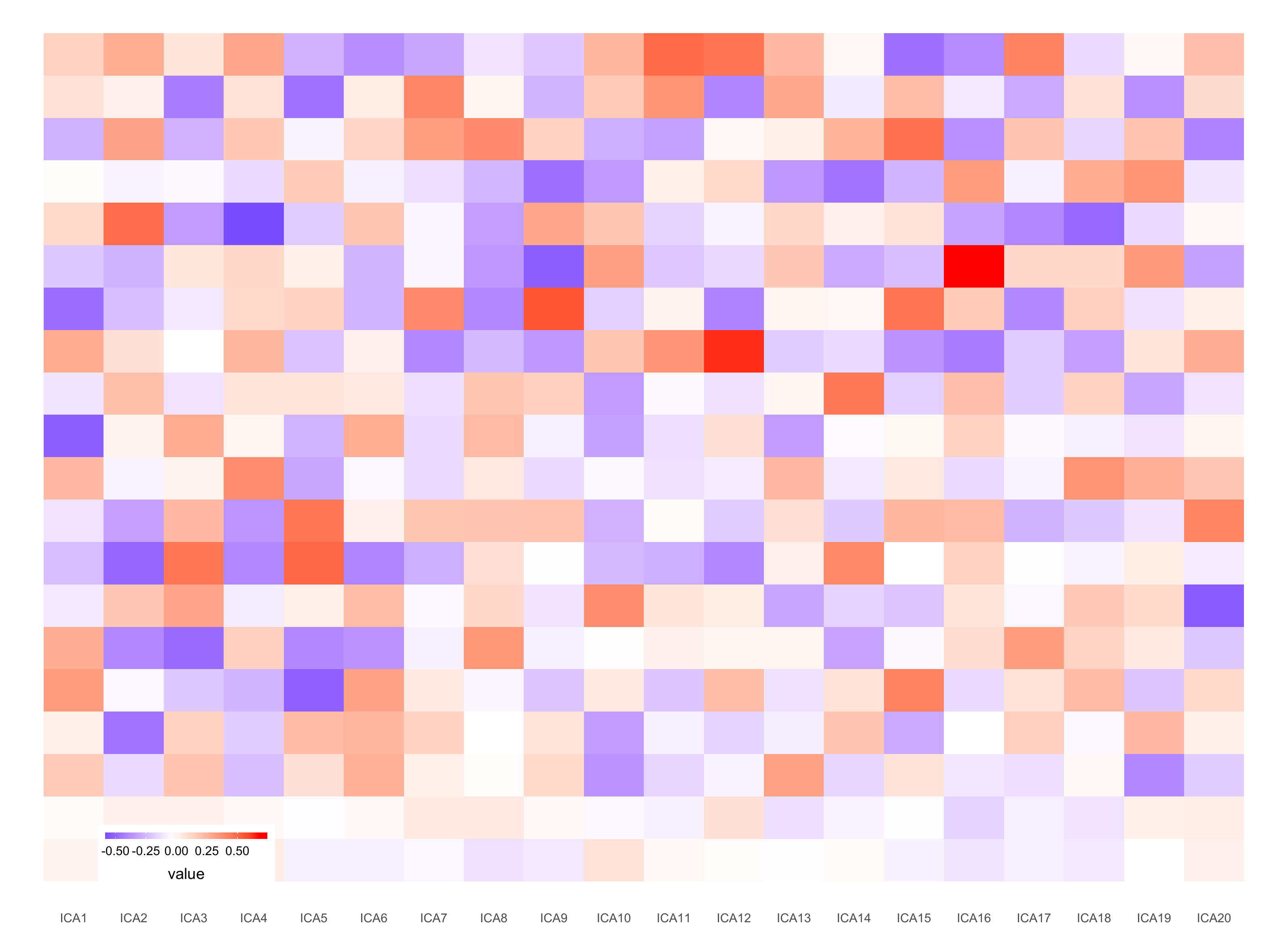}
        \caption{Correlation between BICA and Vanilla ICA}
    \end{subfigure}
    \begin{subfigure}{0.3\textwidth}
        \includegraphics[width=\textwidth]{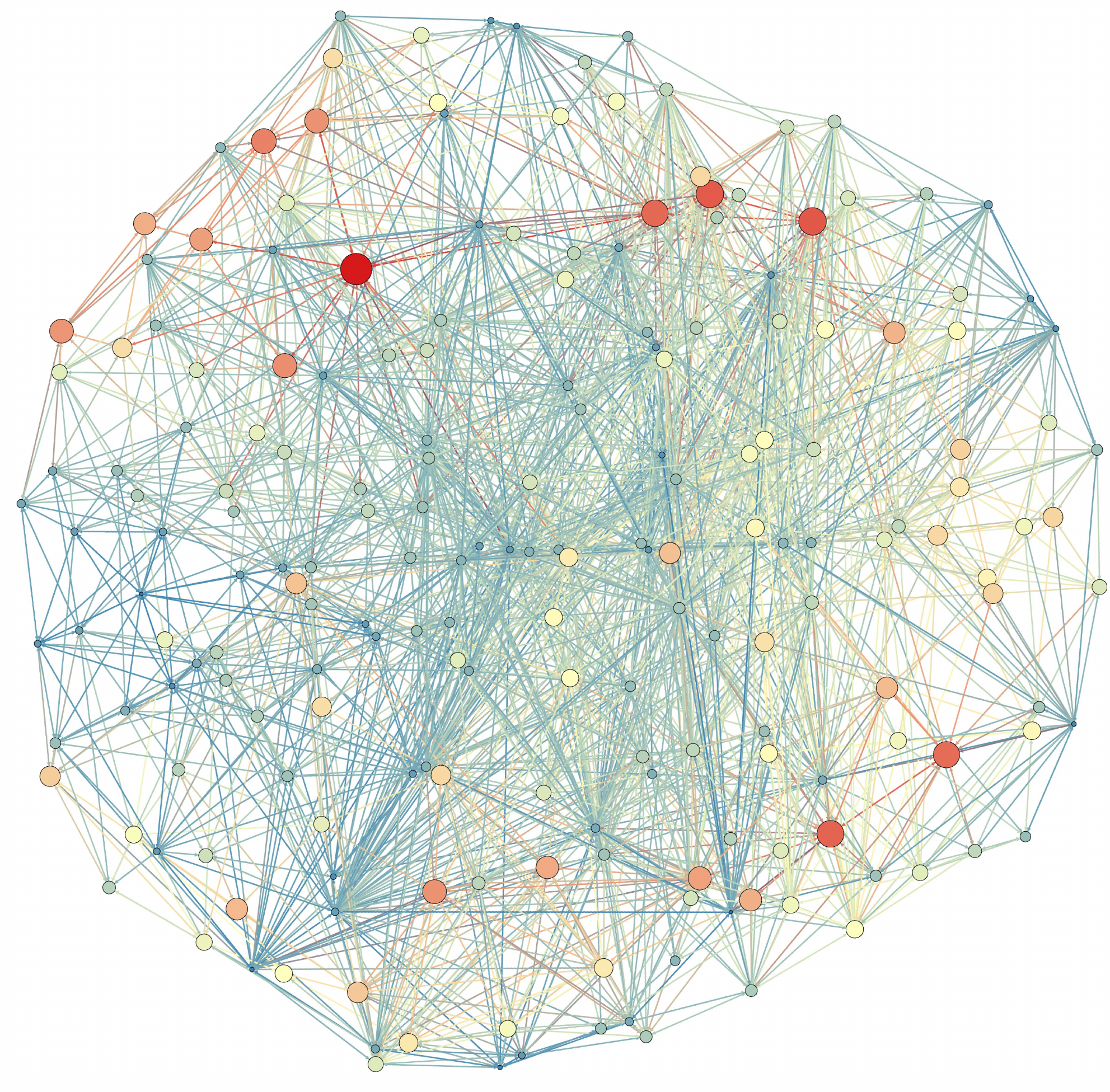}
         \caption{Example of a given component (comp 20)}\label{fig:HNU1:comp20}
    \end{subfigure}
\caption{Results for the HNU1 dataset} \label{fig:HNU1}
\end{figure}

\xhdr{First results} Fig.  \ref{fig:HNU1} shows an example of the components that can be extracted from a given scan (session 1 for the subject 24527, the first of the cohort). As can be seen on Fig. \ref{fig:HNU1:full_model} and \ref{fig:HNU1:comp20} , our Bayesian model manages to capture components that are localized: only a few coefficients are non-zero within each component. By way of comparison, the Vanilla ICA components are much more scattered over the brain (Fig.\ref{fig:HNU1:ica}). The Bayesian model also allows us to quantify the uncertainty associated to each coefficient, as displayed in Fig.\ref{fig:HNU1:credintL}: we see that the right Superior Frontal Gyrus for Subject 24527 is significantly involved in 8 components. In this particular case, the BICA model allows the incorporation of up to 18 components. Our method also allows a characterization of the uncertainty estimates for each node (Fig \ref{fig:HNU1:credintL}): the Right Superior Frontal Gyrus for instance  is only significantly activated in 8 components. All in all, this first visual inspection allows us to check that our method does indeed recover components that are aligned with our original hypotheses: sparse, connected and localized.

\begin{figure}[h]
    \centering
\begin{subfigure}{0.45\textwidth}
    \centering
    \includegraphics[width=\textwidth,height=4cm]{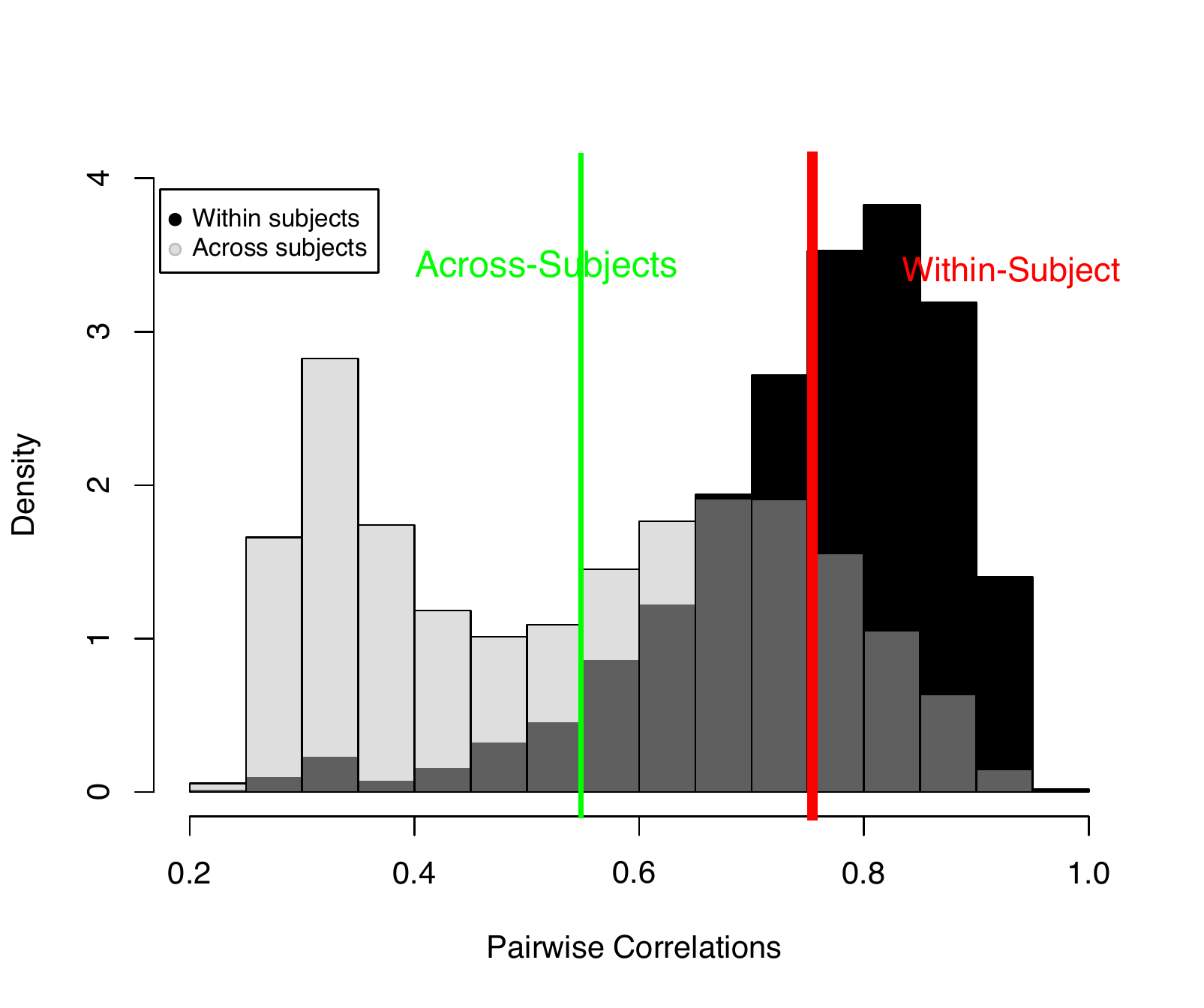}
    \caption{Strongest Pairwise Correlations (Bayesian ICA)}
    \label{fig:hist_corr}
\end{subfigure}
\begin{subfigure}{0.45\textwidth}
    \centering
    \includegraphics[width=\textwidth, height=4cm]{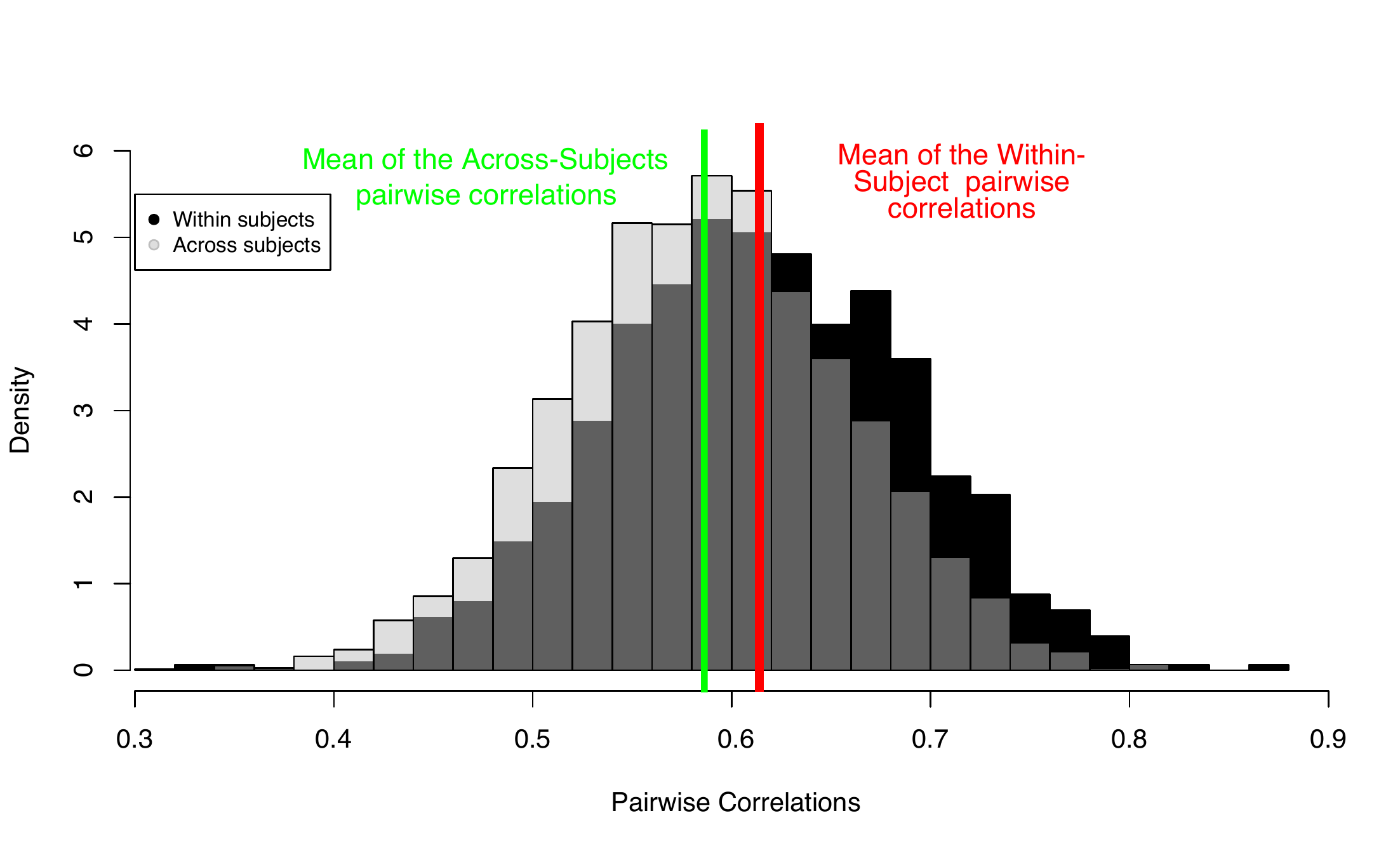}
    \caption{Strongest Pairwise Correlations (Vanilla ICA)}
    \label{fig:hist_corr_ica}
\end{subfigure}
    \caption{Comparison of the distributions of the closest neighbor similarity.}
    \label{fig:multiplegraphs_panlab}
\end{figure}

\xhdr{Robustness of the model} We further quantify the model's performance by evaluating its capacity to identify robust components across scans. To do so, we concatenate all components (as ordered by the $\lambda$s) extracted from all 270 scans--thus yielding a total matrix of size 5400--, and we compute their cross-correlations.\\
\textit{Subject effects.} The first step is to establish the existence (or lackthereof) of subject effects: it seems natural to assume that scans from the same subject should yield components that are generally closer together than across subjects. With this aim in mind, for each component, we find at its $k$-closest neighbors (where $k$ is taken to vary from 1 to 10). We denote as $\mathcal{E}_k$ the set of all such selected pairwise correlations. To create a test statistic, we then compute the cumulative sum of the correlations over all pairs $(i,j) \in \mathcal{E}_k$ belonging to the same subjects:
$$s_k =\sum_{(i,j) \in \mathcal{E}_k} \rho_{ij} \mathbbm{1}_{G_i = G_j}$$
where $G_i$ is the grouping induced by subject $i$.
To compare this against a null distribution, we propose a Friedman-Rafsky test: we permute the labels, and compute  the test statistic $s_k^{(\pi)}$ for the permuted data. In this scenario, a high value for $s_k$ would be indicative of a strong subject effect, since $s_k$ increases with both the strength of the correlation and the inherent number of such pairwise interactions. For $k=1$ (nearest-neighbor test), we observe that:
\begin{itemize}
    \item \textbf{There exists a strong subject effect}: with $s_k=1072.31$ and an average correlation of $0.78$ between nearest neighbor pairs belonging to the same subject (compared to $0.68$ for pairs corresponding to different subject), the Friedman-Rafsky test is overwhelmingly significant: nearest neighbor-pairs are much closer if they belong to the same subject than across subject (Fig.\ref{fig:hist_corr}. We note that this effect is also present--even if less overwhelming---for the Vanilla ICA model (average correlation of 0.60 for within-subject pairs, and 0.56 across pairs, Fig. \ref{fig:hist_corr_ica}). Note that the number itself of within-subject correlation edges does not come out as significant in either model. 
    \item \textbf{The most correlated components are from a different session:} given a set of "same subject" pairwise correlations, we want to establish of these correlations are from the same scan or if they come from different sessions. The later indeed indicates that the model is able to identify robust subject-specific components. We thus permute the labels of the scans for each of these "within-subject" pairwise correlations. Interestingly, we note that the test is significant for our Bayesian ICA model, but not for the Vanilla ICA.
\end{itemize}

\begin{figure}[H]
\centering
\begin{subfigure}{0.45\textwidth}
    \centering
    \includegraphics[width=\textwidth,height=6cm]{FIG_HNU1/LCC_sizes.pdf}
    \caption{Size of the largest connected component}
    \label{fig:LCC:full}
\end{subfigure}
\begin{subfigure}{0.45\textwidth}
    \centering
    \includegraphics[width=\textwidth, height=6cm]{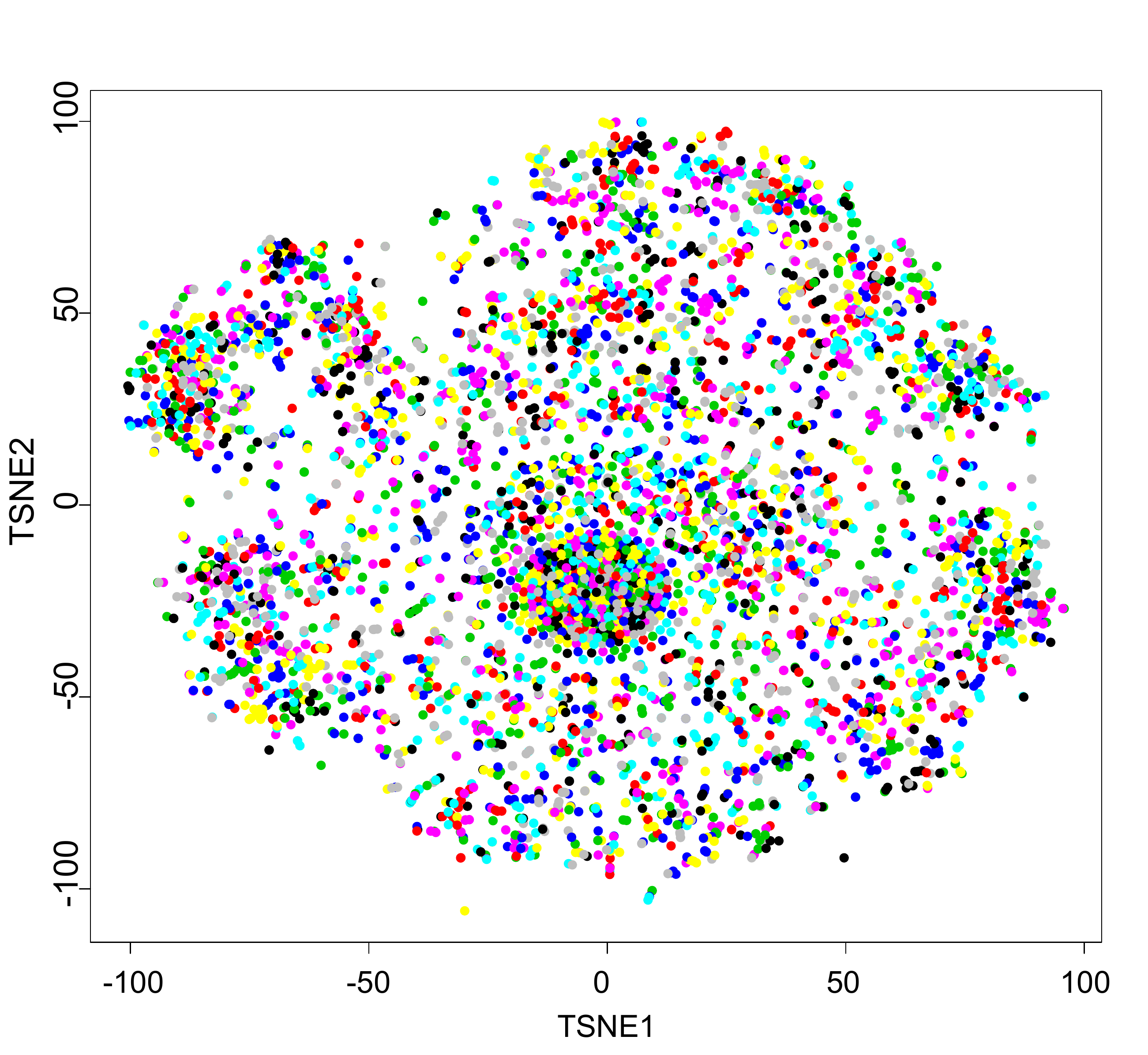}
    \caption{TSNE plot of the recovered loadings, colored by subjects.}
    \label{fig:LCC:tsne}
\end{subfigure}
    \caption{Robustness across sessions.}
    \label{fig:LCC_sizes}
\end{figure}

\begin{figure}[h]
    \begin{subfigure}{0.39\textwidth}
        \includegraphics[width=4.5cm, height=7cm]{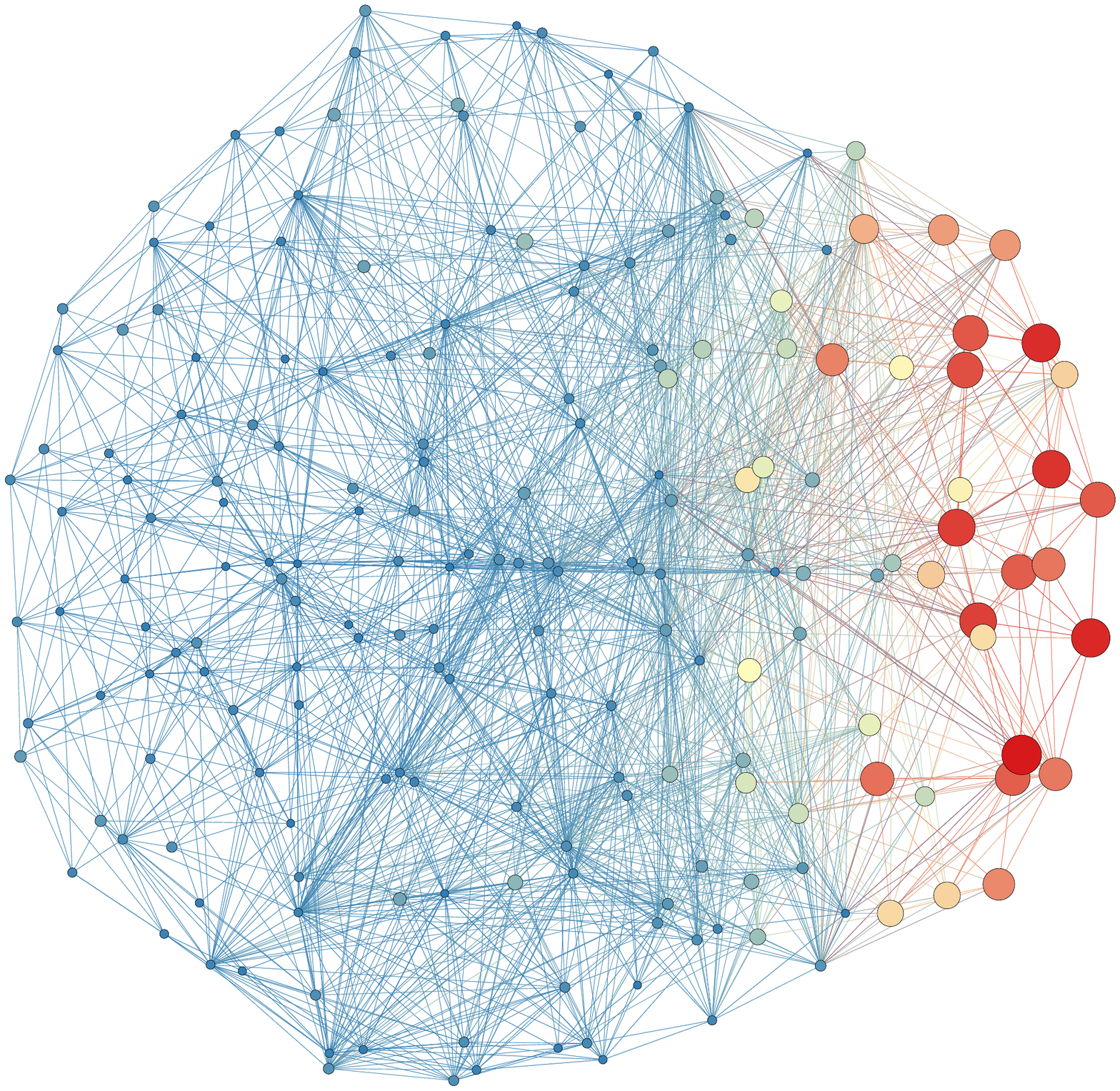}
        \caption{Largest Connected Component}
    \end{subfigure}
    \begin{subfigure}{0.6\textwidth}
        \includegraphics[width=10cm, height=7cm]{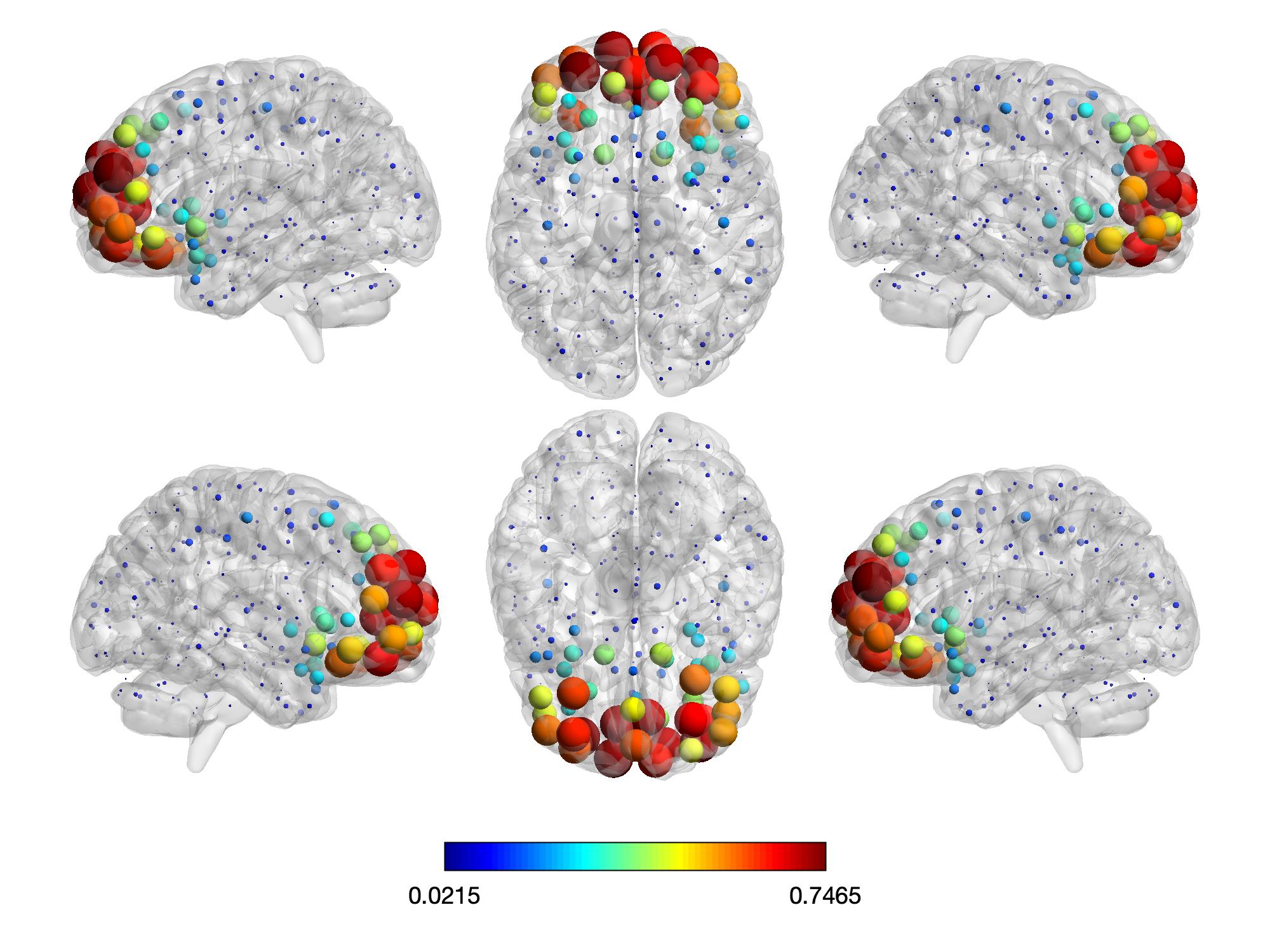}
        \caption{Largest Connected Component (3D view)}
    \end{subfigure}

    \begin{subfigure}{0.39\textwidth}
        \includegraphics[width=4.5cm, height=7cm]{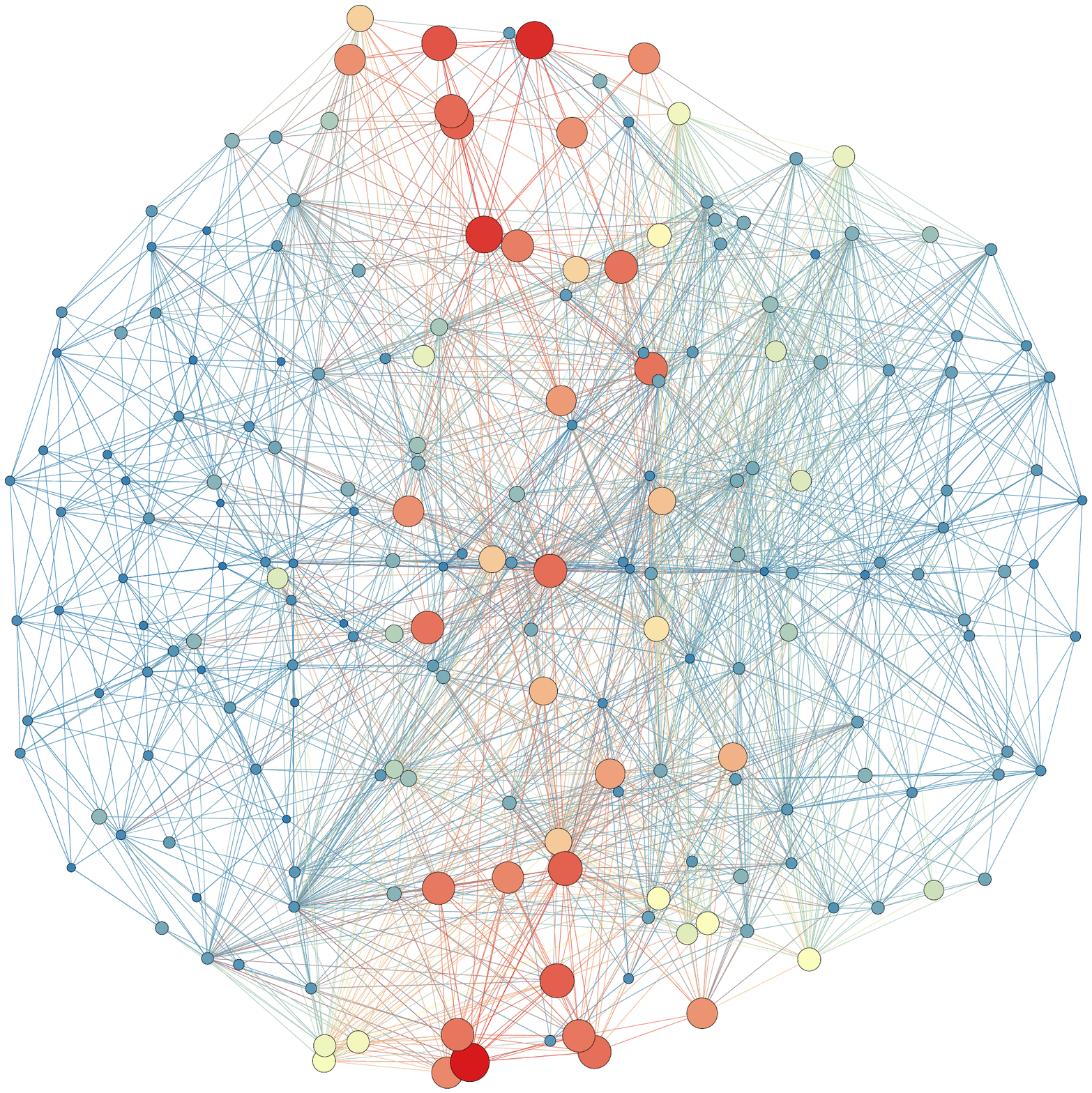}
        \caption{$2^{nd}$ Largest Connected Component}
    \end{subfigure}
    \begin{subfigure}{0.6\textwidth}
        \includegraphics[width=10cm, height=7cm]{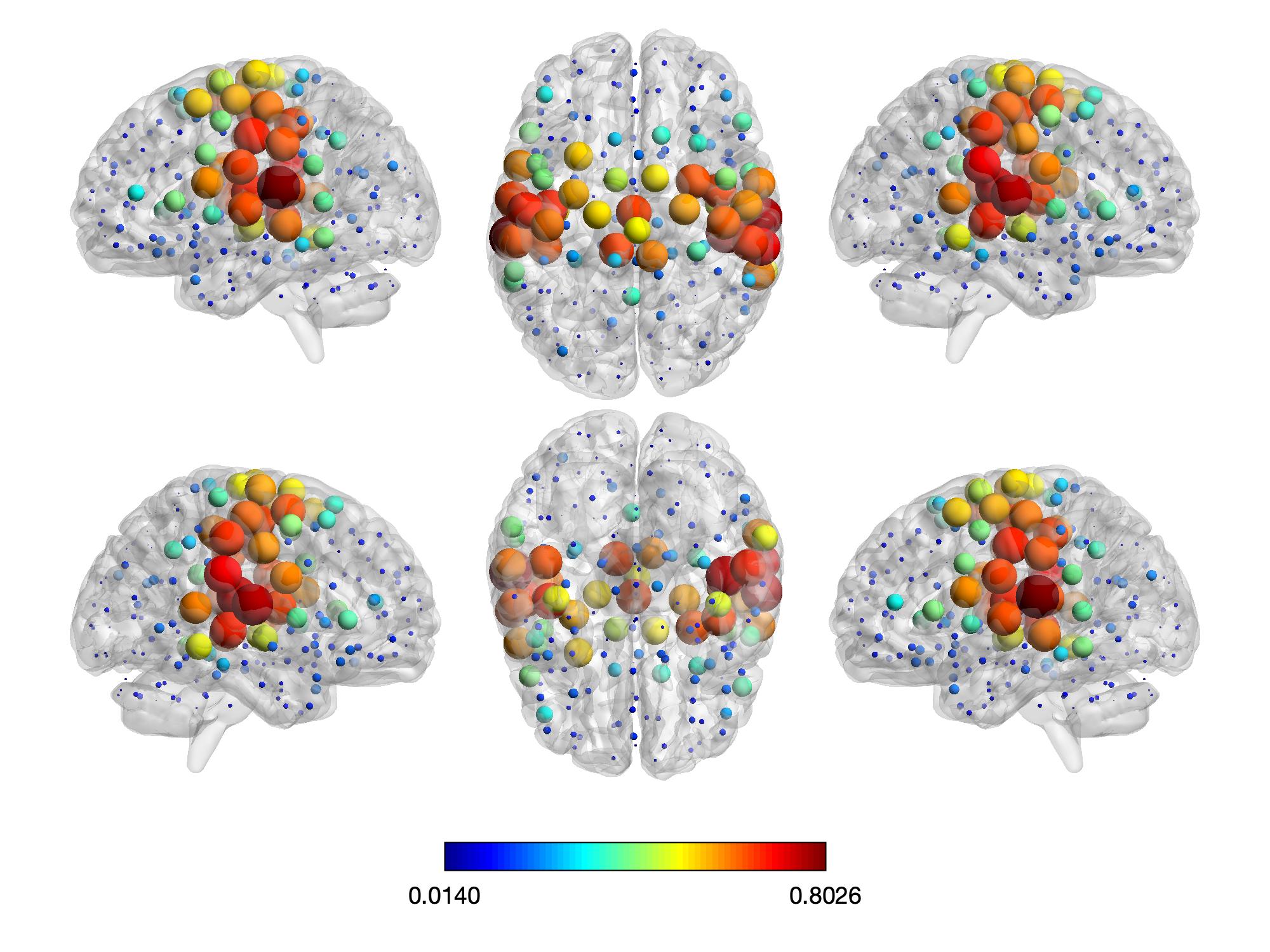}
        \caption{$2^{nd}$ Largest Connected Component (3D view)}
    \end{subfigure}
    \caption{Mean of the first connected components}\label{fig:LCC}
\end{figure}

\begin{figure}[h]
    \begin{subfigure}{0.39\textwidth}
        \includegraphics[width=4.5cm, height=7cm]{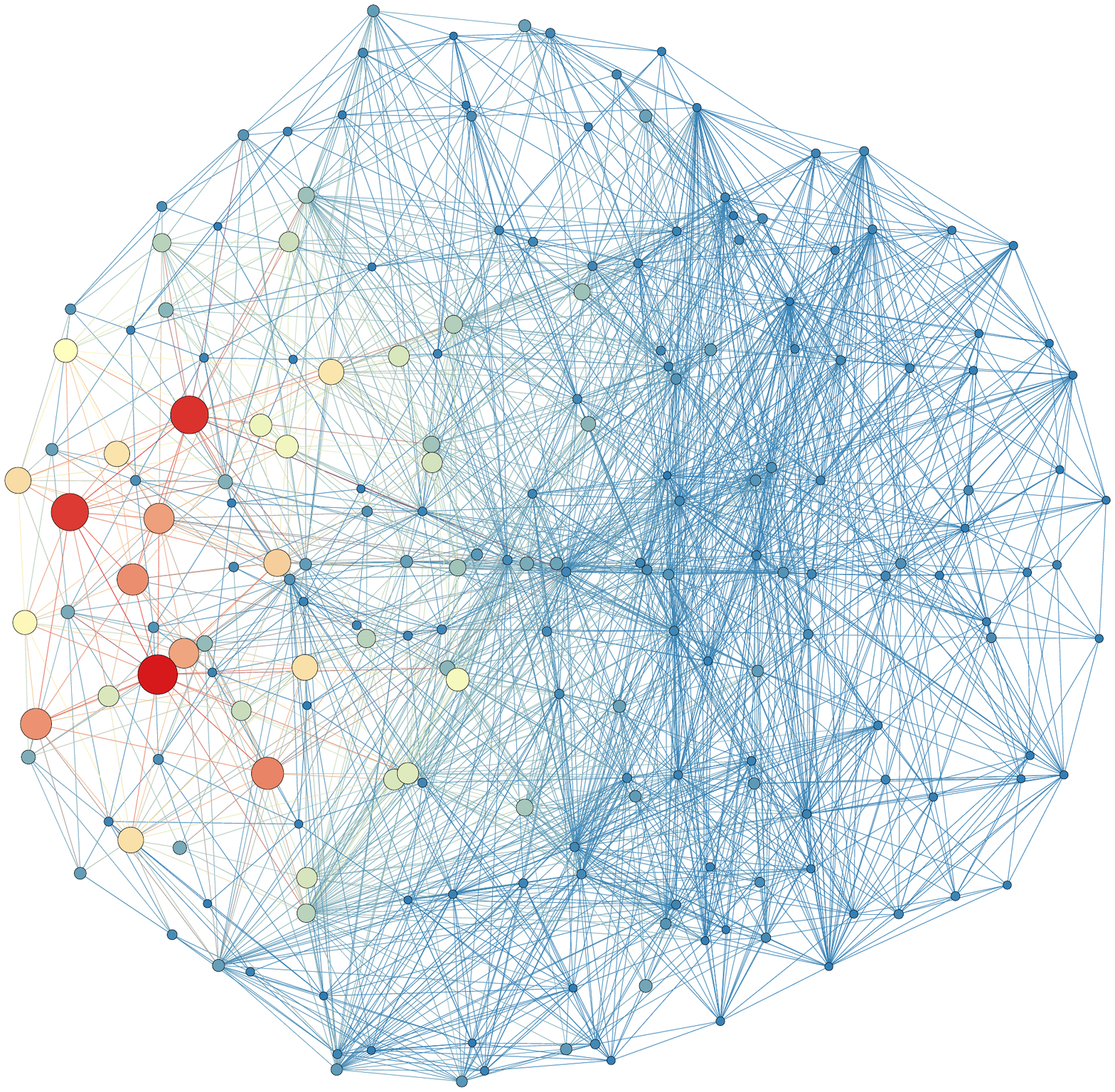}
        \caption{$3^{rd}$ Largest Connected Component}
    \end{subfigure}
    \begin{subfigure}{0.6\textwidth}
        \includegraphics[width=10cm, height=7cm]{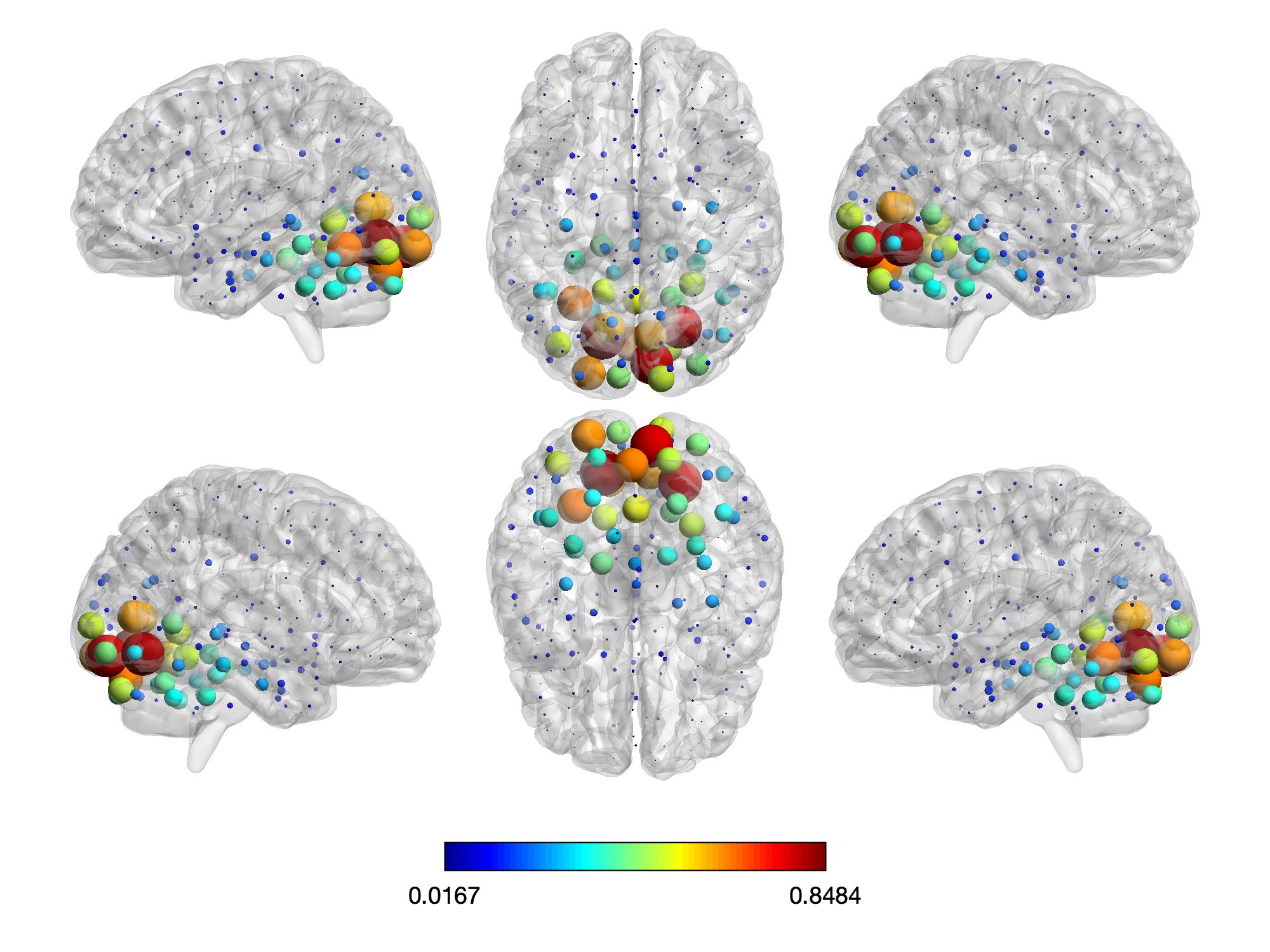}
        \caption{$3^{rd}$ Largest Connected Component (3D view)}
    \end{subfigure}
    \caption{Mean of Component 3}\label{fig:lcc3}

    \begin{subfigure}{0.39\textwidth}
        \includegraphics[width=4.5cm, height=7cm]{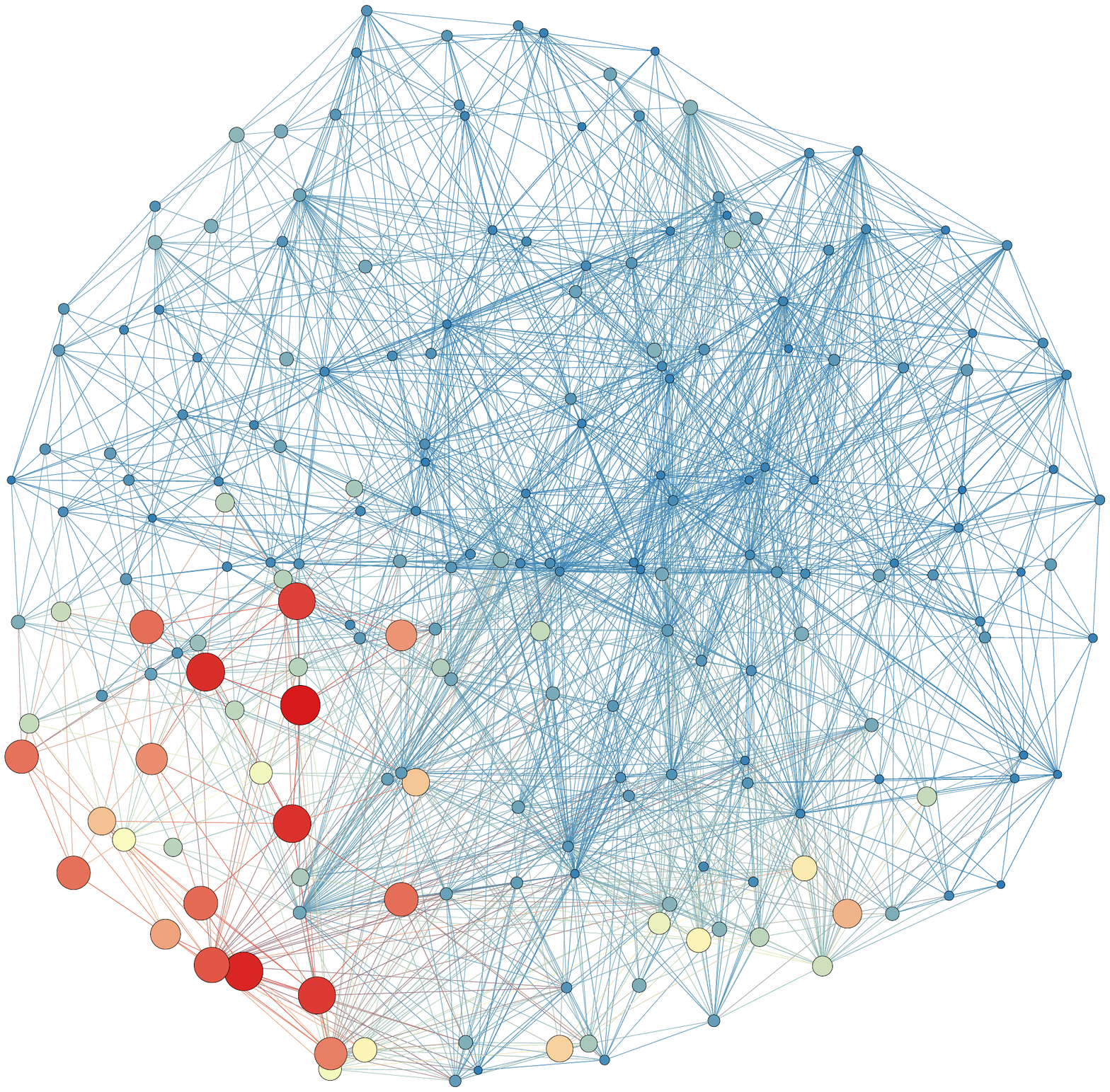}
        \caption{$4^{th}$ Largest Connected Component}
    \end{subfigure}
    \begin{subfigure}{0.6\textwidth}
        \includegraphics[width=10cm, height=7cm]{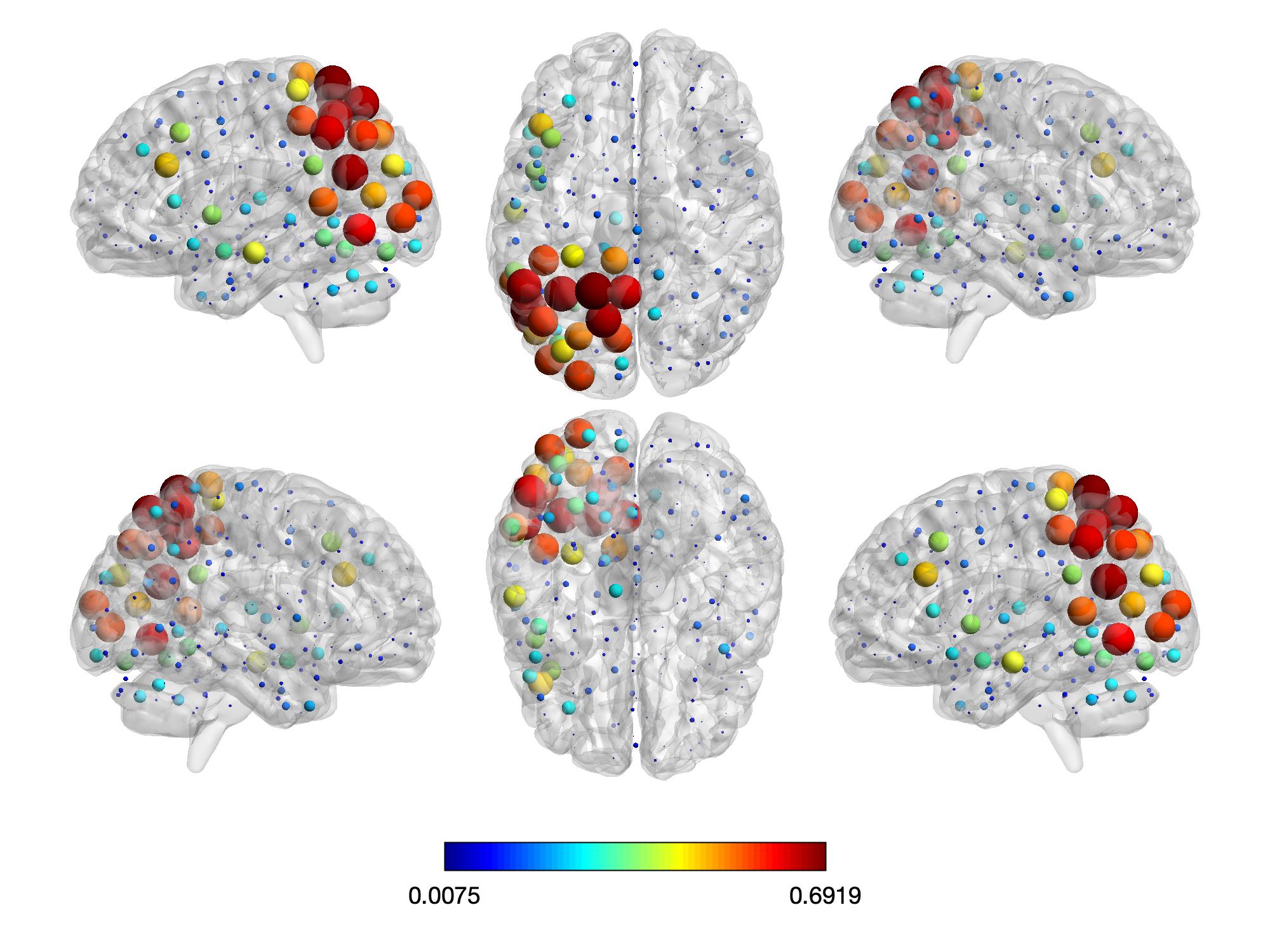}
        \caption{$4^{th}$ Largest Connected Component (3D view)}
    \end{subfigure}
    \caption{Mean of Component 4}\label{fig:lcc4}
\end{figure}

\begin{figure}[h]
    \begin{subfigure}{0.39\textwidth}
        \includegraphics[width=4.5cm, height=7cm]{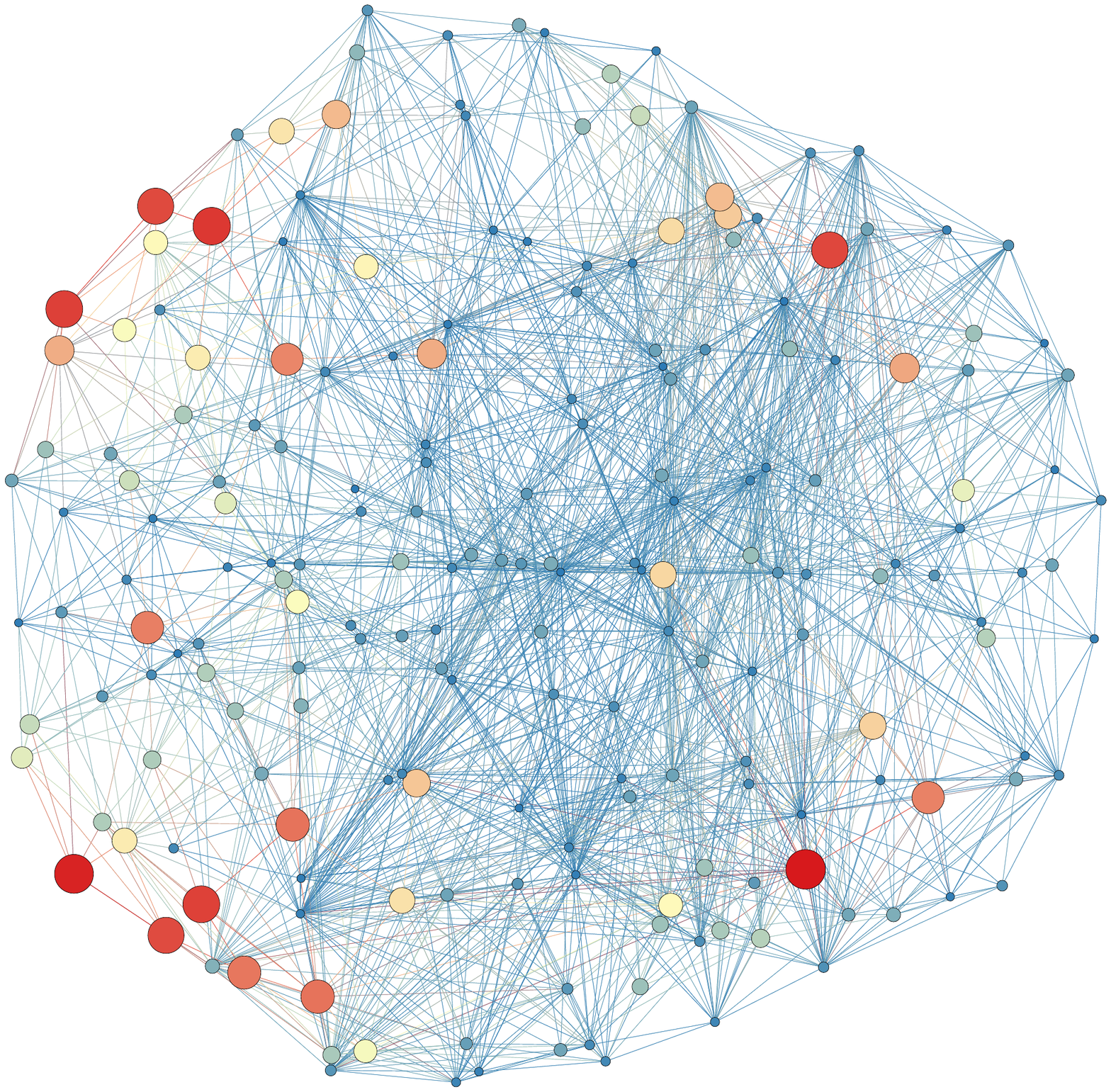}
        \caption{$5^{th}$ Largest Connected Component}
    \end{subfigure}
    \begin{subfigure}{0.6\textwidth}
        \includegraphics[width=10cm, height=7cm]{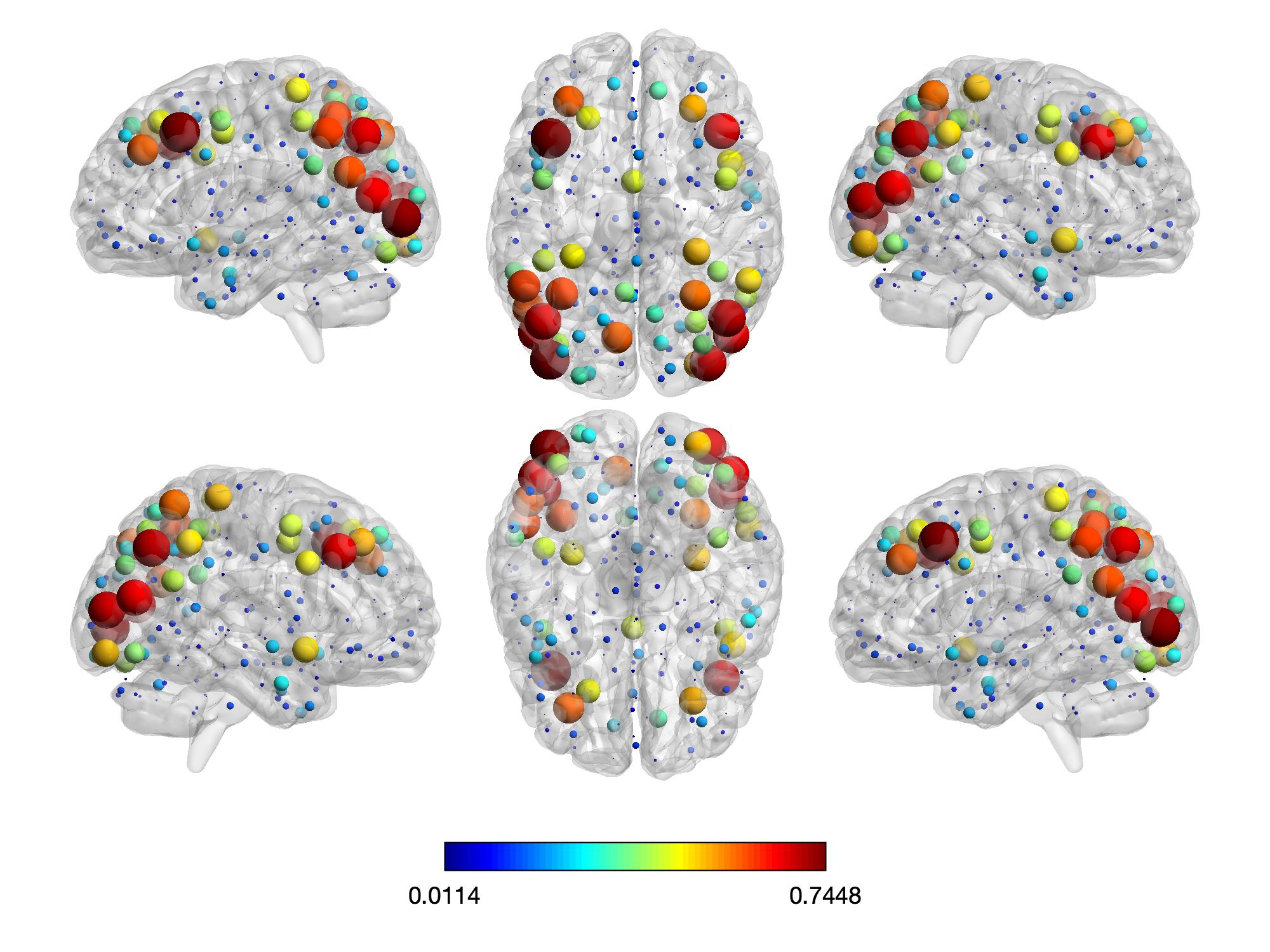}
        \caption{$5^{th}$ Largest Connected Component (3D view)}
    \end{subfigure}
        \caption{Mean of Component 5}\label{fig:lcc5}
\end{figure}

\textit{Robustness across scans.} Finally, we wish to evaluate the model's ability to detect components that are robust across scans and across individuals. To do so, we consider the cross-correlations across components. For a given set of thresholds $\eta \in [0.45,0.95]$, we create a binary similarity matrix between components by setting all correlations above $\eta$ to 1, and below to $0$. We investigate the size of the connected components that this binary similarity yields. We first draw a TSNE plot of the different loadings (Fig. \ref{fig:LCC:tsne}), where each color denotes a subject. We do not observe a clustering by subject, which is reassuring, as the goal of the method is to extract components that generalize across subjects and sessions. The results are displayed in Fig. \ref{fig:LCC_sizes}. Figure \ref{fig:LCC:full}  shows in particular that the largest connected component for very high correlations is big (304 components at threshold 0.8, vs 3 for the Vanilla ICA). This shows that the Bayesian model has successfully identified robust components across scans and subjects. We represent each of these connected communities of loadings by its mean: the average top-5 connected components at threshold 0.8 are displayed in Fig.\ref{fig:LCC}-\ref{fig:lcc5}. Interestingly, we notice that these average most robust loadings are extremely localized: the first one is located in the pre-frontal cortex, the second in the sensory motor cortex. Interestingly, the fifth component defines a network that appears to be in line with the robust connectome findings in \cite{tozzi2019short}.

{}

{}

{}

%% file: conclusion.tex

We have proposed here a version of Bayesian ICA that allows the recovery of biologically interpretable, connected subcomponents from time series that lie on a graph, with automatic self-tuning and minimal user input while providing uncertainty estimates on the recovered loadings. We have also provided a fast numerical approximation of point estimates. Our experiments on synthetic graphs validate the theoretical framework and show that submodules of co-activated nodes can be accurately recovered. The application of our method to real-life data seems to indicate that such an approach does indeed promote reliability and reproducibility among different connectomes, as well as biologically connected components-- a necessary ingredient in brain connectome studies, where the main objective is to relate such subnetworks to cognitive processes. 

%% file: EM_full.tex

We begin by providing a numerical approximation to the constrained ICA approach. While this approach does not allow us to get confidence intervals for the parameters that we are estimating, it will nonetheless produce ICA components that are consistent with our 3 hypotheses of section \ref{sec:bayesian_ica}: non-negative, sparse and localized. This method can be used either for warm-starting the Bayesian HMC algotihm, or by itself if point estimates are sufficient to the analysis.

To do so, we take a step back from the Bayesian model, which requires to estimate both $S$ and $A$, while only $A$ is of true importance in the subnetwork discovery problem. Since  by assumption, $\frac{1}{T}\mathbb{E}[S^TS] =I$, we propose integrating $S$ out of the model altogether, and focusing on finding a solution for $A$ that reconstructs the correlation of the observed time series:
$$ \frac{1}{T} Y^TY  \approx A^T \frac{S^TS}{T}A + \text{Diag}(\gamma^2) \approx  A^TA + \text{Diag}(\gamma^2) $$

This leads us to consider the Stein loss associated to the matrix $A^TA$. Following \cite{naul2017sparse}, the Stein loss between two Positive Definite Symmetric (PDS) matrices $\Sigma$ and $\hat{\Sigma} \in \mathbb{R}^{N \times N}$ is defined as:
$$ L(\hat{\Sigma},\Sigma) =  tr(\hat{\Sigma} \Sigma^{-1}) - \log(| \hat{\Sigma} \Sigma^{-1}|) -N$$
Using $\Sigma=S = \frac{Y^TY}{T}$ the sample correlation matrix, and $ \hat{\Sigma} = A^TA + \text{diag}(\gamma_i^2)  $, we want to minimize:
\begin{equation}
    \begin{split}
        \text{Minimize }_{\hat{\Sigma}}& tr(\hat{\Sigma} S^{-1}) - \log(| \hat{\Sigma} S^{-1}|) -N\\
        \text{ such that } &  \hat{\Sigma} = A^TA + \text{diag}(\gamma_i)\\
        & \text{diag}(\hat{\Sigma}) = 1\\
        & \forall i \leq N, \quad \gamma_i \in (0,1) \\
        & ||A||_1 \leq s \hspace{1.15cm} \text{(sparsity)}\\
        &\forall k, A_kLA_k^T \leq \rho \hspace{0.2cm} \text{(connectedness)}\\
    \end{split}{}
\end{equation}
Formulated as such, this problem amounts to finding an optimal low-rank decomposition under Stein loss that is also aligned with our original sparsity and localization assumptions. However, this loss is convex as a function of $\hat{\Sigma}$ but not a a function of $A$. To solve this new problem, following \cite{fletcher1980first}, we use a composite algorithm  \citep{fletcher1982model,duchi2018stochastic}, which linearizes the previous loss and provides a series of convex objectives to sequentially optimize:
\begin{equation}
    \begin{split}
        \text{Minimize}_{\hat{\Sigma}} & tr(\hat{\Sigma} \Sigma^{-1}) - \log(| \hat{\Sigma} \Sigma^{-1}|) -p\\
        \text{ such that } &  \hat{\Sigma} = A^{(k)T}A^{(k)} + A^{(k)T} X + X^T A^{(k)}   + \text{diag}(\gamma_i)\\
        & \text{diag}(\hat{\Sigma}) = 1\\
          & \forall i \leq N, \quad \gamma_i \in (0,1) \\
         &  \forall j \leq K, \quad || X_{j\cdot} ||_2 \leq \delta \\
         & A^{(k+1)} =  A^{(k)} + X \geq 0 \hspace{1.5cm} \text{(non negative entries)} \\
        & ||A^{(k+1)}||_1 \leq s \hspace{3cm} \text{(sparsity)}\\
        &\forall k, A^{(k+1)}LA^{(k+1)T} \leq \rho \hspace{1.5cm} \text{(connectedness)}\\
    \end{split} \label{eq:linearized} 
\end{equation}
Each iteration in Eq. \ref{eq:linearized} is now convex and we thus know that there exists a global optimal solution. Note that while each iteration is convex, the overall problem remains non-convex, and the solution might thus heavily depend on global initialization. We suggest use as warm start the absolute value of the Vanilla ICA loadings, as these can typically be very efficiently computed and provide good starting points\footnote{The code is publicly available at \url{https://github.com/donnate/ConstrainedICA}}.\\

The algorithm operates as follows:
\vspace{-0.1cm}
\begin{enumerate}
    \item Optimize Eq. \ref{eq:linearized} with respect to $X$ and $\gamma^2$
    \item Solve for $A^{(k+1)}$
    \item Iterate until convergence
\end{enumerate}

\xhdr{(A) Accurate resolution of the problem using CVXPY} As a proof of concept, we solve the previous problem using CVXPY \citep{diamond2016cvxpy}. CVXPY allows to find accurate solutions to the previous problem, and we use it here to check if the algorithm works efficiently on our synthetic example of section \ref{sec:exp}.

\begin{figure}[H]
    \centering
    \begin{subfigure}{0.49\textwidth}
    \includegraphics[width=\textwidth]{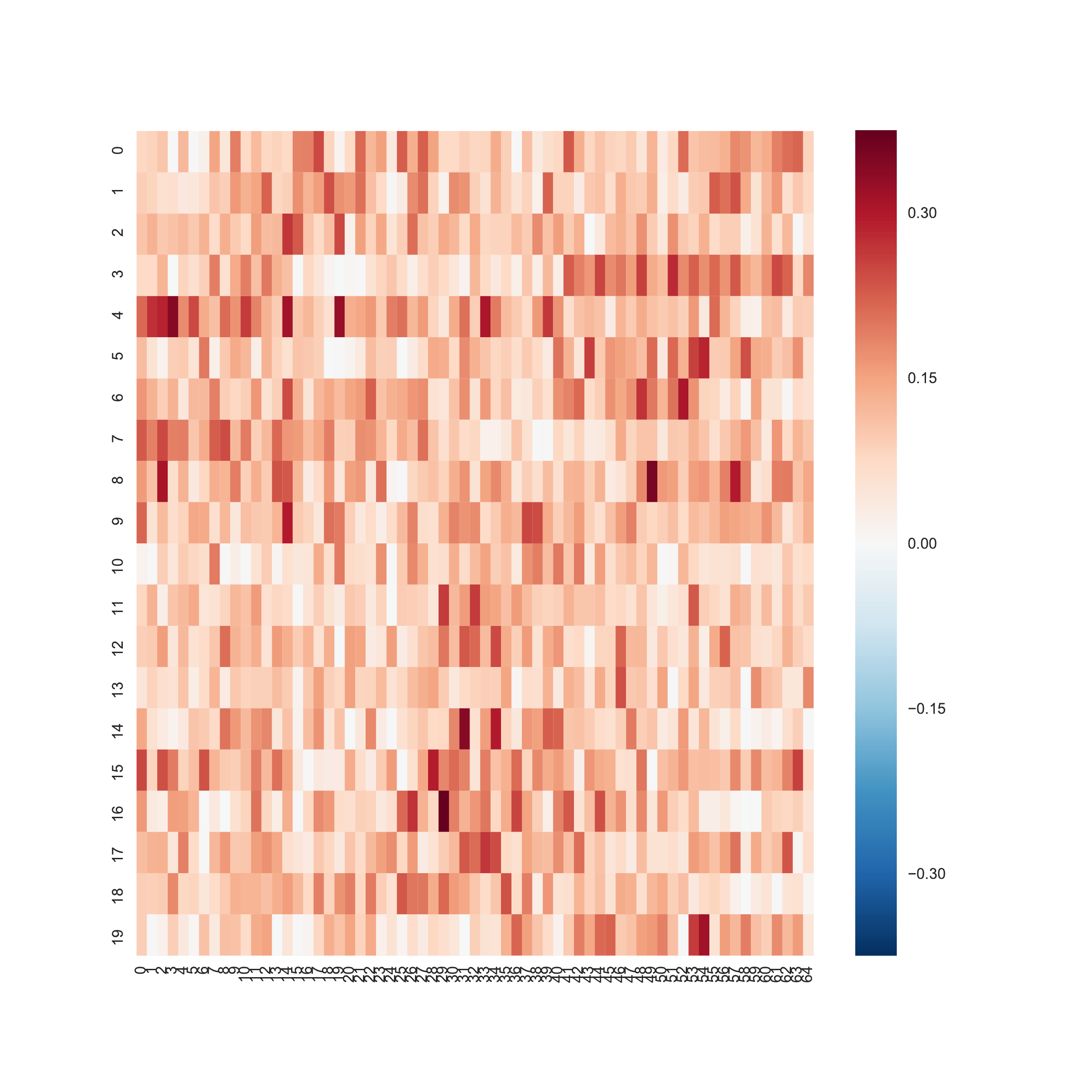}
    \caption{Random initialization of the Components}
    \label{fig:CVXPY:1}
    \end{subfigure}
        \begin{subfigure}{0.49\textwidth}
    \includegraphics[width=\textwidth]{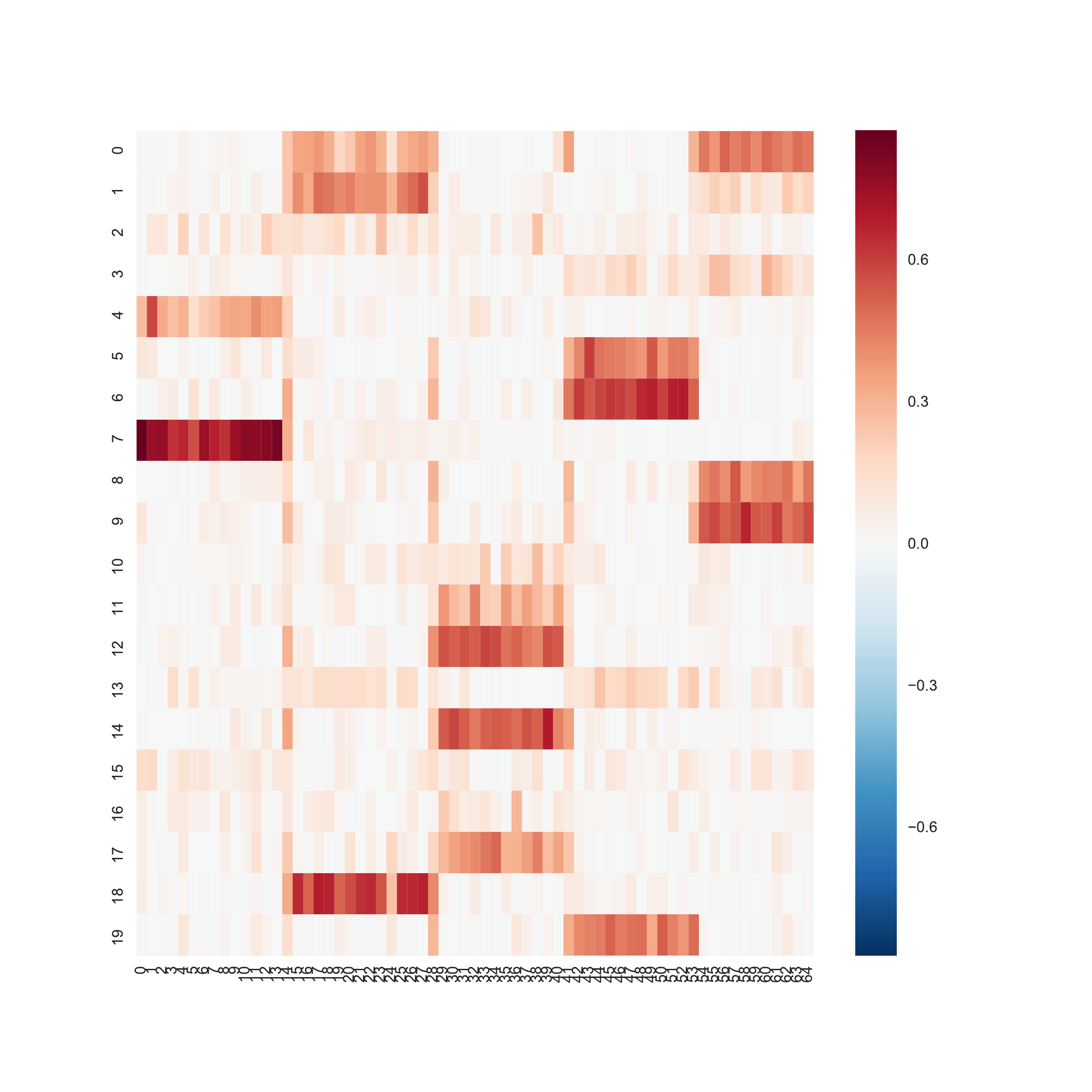}
    \caption{Components at iteration 10}
    \label{fig:CVXPY:2}
    \end{subfigure}
    
    \begin{subfigure}{0.49\textwidth}
    \includegraphics[width=\textwidth]{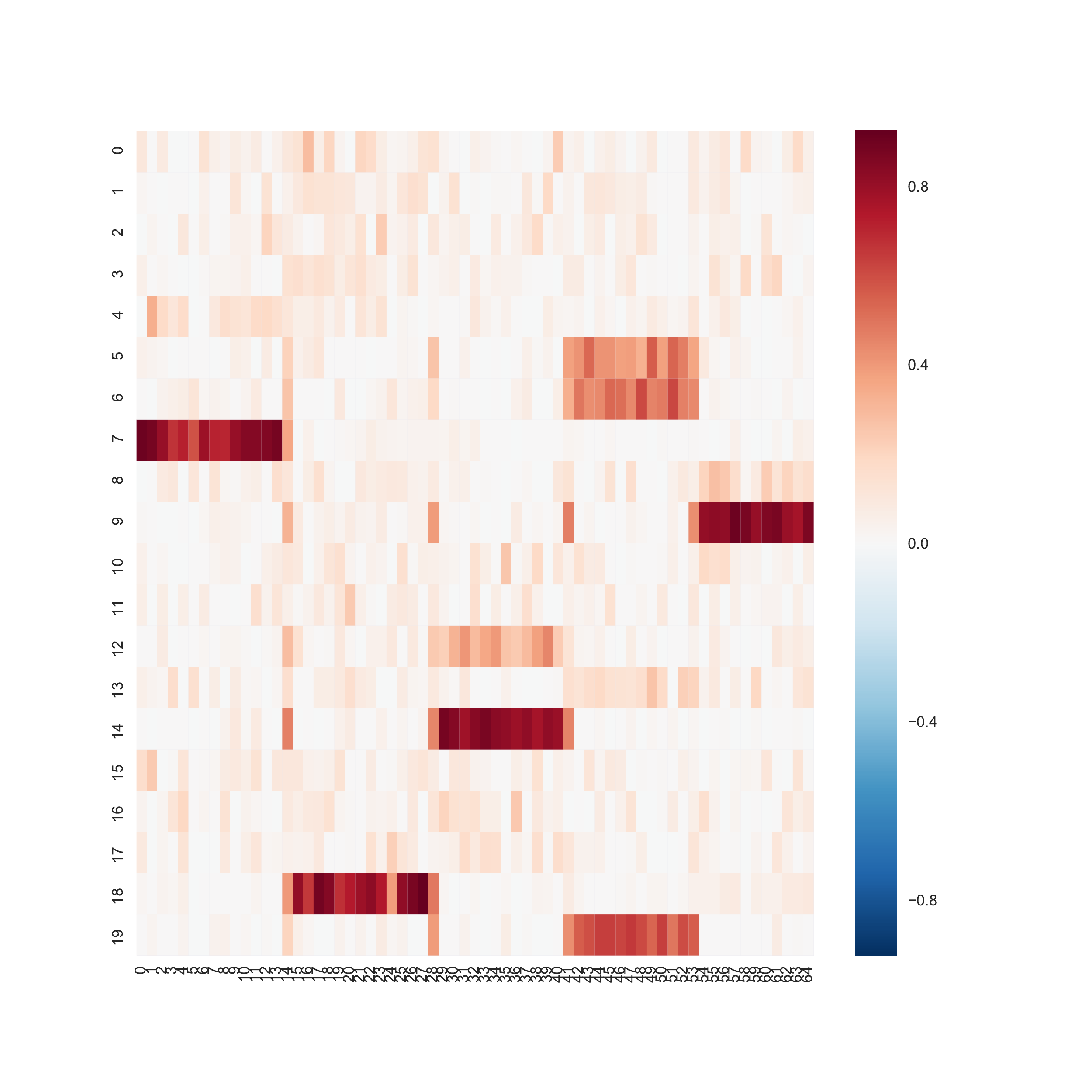}
    \caption{Components at iteration 75}
    \label{fig:CVXPY:3}
    \end{subfigure}
        \begin{subfigure}{0.49\textwidth}
    \includegraphics[width=\textwidth]{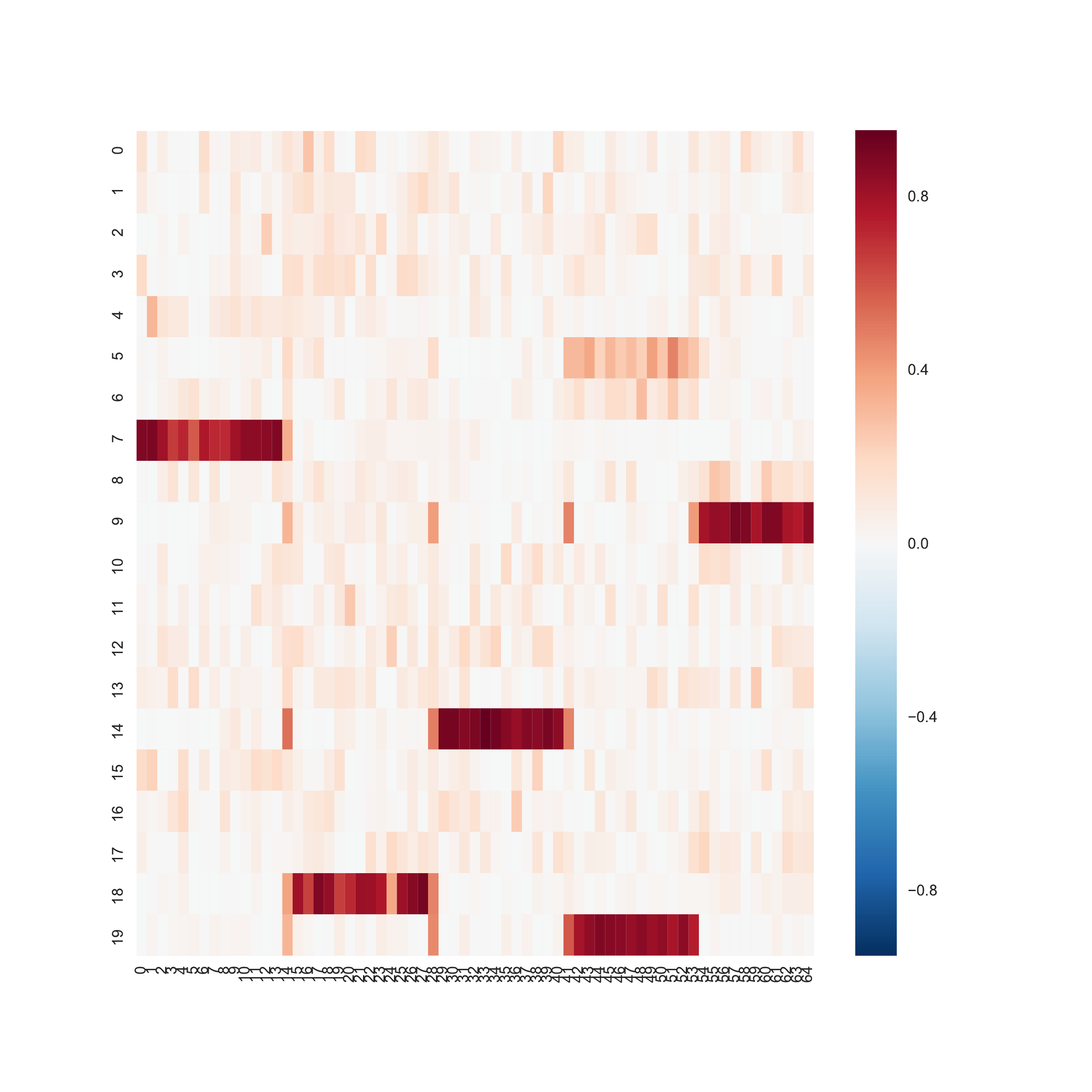}
    \caption{Components at iteration 200}
    \label{fig:CVXPY:4}
    \end{subfigure}
\caption{Visualization of the recovered components at various steps of the optimization process.}\label{fig:CVXPY}
\end{figure}

Fig. \ref{fig:CVXPY} allows us to validate the method: we see that, as the number of iteration increases, the recovered components become sparser and sparser. In particular, it is interesting to note that after only 10 iterations, the Ground-Truth components are recovered (the colored  blocks in \ref{fig:CVXPY:2} match the localized ground truth components that we expect to see). However, the components are redundant. As the number of iteration increases, redundant components fade away and the algorithm recovers both the number of ground truth components (5, in this case) and  the ground truth components themselves accurately (Fig. \ref{fig:CVXPY:3} and \ref{fig:CVXPY:4}).

\vspace{-0.5cm}

\xhdr{(B) Scalable resolution of the problem} While, for smaller problems, this can accurately be solved using numerical solvers such as CVXPY, for large problems, we need to find more efficient solvers, and potentially compromise between accuracy and scalability. The previous paragraph has highlighted the potential of the algorithm to extract accurate components, and we now turn to a more scalable derivation of the updates of our two-step algorithm. \\
\textbf{Step 1: Gradient descent to solve for $X$ and $\gamma^2$.}The first step focuses on solving the following objective:
\begin{equation}
    \begin{split}
        \text{Minimize }_{\hat{\Sigma}}& tr(\hat{\Sigma} S^{-1}) - \log(| \hat{\Sigma}|) \\
        \text{ such that } &  \hat{\Sigma} = A^{(k)T}A^{(k)} + A^{(k)T}X  +X^TA^{(k)T} + \text{diag}(\gamma_i)\\
        & \text{diag}(\hat{\Sigma}) = 1\\
                  & \forall i \leq N, \quad \gamma_i \in (0,1) \\
        & \forall i \leq N, \quad  ||X_i|| \leq \delta\\
    \end{split} \label{eq:linearized} 
\end{equation}

\begin{equation}
    \begin{split}
     \iff   \text{Minimize }_{X, \gamma}& 2 tr( X S^{-1}A^{(k)T}) +tr( \text{Diag}(\gamma) S^{-1})\\
     &- \log\Big(|A^{(k)^T}A^{(k)}  + A^{(k)^T}X  +X^TA^{(k)} +\text{Diag}(\gamma) |\Big) \\
        \text{ such that } &  \text{diag}\Big((A^{(k)})^TA^{(k)}  + A^{(k)^T}X  +X^TA^{(k)} + \text{diag}(\gamma_i) \Big) =1\\
        & \forall i \leq N,\quad  ||X_i||_2 \leq \delta \\
                          & \forall i \leq N, \quad \gamma_i \in (0,1) \\
    \end{split} \label{eq:linearized2} 
\end{equation}
We solve the previous problem using a gradient descent method. In particular, here, we propose using FISTA \citep{beck2009fast} updates, which provide an optimal step size (thus foregoing the need for the practioner to select it) and are typically fast to converge. Derivations of the sequential steps are provided in the subsequent paragraphs.\\

\textit{Computing the gradient with respect to $X$.} The objective in Eq. \ref{eq:linearized2} are convex with respect to $X$, and each column of $X$ is in the ball of radius $\delta$. All we need to show now if that the gradients are Lipschitz in order to apply the FISTA updates.\\

We begin by deriving the gradients and updates for $X$ and $\gamma$ in Eq. \ref{eq:linearized2}
\begin{equation}
    \begin{split} 
    &\log(| (A^TA + A^T (X + tH) +(X + tH)^TA + \text{Diag}(\gamma) )|)\\
   = &\log(| M  + t(H^TS + A^TH) S^{-1}|) \quad {\footnotesize{\text{ where } M = (A^TA + A^TX +X^TA + \text{Diag}(\gamma)})}\\
   = &\log(| M|) +\log(|I  + tM^{-1}(H^TS + A^TH) |)  +o(t)\\
   = &\log(| M|) +t\text{Trace}\big[M^{-1}(H^TA + A^TH)\big] +o(t)\\
   = &\log(| M|) +t\text{Trace}\big[H^TAM^{-1} + M^{-1}A^TH\big] +o(t)\\
      = &\log(| M|) +2t\text{Trace}\big[H^TAM^{-1}\big] +o(t)\\
        \end{split} \label{eq:derivatives2} 
\end{equation}
Thus, noting that $M$ is symmetric:
$$\nabla_X (- \log(| A^TA + 2A^T (X + tH) |))  = -2(M^{-1}A^T)^T = -2AM^{-1},$$
and the full gradient with respect to $X$ is thus given by:
$$ \nabla_{X} L = 2A(S^{-1} - M^{-1}) $$
Now, defining $K = A^TA +\text{Diag}(\gamma)$, we also have:
\begin{equation*}
    \begin{split}
       M &= A^TA +X^TA + A^TX + \text{Diag}(\gamma^2)\\
       &=K^{1/2}[ I_N +  K^{-1/2}(X^TA + A^TX ) K^{-1/2}   ]K^{1/2} \\ 
       \implies M^{-1} &\approx K^{-1/2}[ I_N -  K^{-1/2}(X^TA + A^TX ) K^{-1/2}   ]K^{-1/2}\\
              \implies AM^{-1} &\approx AK^{-1/2}[ I_N -  K^{-1/2}(X^TA + A^TX ) K^{-1/2}   ]K^{-1/2}\\
    \end{split}
\end{equation*}
Hence:
$$  || \nabla_X L(X_1,\gamma) - \nabla_X L(X_2,\gamma) ||\leq 2 ||AK^{-1}(X_1 -X_2)^T A K^{-1} + AK^{-1}A^T(X_1 -X_2) K^{-1}||$$
$$||\nabla_{X} L(X_1) - \nabla_{X} L(X_2)|| \leq  \sqrt{8}(||AK^{-1}|| \times  ||K^{-1}| \times || A|| ) \times ||X_2-X_1||$$
Thus the gradient of the loss in Eq. \ref{eq:linearized2} is Lipschitz with a constant upper-bounded by  $ L_X = \sqrt{8}(||AK^{-1}|| \times  ||K^{-1}| \times || A|| )$.

\noindent \textit{Computing the gradient with respect to $\gamma$.} Similarly:
\begin{equation}
    \begin{split} 
    &\log(| A^TA + X^TA + A^T X   + \text{Diag}((\gamma +th)) |)\\
    =&\log(| M  + t  \text{Diag}(h)|)\\
     =&\log(| M|) +\log(|I  + tM^{-1}\text{Diag}(h )|)\\
 =& \log(|M|) + t Tr(M^{-1}\text{Diag}(h )) +o(t)
        \end{split} \label{eq:derivatives2} 
\end{equation}
$$ \nabla_{\gamma} L = \text{diag}(S^{-1}) -\text{diag}(M^{-1}) =\text{diag} (S^{-1} -M^{-1} ) $$
The gradient of the loss with respect to $\gamma$ is thus also Lipschitz, with constant upper-bounded by $L_{\gamma} = \max \{ \text{Diag}(  K_2^{-1}) \}^2$:
$$ \Big( \nabla_{\gamma} L(X, \gamma_1)  -  \nabla_{\gamma} L(X, \gamma_2) \Big) \leq \max \{ \text{Diag}(  K_2^{-1}) \}^2 ||\gamma_1 -\gamma_2|| $$
where $K_2 = A^TA + A^TX + X^T A$. 
The FISTA update for X are summarized in Alg. \ref{alg:fistaX}.
\begin{minipage}{0.49\textwidth}
\begin{algorithm}[H]
\SetAlgoLined
\KwResult{Updated $X^{(k)}$}
 \While{not converged}{
 $X_k = p_{L_X}(Y_X)$\;
 $X_k = \Pi_{\mathcal{B}_{\delta}}\big(Y_X - \frac{1}{L_X} \nabla_X L \big) $\;
 $t_{k+1} = \frac{1 + \sqrt{1 + 4t_k^2}}{2}$\;
  $Y_{X} = {X}_k + \frac{t_k -1}{t_{k+1}} (X_k-{X}_{k-1})$\;
 }
 \caption{FISTA Updates for $X$}\label{alg:fistaX}
\end{algorithm}
\end{minipage}
\begin{minipage}{0.49\textwidth}
\begin{algorithm}[H]
\SetAlgoLined
\KwResult{$\gamma^{(k)}$}
 \While{not converged}{
 $\gamma_k = p_{L_\gamma}(Y_{\gamma})$\;
 $\iff \gamma_k = \Pi_{[0,1]}\big(Y_\gamma - \frac{1}{L_\gamma} \nabla_\gamma L \big) $\;
 $t_{k+1} = \frac{1 + \sqrt{1 + 4t_k^2}}{2}$\;
  $Y_{\gamma} = {\gamma}_k + \frac{t_k - 1}{t_{k+1}} ({\gamma}_k-{\gamma}_{k-1})$\;
 }
 \caption{FISTA Updates for $\gamma$}\label{alg:fistaG}
\end{algorithm}
\end{minipage}

\xhdr{Step 2: Updating $A^{(k+1)}$}
Step 2 can be easily solved by introducing the augmented Lagrangian of the problem and solving:

\begin{equation}
    \begin{split} 
\min_A & \frac{1}{2} || (A^{(k)} + X) - A||^2  + s || A||_1 + \frac{\rho}{2} Trace(ALA^T)\\
          \text{such that }      & \text{Diag}(A)  + \text{Diag}(\gamma) =1\\
                           \end{split} 
\end{equation}
We propose to solve this problem using FISTA updates, and projecting onto the subspace $\Delta = \{ M \in \mathbb{R}^{N \times N}: \forall i, M_{ii}=1 \}$ of matrices with diagonal equal to 1. The gradient here is given by:
$$ \nabla L_2 = (A -  (A^{(k)} + X )) + \rho AL  $$
which is thus Lipschtiz with constant  $L_A = || I + \rho * L||$
The FISTA updates for solving Eq. \ref{eq:linearized}, combined with those for $X$ and $\gamma$  are summarized in Algorithm \ref{al:FISTA}.

\begin{algorithm}[H]
\SetAlgoLined
\KwResult{subnetworks $A$}
$A = |A_{\text{Vanilla ICA}}|$\;
 \While{not converged}{
 $X_k = \Pi_{\mathcal{B}_{\delta}}\big(Y_X - \frac{1}{L_X} \nabla_X L \big) $\;
 $ \gamma_k = \Pi_{[0,1]}\big(Y_\gamma - \frac{1}{L_\gamma} \nabla_\gamma L \big)$\;
 $t_{k+1} = \frac{1 + \sqrt{1 + 4t_k^2}}{2}$\;
 $Y_X = X_k + \frac{t_k -1}{t_{k+1}} (X_k-X_{k-1})$\;
  $Y_{\gamma} = {\gamma}_k + \frac{t_k -1}{t_{k+1}} ({\gamma}_k-{\gamma}_{k-1})$\;
 $ A_k = \Pi_{\Delta}(Y_{A} - \frac{1}{L_A} \nabla_A L )$\;
 $Y_A = A_k + \frac{t_k -1}{t_{k+1}} (A_k-A_{k-1})$\;
 }
 \caption{FISTA-based Numerical ICA for localized and sparse components}\label{al:FISTA}
\end{algorithm}


%% file: appendix_scan.tex

\xhdr{HNU - Hangzhou Normal University- dataset}\footnote{All the information in this subsection was gathered on the study's \hyperlink{http://fcon_1000.projects.nitrc.org/indi/CoRR/html/hnu_1.html}{{\color{blue}website}}.}

This sample includes 30 healthy adults, aged 20 to 30. Each participant received ten scans across one month, one scan every three days. Five modalities (EPI/ASL/T1/DTI/T2) of brain images were acquired for all subjects. During functional scanning, subjects were presented with a fixation cross and were instructed to keep their eyes open, relax and move as little as possible while observing the fixation cross. Subjects were also instructed not to engage in breath counting or meditation. All imaging data was collected on 3T GE Discovery MR 75 using an 8-channel head coil \citep{kong2018spatial}.

\subsection{Functional fMRI Pre-processing}\label{appendix:functional}
The fMRI data used in this paper is the time series corresponding to the \hyperref[http://preprocessed-connectomes-project.org/abide/Pipelines.html]{Craddock-200 atlas}. Functional parcellation was accomplished using a two-stage spatially-constrained functional procedure applied to preprocessed and unfiltered resting state data corresponding to 41 individuals from an independent dataset (age: 18–55; mean 31.2; std. dev. 7.8; 19 females). A grey matter mask was constructed by averaging individual-level grey matter masks derived by automated segmentation. Individual-level connectivity graphs were constructed by treating each within-gm-mask voxel as a node and edges corresponding to super-threshold temporal correlations to the voxels’ 3D (27 voxel) neighborhood. Each graph was partitioned into 200 regions using normalized cut spectral clustering. Association matrices were constructed from the clustering results by setting the connectivity between voxels to 1 if they are in the same ROI and 0 otherwise. A group-level correspondence matrix was constructed by averaging the individual level association matrices and subsequently partitioned into 200 regions using normalized cut clustering. The resulting group-level analysis was fractionated into functional resolution using nearest-neighbor interpolation. \\

The pre-processing was done according to the Conﬁgurable Pipeline for the Analysis of Connectomes (\hyperlink{http://fcp-indi.github.com}{C-PAC}). This python-based pipeline tool makes use of AFNI, ANTs, FSL, and custom python code.

As per the \hyperlink{http://preprocessed-connectomes-project.org/abide/cpac.html}{ABIDE's detailed explanations}, this pipeline includes:
\begin{itemize}
    \item {\it \bf Structual Preprocessing}
    \begin{enumerate}
        \item Skull-stripping using AFNI’s 3dSkullStrip.
        \item Segment the brain into three tissue types using FSL’s FAST,
        \item Constrain the individual subject tissue segmentations by tissue priors from standard space provided with FSL.
        \item Individual skull stripped brains were normalized to Montreal Neurological Institute (MNI)152 stereotactic space (1 $mm^3$ isotropic) with linear and non-linear registrations using ANTs.
    \end{enumerate}
    \item {\it \bf  Functional Preprocessing}
       \begin{enumerate}
      \item Slice time correction using AFNI’s 3dTshift
       \item Motion correct to the average image using AFNI’s 3dvolreg (two iterations)
       \item Skull-strip using AFNI’s 3dAutomask
         \item Global mean intensity normalization to 10,000
        \item  Nuisance signal regression was applied including motion parameters:
        \begin{itemize}
            \item 6 head motion parameters, 6 head motion parameters,  and the 12 corresponding squared items

            \item top 5 principal components from the signal in the white-matter and cerebro-spinal fluid derived from the prior tissue segmentations transformed from anatomical to functional space
    \item linear and quadratic trends
        \end{itemize}
    \item Band-pass filtering (0.01-0.1Hz) was applied for only for one set of strategies
    \item Functional images were registered to anatomical space with a linear transformation and then a white-matter boundary based transformation using FSL’s FLIRT and the prior white-matter tissue segmentation from FAST
    \item The previous anatomical to standard space registration was applied to the functional data in order to transform them to standard space.
    \end{enumerate}
\end{itemize}

\begin{figure}
    \centering
    \includegraphics[width=\textwidth]{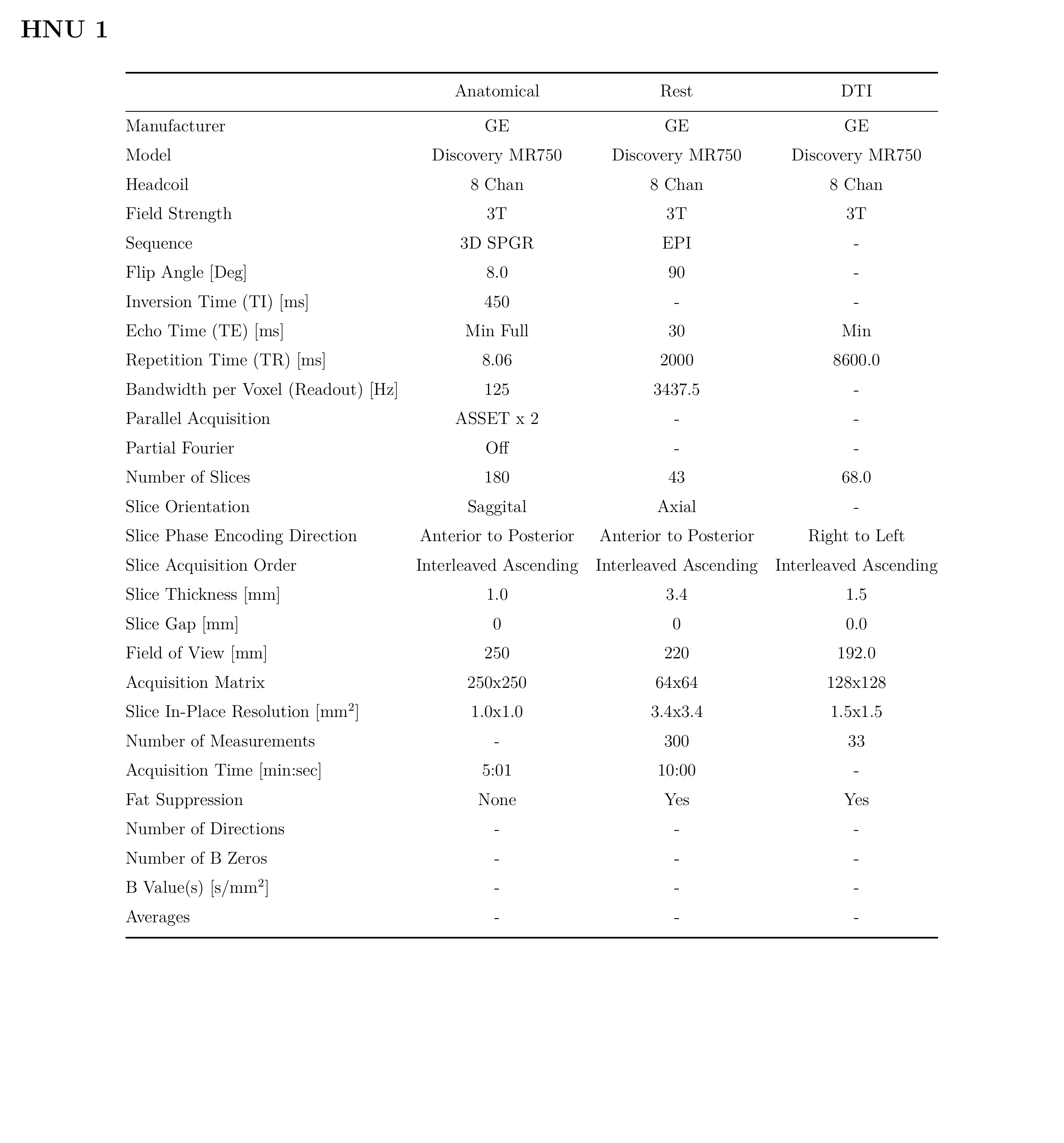}
    \caption{Scan Parameters for the HNU1 dataset.}
    \label{fig:scaHNU}
\end{figure}

\begin{figure}[H]
    \centering
    \begin{subfigure}{0.49\textwidth}
    \includegraphics[width=\textwidth]{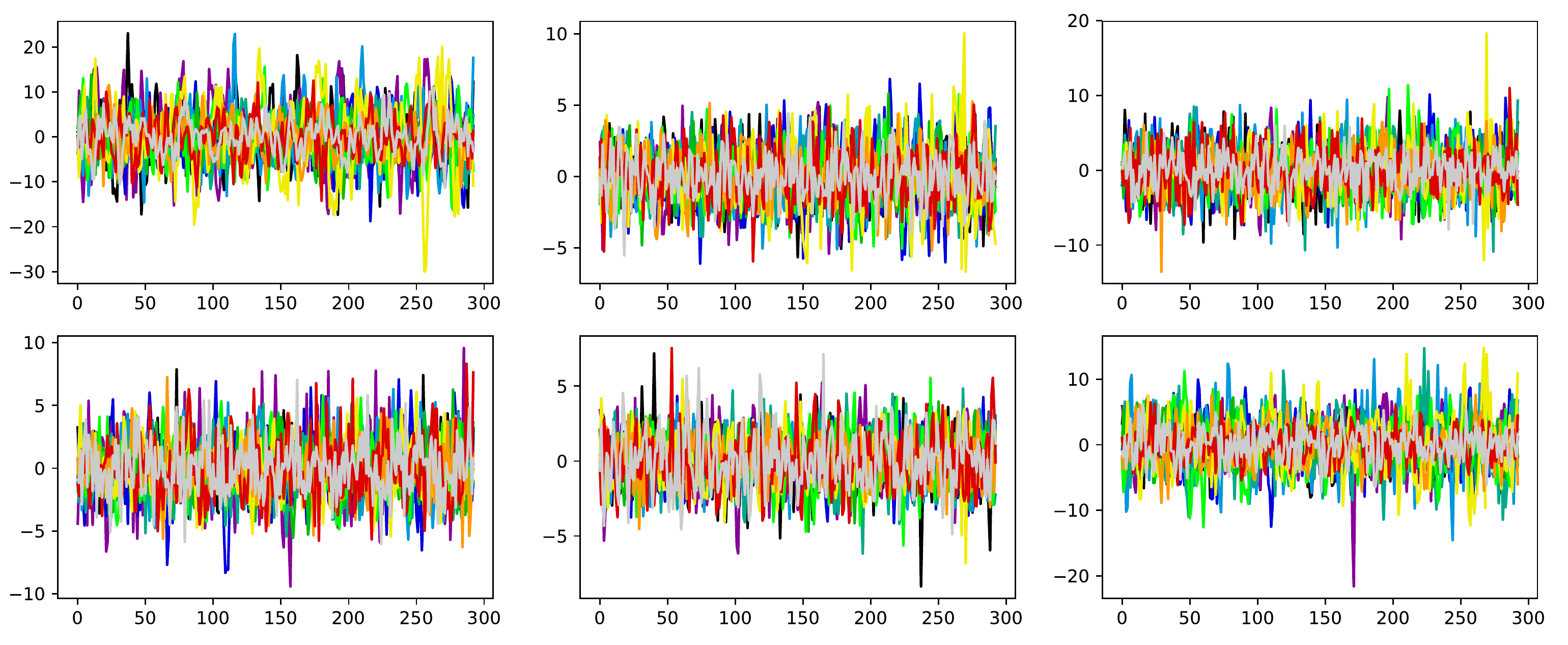}
    \caption{Time Series for a randomly selected subset of nodes across different subjects and scans}
    \label{fig:TS_examples}
    \end{subfigure}
    \begin{subfigure}{0.49\textwidth}
    \includegraphics[width=\textwidth]{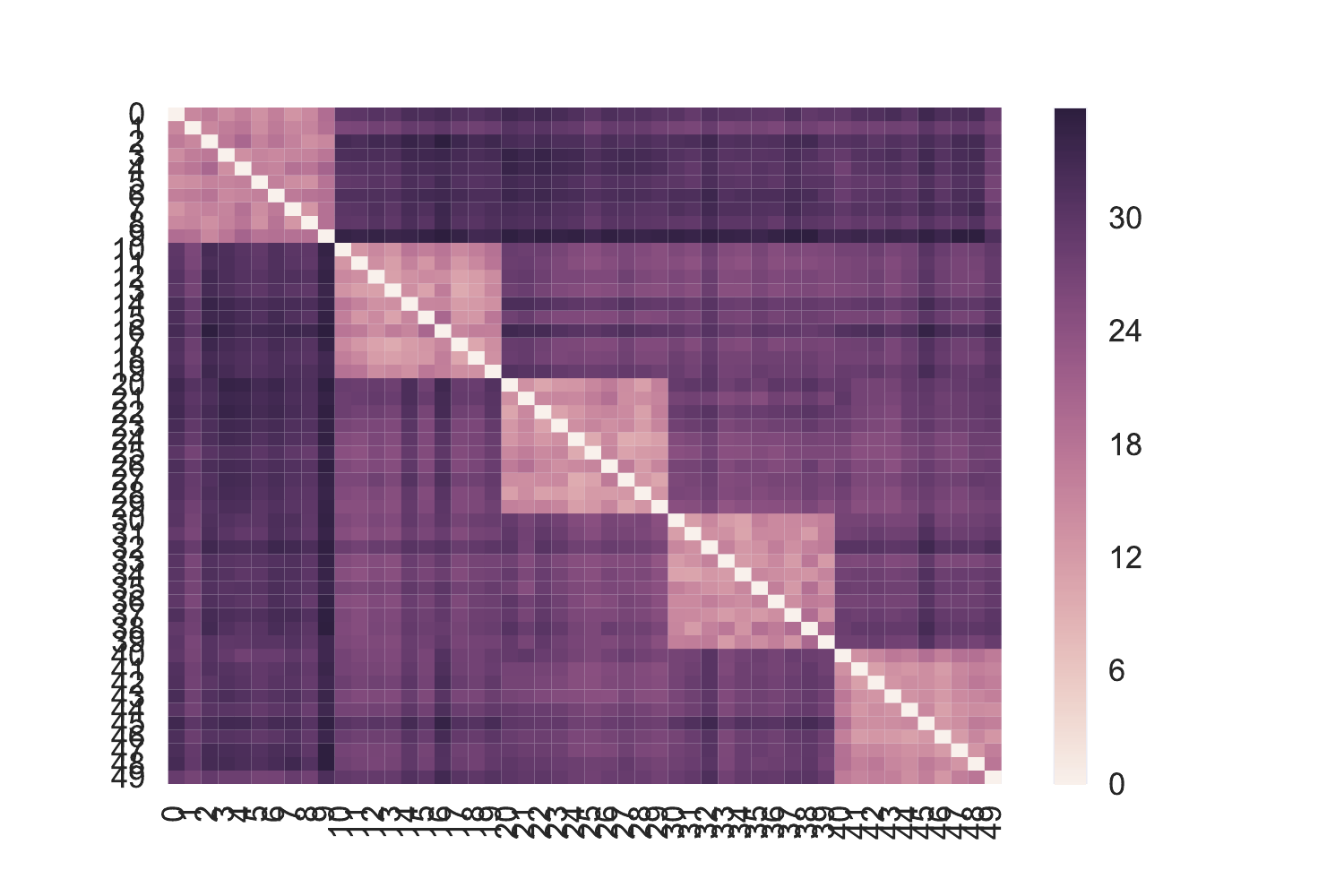}
    \caption{Distances between adjacency matrices for 5 scans across 3 different subjects.}
    \label{fig:adj_examples}
    \end{subfigure}
    \caption{A few visualization of the data}\label{fig:data_vis}
\end{figure}

Fig. \ref{fig:data_vis} shows the raw fMRI time series of the data, as well as the distance matrix between a few adjacency matrices. As indicated by the block diagonal structure in Fig. \ref{fig:adj_examples}, adjacency matrices corresponding to the anatomical white matter connections between brain regions exhibit a strong subject effect: structural connectomes are much similar within each subject.

\begin{figure}[H]
    \centering
    \begin{subfigure}{0.49\textwidth}
    \includegraphics[width=\textwidth]{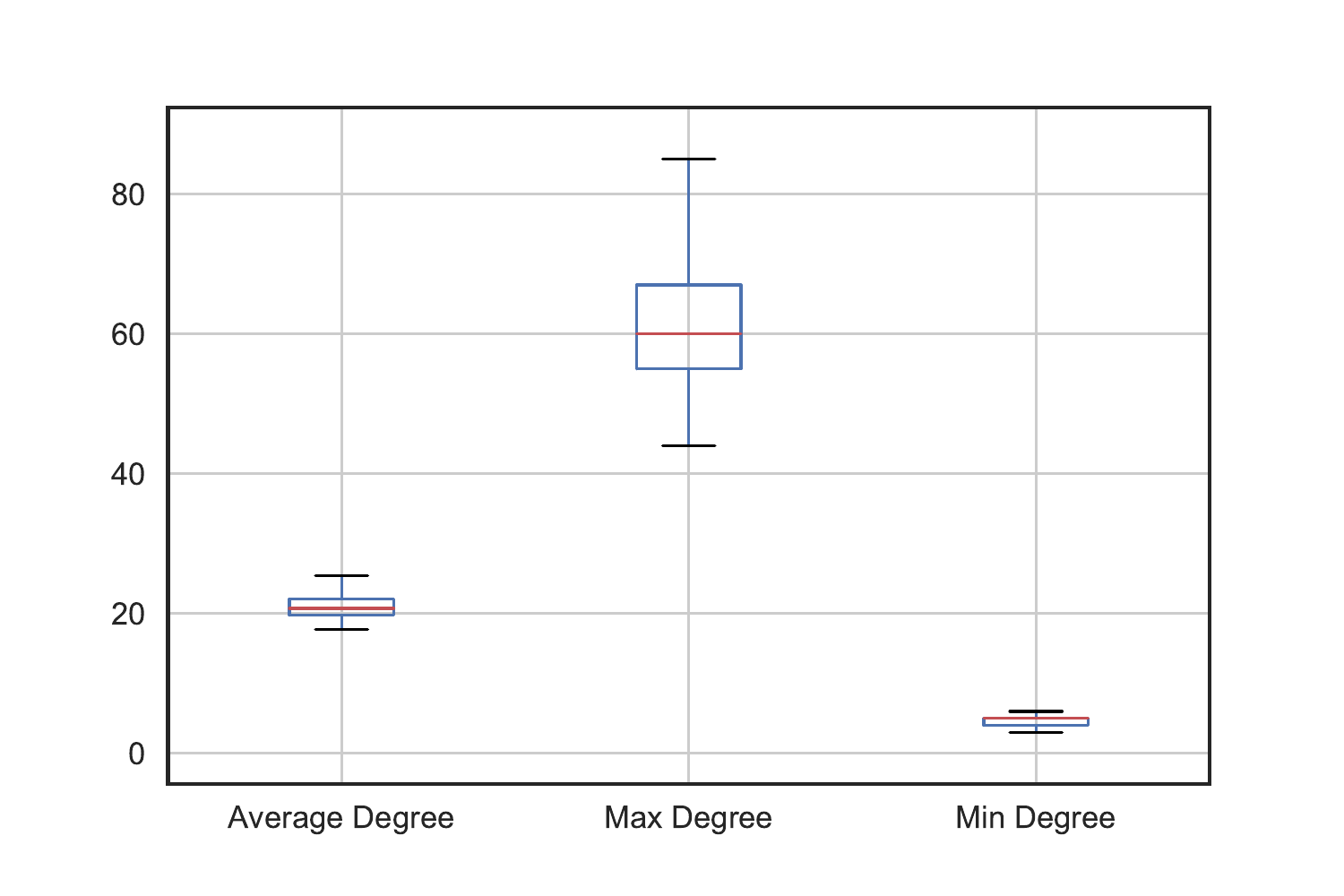}
    \caption{Properties of the node degree distributions}
    \label{fig:TS_examples}
    \end{subfigure}
    \begin{subfigure}{0.49\textwidth}
    \includegraphics[width=\textwidth]{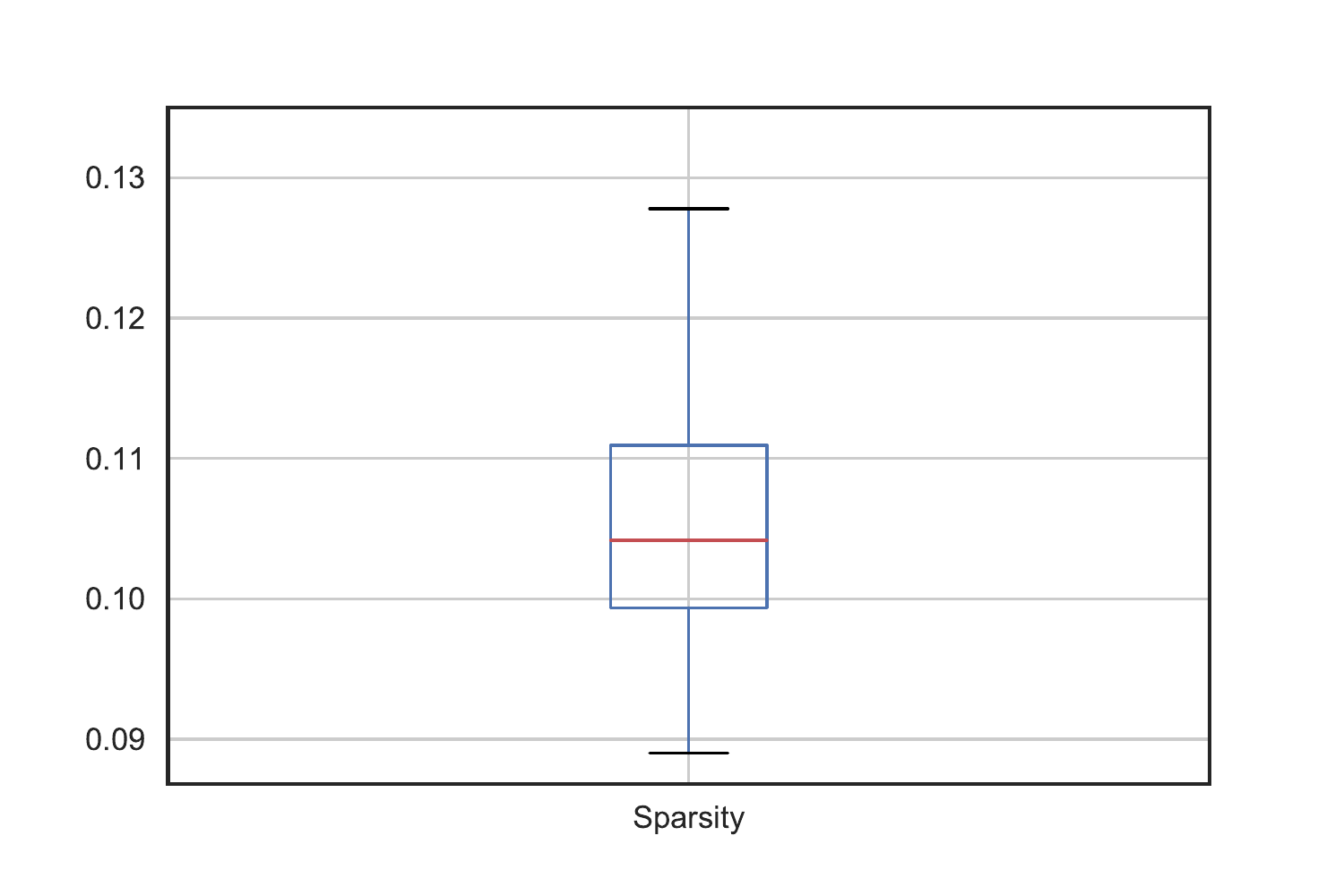}
    \caption{Average sparsity in the structural connectomes}
    \label{fig:adj_examples}
    \end{subfigure}
    \caption{Structural Graph properties}\label{fig:data_vis2}
\end{figure}

Fig. \ref{fig:data_vis2} also highlights some of the properties of the graphs that we have at hand. In particular, the structural sparsity in this dataset (that is, the number of edges in the graph over the total number of possible edges) is 0.113. The structures that we work with are thus relatively sparse.

\subsection{Structural Processing Pipeline: NDMG}\label{appendix:structural}
The structural connectomes that we use in this paper were obtained from the \href{https://neurodata.io/}{neurodata}'s website and have been processed through \textbf{NDMG}. As per the package's \href{https://neurodata.io/ndmg/}{website}, the Python pipeline \textbf{NDMG} \citep{kiar2017comprehensive}  provides end-to-end ``robust and reliable estimates of MRI connectivity at 1mm resolution [..] from 48 to 72,000 nodes, all in MNI152 standard space." The details can be found in the associated paper \citep{kiar2017comprehensive}, but, for the sake of clarity and completeness, we summarize here the four key components of this pipeline.

\xhdr{(1) Registration}  \textbf{NDMG} begins by using FSL to perform  a series of linear ``standard'' registrations on  minimally preprocessed DWI and T1W images and aligns them to the MNI152 atlas. 

\xhdr{(2) Tensor Estimation} As per \cite{kiar2017comprehensive}, the MNI152-aligned ``diffusion volumes and b-values/b-vectors files are transformed into a 6-dimensional tensor volume. A fractional anisotropy map of the tensors is provided for QA, again using multiple depths in each of the three canonical image planes.''

\xhdr{(3) Tractography} As per \cite{kiar2017comprehensive},  a deterministic tractography algorithm (Dipy’s EuDX \citep{garyfallidis2012quickbundles}) is used to generate streamlines. Each voxel at the boundary of the brain mask ``is used as a seed-point in EuDX and fibers are produced and then pruned based on their length".

\xhdr{(4)  Graph Generation}  Fibers are traced through predefined parcellations and ``an undirected edge is added to the graph for every pair of regions along the path, where the weight of the edge is the cumulative number of fibers between two regions"\citep{kiar2017comprehensive}.

Here, consistently with the choice of the parcellation for the fMRI pre-processing, we have chosen the structural connectomes for the Craddock 200 atlas.